\documentclass[final,aps,prd,twocolumn,showpacs,eqsecnum,amsmath,amssymb,amsfonts,english,twoside]{revtex4}

\usepackage{babel}
\usepackage{graphicx}

\begin{document}


\title{Fields of accelerated sources: Born in de~Sitter\footnote{%
Published in J. Math. Phys. \textbf{46}, 102504 (2005).\\
This version differs only by a more compact formatting.}}

\author{Ji\v{r}\'{\i} Bi\v{c}\'ak}
\email[]{bicak@mbox.troja.mff.cuni.cz}

\author{Pavel Krtou\v{s}}
\email[]{Pavel.Krtous@mff.cuni.cz}

\affiliation{%
  Institute of Theoretical Physics,
  Charles University,\\
  V Hole\v{s}ovi\v{c}k\'{a}ch 2, 180 00 Prague 8,
  Czech Republic\\
  }
\affiliation{%
  Max-Planck Institute for Gravitational Physics (Albert Einstein Institute),\\
  14476 Golm,
  Germany
  }


\received{8 June 2005}
\accepted{6 July 2005}
\published{27 October 2005}

\begin{abstract}
This paper deals thoroughly with the
scalar and electromagnetic fields of uniformly accelerated
charges in de~Sitter spacetime. It gives details and makes
various extensions of our Physical Review Letter from
2002. The basic properties of the classical Born solutions
representing two uniformly accelerated charges in flat
spacetime are first summarized. The worldlines of
uniformly accelerated particles in de~Sitter universe are
defined and described in a number of coordinate frames,
some of them being of cosmological significance, the other are
tied naturally to the particles. The scalar and
electromagnetic fields due to the accelerated charges are
constructed by using conformal relations between Minkowski
and de~Sitter space. The properties of the generalized
\vague{cosmological} Born solutions are analyzed and
elucidated in various coordinate systems. In particular, a
limiting procedure is demonstrated which brings the
cosmological Born fields in de~Sitter space back to the
classical Born solutions in Minkowski space. In an
extensive Appendix, which can be used independently of the
main text, nine families of coordinate systems in de~Sitter
spacetime are described analytically and illustrated
graphically in a number of conformal diagrams.
\end{abstract}



\maketitle

\topmargin-24pt
\pagestyle{jmpheadings}
\markboth%
{J. Bi\v{c}\'ak and P. Krtou\v{s}\hfill J. Math. Phys. \textbf{46}, 102504 (2005) [reformatted]}%
{Fields of accelerated sources: Born in de~Sitter\hfill J. Math. Phys. \textbf{46}, 102504 (2005) [reformatted]}

\section{Introduction}
\label{sc:Intro}

In 1969, on the sixtieth anniversary of Max Born's \cite{Born:1909} first
analysis of the field of a uniformly accelerated charge, Ginzburg,
Nobelist in 2003, reanalyzed \cite{Ginzburg:1970,Ginzburg:book1979,Ginzburg:book1989} this---what he
called---``perpetual problem of classical physics,'' with the conclusion that the
problem ``is already clear enough not to be regarded as perpetual.''
Ginzburg confirmed the presence of radiation and emphasized that the vanishing
of the radiation reaction force during the uniformly accelerated motion of the
charge ``is in no way paradoxical, in spite of the presence of
radiation,'' since ``a non-zero total energy flux through a surface
surrounding a charge at a zero radiation force is exactly equal to the
decrease of the field energy in the volume enclosed by this surface.'' Despite
Ginzburg's view, however, the problem does not seem to lose its
\vague{perpetuity.} A number of distinguished physicists who dealt with it
before Ginzburg like Sommerfeld, Schott, von Laue, Pauli and others have,
after Ginzburg, been followed by such authors as, for example, Bondi
\cite{Bondi:1981}, Boulware \cite{Boulware:1980}, Peierls \cite{Peierls:1979},
Thirring \cite{Thirring:book2} and others
\cite{Herrera:1983,HarpazSoker:1998,GuptaPadmanabhan:1998,ShariatiKhorrami:1999}.

The fields and radiation patterns from uniformly accelerated general multipole
particles were also studied \cite{BicakMuschall:1990}. The December 2000 issue
of Annals of Physics contains three papers by Eriksen and Gr{\o}n
\cite{EriksenGron:2000,EriksenGron:2000b,EriksenGron:2000c}
with numerous references on \vague{electrodynamics of
hyperbolically accelerated charges}.
(Yet, except for \cite{Born:1909} and \cite{Boulware:1980}, the explicit
citations above are not contained in \cite{EriksenGron:2000,EriksenGron:2000b,EriksenGron:2000c}.)

Spacetimes describing \vague{uniformly accelerated particles or black holes} play
fundamental role in general relativity. They are the only explicit solutions
of Einstein's field equations known which are radiative and represent the
fields of finite sources. Born fields in electrodynamics are
produced by two charges moving along an \vague{axis of symmetry} in opposite
directions with uniform accelerations of the same magnitude.
They have two symmetries: they are axially
symmetric and symmetric with respect to the boosts along the
axis of symmetry.
Their general-relativistic counterparts, the boost rotation symmetric spacetimes,
are unique because of a theorem which roughly states that in
axially symmetric, locally asymptotically flat spacetimes the only additional
symmetry that does not exclude radiation is the boost symmetry. The
boost-rotation symmetric spacetimes have been used in gravitational radiation
theory, quantum gravity, and as test beds in numerical relativity; their
general structure is described in \cite{BicakSchmidt:1989}, their applications
and new references are given in the reviews
\cite{Bicak:inEhlers,BicakKrtous:2003,Pravdovi:2000}.
One of the best known examples, the so-called C-metric, describing uniformly accelerated
black holes, is the only boost-rotation symmetric solution known also for a
nonvanishing cosmological constant $\Lambda$.
Asymptotically this \vague{generalized} C-metric approaches
de~Sitter spacetime if ${\Lambda > 0}$. It is well-known
from the classical work of Penrose \cite{Penrose:1965} on the asymptotic
properties of fields and spacetimes that, in contrast to asymptotically
Minkowskian spacetimes with null (lightlike) conformal infinities ${\scri^\pm}$,
asymptotically de~Sitter vacuum spacetimes have two disjoint conformal
infinities, past and future, which are both \emph{spacelike}.
When ${\Lambda < 0}$, as in anti-de~Sitter space, the conformal infinity is timelike,
and it is not disjoint. (In the analytically extended C-metrics, there is an
infinite number of such infinities which can be reached by going
\vague{through} black holes like with a Reissner-Nordtr\"om black hole, but this is
not pertinent to the present work.)

The importance of de~Sitter spacetime in the history of modern cosmology seems
to grow steadily. The \vague{flat} de~Sitter universe became the standard
cosmological model in steady state theory, more recently, as the \vague{first
approximation} of inflationary models, and today, with indications that
${\Lambda > 0}$ in our Universe, it is an asymptote of all indefinitely
expanding Friedmann-Robertson-Walker models with ${\Lambda > 0}$. In fact much
more general cosmological models with $\Lambda > 0$ approach de~Sitter model
asymptotically in time. This manifestation of the validity of the
\vague{cosmic no-hair conjecture} \cite{Maeda:1989}, \cite{Rendall:2004} will
also be noticed in the properties of the fields analyzed in this work.

Motivated by the role of the Born solution in classical electrodynamics, by the
importance of the boost-rotation symmetric spacetimes in general relativity,
and by the relevance of de~Sitter space in contemporary cosmology, we have
recently generalized the Born solution for scalar and electromagnetic fields to
the case of two charges uniformly accelerated in de~Sitter universe
\cite{BicakKrtous:2001}. In the present paper we give calculations and
detailed proofs of the results and statements briefly sketched  in our paper
\cite{BicakKrtous:2002}. In addition, we investigate the character of the
field in a number of various coordinate systems which are relevant either in a
general-relativistic context or from a cosmological perspective.

The appropriate coordinates and corresponding tetrad fields were important in
finding our recent results on a general asymptotic behavior of fields in
the neighborhood  of future infinity ${\scri^+}$ in asymptotically de~Sitter
spacetimes \cite{KrtousPodolskyBicak:2003}. In obtaining these results
we were inspired by the inspection of the electromagnetic fields from
uniformly accelerated charges in de~Sitter universe.

It was known from the work of Penrose  since late 1960's that the radiation
field is \vague{less invariantly} defined when ${\scri^+}$ is
spacelike---that it depends on the direction in which ${\scri^+}$ is approached.
However, no explicit models were available. The investigation of the test
fields of accelerated charges in de~Sitter universe has served as a useful
example; it was then generalized also to the study of asymptotic and radiative
properties of the C-metric with ${\Lambda > 0}$ \cite{KrtousPodolsky:2003}, as
well as to the case of the C-metric with ${\Lambda < 0}$ when infinity is
timelike \cite{PodolskyOrtaggioKrtous:2003}. (For other recent works on the
\vague{cosmological} C-metric, see, e.g., \cite{DiasLemos:2003b,DiasLemos:2003a}.)
These studies led to more general conclusions
\cite{KrtousPodolskyBicak:2003}: the directional pattern of gravitational and
electromagnetic radiation near de~Sitter-like conformal infinity has a
universal character, determined by the algebraic (Petrov) type of a solution
of the Maxwell/Einstein equations considered. In particular, the radiation
field vanishes along directions opposite to principal null directions. Very
recently analogous conclusions have been obtained for spacetimes with
anti-de~Sitter asymptotics \cite{KrtousPodolsky:2004a}.

Since past and future infinities are spacelike in de~Sitter spacetime, there
exist particle and event horizons. Under the presence of the horizons, purely
retarded fields (appropriately defined) become singular or even cannot be
constructed at the \vague{creation light cones}, i.e., at future light cones of
the \vague{points} at ${\scri^-}$ at which the sources \vague{enter} the
universe. In \cite{BicakKrtous:2001} we analyzed this phenomenon in detail and
constructed smooth (outside the sources) fields involving both retarded and
advanced effects. As demonstrated in \cite{BicakKrtous:2001}, to be
\vague{born in de~Sitter} is quite a different matter than to be \vague{born
in Minkowski}. This reveals the double meaning of the second---perhaps
somewhat enigmatic---part of the title of this paper.

Its plan is as follows. In order to gain an understanding of the generalized
Born solution in de~Sitter space it is advantageous to be familiar with some
details of the classical Born solution in Minkowski space. Hence, its
properties most relevant for our purpose are summarized in Section~\ref{sc:BornM}. Here we
also discuss why in Minkowski space problems with purely retarded fields of
uniformly accelerated particles do not arise.

There exists vast literature on
de~Sitter space in which various types of coordinates are employed. We shall
construct fields in de~Sitter space by using its conformal relations to
Minkowski space. For our aim coordinate systems on conformally compactified
spaces and their properties will be particularly useful. These, together with
several \vague{cosmological} and \vague{static} coordinate systems, will be
described and graphically illustrated in conformal diagrams in Section~\ref{sc:Coor}.
What is meant by \vague{uniformly accelerated particles in de~Sitter space} is
defined and the properties of the corresponding worldlines are studied in
Section~\ref{sc:AccPart}. For technical reasons it is more advantageous to consider
particles which asymptotically start and end at the poles of coordinates
covering de~Sitter space, i.e., particles \vague{born at the poles} (Section~\ref{ssc:patpoles}). In order
to find a direct relation between the standard form of the Born solution
produced by two charges at each time located symmetrically with respect to the
origin of Minkowski space and the generalized Born solution in de~Sitter space,
it is necessary to construct also worldlines of uniformly accelerated
particles which are \vague{born at the equator} (Section~\ref{ssc:patequat}).

With the worldlines of accelerated particles available, it is advantageous to
consider coordinates in de~Sitter space which are centered on these
worldlines. These \vague{accelerated coordinates} and \vague{Robinson-Trautman
coordinates} are obtained, in a constructive manner, in Section~\ref{sc:AccCoor}.

Section~\ref{sc:Fields} is devoted to the fields from particles \vague{born at the
poles}. Here we also study in detail their properties in various coordinate
systems introduced before. The fields of particles \vague{born at the equator}
are found in Section~\ref{sc:BorndS} by a simple rotation. Starting from these fields we
demonstrate by means of which limiting procedure the standard Born field in
Minkowski space can be regained. Finally, we conclude by few remarks in
Section~\ref{sc:Concl}.

The paper contains a rather extensive Appendix in which nine families of
coordinate systems employed in the main text are described in detail,
illustrated graphically, their relations are given, and corresponding metric
forms as well as orthonormal tetrads are presented. We believe the Appendix
can be used as a general-purpose catalogue in other studies of physics in de~Sitter spacetime.

\section{Born in Minkowski}
\label{sc:BornM}

\begin{figure*}
\includegraphics{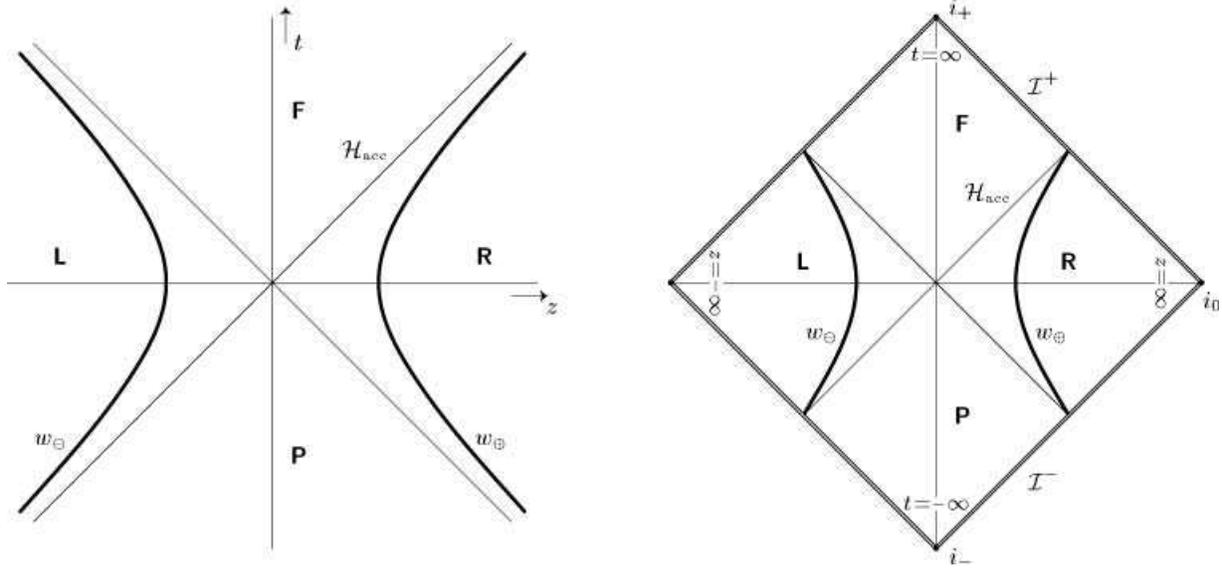}
\caption{\label{fig:Mink}%
A pair of uniformly accelerated charges in Minkowski spacetime
(with the conformal diagram on the right).
The boost Killing vector is timelike in
regions ${\Ldom}$ and ${\Rdom}$;
it is spacelike in ${\Fdom}$ and ${\Pdom}$.
The charges are causally disconnected
by null hypersurfaces (\vague{the roof})
${-\MOt^2+\MOz^2=0}$. These hypersurfaces
represent the acceleration horizon for
uniformly accelerated observers with respect
to which the charges are at rest.
}
\end{figure*}

It was Einstein in 1908, inspired by a letter from Planck,
who first defined a \defterm{uniformly accelerated} motion
in special relativity \cite{Einstein:1907,Einstein:1908}. A particle is in
uniformly accelerated motion if its acceleration has a fixed
constant value in instantaneous rest frames of the particle.
This can be stated in a covariant form (see,
e.g., \cite{Rohrlich:book}) as
\begin{equation}\label{unacc}
  \tens{P}^\alpha_\mu\, \dot{\tens{a}}^\mu =
  \dot{\tens{a}}^\alpha - (\tens{a}^\mu\, \tens{a}_\mu)\,
  \tens{u}^\alpha = 0\commae
\end{equation}
$\tens{u}^\alpha$ being four-velocity,
${\dot{}\, \equiv \tens{u}^\mu\covd_\mu}$ covariant
derivative with respect to  proper time,
${\tens{a}^\alpha = \dot{\tens{u}}^\alpha}$ four-acceleration, and
${\tens{P}^\alpha_\mu=\tens{\delta}^\alpha_\mu+\tens{u}^\alpha\, \tens{u}_\mu}$ is the
projection tensor into the hypersurface orthogonal to
$\tens{u}^\alpha$. Eq.~\eqref{unacc} implies
${\dot{\tens{a}}^\mu \tens{a}_\mu=0}$ so that the condition of uniform
acceleration guarantees that the magnitude of the
four-acceleration is constant,
\begin{equation}
   a_\Mink = \sqrt{\tens{a}^\mu\, \tens{a}_\mu} = \text{constant},
\end{equation}
although ${\dot{\tens{a}}^\mu\neq0}$. Integrating Eq.~\eqref{unacc} in Minkowski spacetime,
one finds that the worldline of a uniformly accelerated
particle is a hyperbola \cite{Hill:1945,Hill:1947}. One
can then choose an inertial frame, in which the initial
three-velocity  and three-acceleration are parallel; in such frames
the motion is spatially 1-dimensional.
It can be produced by putting a
test charged particle into a homogenous electric field
with initial velocity aligned with the field.
The motion along the $\MOz$~axis
is illustrated in Fig.~\ref{fig:Mink}. There, in fact,
\emph{two} particles uniformly accelerated in opposite
directions are shown, the one moving along the positive
(${\pmpart=+1}$ for particle $w_\ppart$ in the figure)
and the second one along the negative $\MOz$~axis
(${\pmpart=-1}$ for particle ${w_\mpart}$); their worldlines parametrized by
proper time $\prt_\Mink$ are
\begin{equation}\label{wlMink}
 \MOz=\pmpart\,\Bo\cosh\frac{\prt_\Mink}\Bo\comma
 \MOt=\Bo\sinh\frac{\prt_\Mink}\Bo\comma
 \MOx=\MOy=0\commae
\end{equation}
or
\begin{equation}\label{wlMinkzt}
\MOz=\pmpart\sqrt{\MOt^2+\Bo^2}\period
\end{equation}
Here we have chosen the particles to be at rest at ${\MOz=\pmpart\Bo}$
at ${\MOt=0}$. Then their three-acceleration at initial moment ${\MOt=0}$ is
${a_\Mink = \abs{{d^2 \MOz}/{d\MOt^2}} = 1/\Bo}$. As
${\MOt\to\infty}$, the three-velocity
${v_\Mink=\abs{d\MOz/d\MOt}=\MOt/\sqrt{\MOt^2+\Bo^2}}$
approaches the velocity of light. This is the well-known
\defterm{hyperbolic motion}.

The worldlines of the particles coincide with the orbits of the
boost Killing vector in the $\MOt\textdash\MOz$ plane,
\begin{equation}\label{KillVec}
   \tens{\xi}_{\txtlab{boost}}=\MOz\cv{\MOt}+\MOt\cv{\MOz}\period
\end{equation}
These orbits, given by ${-\MOt^2+\MOz^2=\text{constant}}$,
${\MOx,\,\MOy=\text{constant}}$, are timelike
at ${-\MOt^2+\MOz^2>0}$, but they are
spacelike at ${-\MOt^2+\MOz^2<0}$. The fields (scalar,
electromagnetic, higher-spin) produced by
charged particles in the hyperbolic motion
will have boost-rotational symmetry. They
are thus static in the region ${-\MOt^2+\MOz^2>0}$---\vague{below the roof}
as introduced in \cite{BicakSchmidt:1989}, however, we can
expect them to be radiative in the region ${-\MOt^2+\MOz^2<0}$---\vague{above the
roof}.

Consider a massless scalar field $\SF$ with the scalar
charge source $\SFsource$ satisfying, in a general
4-dimensional spacetime, the wave equation
\begin{equation}
  \bigl[\dalamb-{\textstyle\frac16}\sccurv\bigr]\,\SF=\SFsource\commae
\end{equation}
in which ${\dalamb\equiv\mtrc^{\mu\nu}\covd_\mu\covd_\nu}$ is
the curved-space d'Alambertian, and $\sccurv$ is the scalar
curvature (of course, in Minkowski space $\sccurv=0$).
We are interested in a field due to two monopole particles
with the same constant scalar charge of magnitude $\SFcharge$ moving
along hyperbolae \eqref{wlMink}. The source at a spacetime point $x$
is thus given by
\begin{equation}\label{SFsource}
  \SFsource = \SFsource_\ppart + \SFsource_\mpart\comma
  \SFsource_\pmpart =
  \SFcharge\int\deltafc(x-w_\pmpart(\prt_\Mink))\,d\prt_\Mink\commae
\end{equation}
where $w_\pmpart(\prt_\Mink)$ denotes the worldlines of the particles.
The resulting fields may be written as
\begin{equation}
  \SF= \SF_\ppart + \SF_\mpart\commae
\end{equation}
where $\SF_\pmpart$ is produced by $\SFsource_\pmpart$. The
retarded and advanced fields of these sources
are constructed and analyzed in detail in
Ref.~\cite{BicakSchmidt:1989}.
It can be demonstrated that the retarded and advanced fields
due to the particle $w_\ppart$ or $w_\mpart$
are all given by exactly \emph{identical}
expression
\begin{equation}\label{SFsymM}
   \SF_\MBorn = \frac{\SFcharge}{4\pi}\frac1\Rfact\commae
\end{equation}
which, however, is valid in different regions of spacetime.
Namely,
\begin{equation}\label{SFraM}
   \SF_{\ret/\adv\,\pmpart} =
   \frac{\SFcharge}{4\pi}\frac1\Rfact\;
   \stepfc(\pmpart\MOz\pm\MOt)\commae
\end{equation}
$\stepfc$ being the step function and upper/lower sign is valid
for retarded/advanced case.
The quantity $\Rfact$ in the denominator is given by
\begin{equation}\label{Rfact}
  \Rfact=\frac1{2\Bo}\Bigl(\bigl(\Bo^2+\MOt^2-\MOr^2\bigr)^2
  +4\Bo^2\MOr^2\sin^2\tht\Bigr)^{\frac12}\period
\end{equation}
It has the meaning of a \defterm{retarded} or
\defterm{advanced distance}---it is a spatial distance of the
\vague{observation} (field) point
from the position of the source at retarded or advanced
time. Here, as usual,
$\MOx=\MOr\sin\tht\cos\ph$,
$\MOy=\MOr\sin\tht\sin\ph$,
$\MOz=\MOr\cos\tht$.
The fields \eqref{SFsymM}, as well as \eqref{SFraM}, are,
at first glance, axially (rotationally) symmetric.
They are also unchanged under the boost along the $\MOz$~axis.

The field $\SF_\MBorn$ can, in fact, be viewed as the field
due to both accelerated particles, i.e., as the field
corresponding to the source \eqref{SFsource}.
Inspecting regions at which the retarded and advanced
fields \eqref{SFraM} are
non-vanishing we discover that $\SF_\MBorn$
admits the interpretation as arising from 1-parametric
combination of retarded and advanced effects
from both particles:
\begin{equation}\label{racomb}
  \SF_\MBorn = \xi\SF_{\ret\ppart} + (1-\xi)\SF_{\adv\ppart}
  +(1-\xi)\SF_{\ret\mpart} + \xi\SF_{\adv\mpart}\commae
\end{equation}
where $\xi\in\realn$ is an arbitrary constant parameter. In
particular, choosing ${\xi=\frac12}$, the field $\SF_\MBorn$ arises
from ${\frac12(\SF_\ret+\SF_\adv)}$ from both particles. With
${\xi=1}$, the field can be interpreted as being caused by
purely retarded effects from particle $w_\ppart$ in region
${\MOz+\MOt>0}$, and by purely advanced effects from particle
$w_\mpart$ in region ${\MOz+\MOt<0}$.

The case of electrodynamics is very similar. The solution
corresponding to the scalar field \eqref{SFsymM} was found
by Born in 1909 \cite{Born:1909}.
It is customarily given in cylindrical coordinates (see, e.g.,
\cite{Rohrlich:book,FultonRohrlich:1960,EriksenGron:2000}),
however, in order to compare it with its generalization to
de~Sitter universe, it
is more convenient to write it down in spherical coordinates:
\begin{equation}\label{EMBornM}
\begin{split}
  \EMF_\MBorn =
  -\frac{\EMcharge}{4\pi}&\frac{1}{2\Bo}\,\frac{1}{\Rfact^3}\\
  \times\Bigl(\!
  & -(\Bo^2+\MOt^2-\MOr^2)\,\cos\tht\,\grad\MOt\wedge\grad\MOr\\
  & +(\Bo^2+\MOt^2+\MOr^2)\,\MOr\sin\tht\,\grad\MOt\wedge\grad\tht\\
  & -2\, \MOt\,\MOr^2\sin\tht\,\grad\MOr\wedge\grad\tht
  \Bigr)\period
\end{split}
\end{equation}
The field can be obtained from the Li\'{e}nard-Wiechert retarded
and advanced potentials of two charged
particles moving along hyperbolae \eqref{wlMink},
however, in contrast to the scalar case
when charges are exactly the same, the electric charges
have \emph{opposite} signs.
Similarly to the scalar case, the field is smooth
everywhere, except for the places where the particles occur.
$\EMF_\MBorn$ can be interpreted in the
precisely same way as the scalar field \eqref{SFsymM}, i.e.,
as the 1-parametric combination of retarded and advanced
effects from both charges, analogously to
Eq.~\eqref{racomb}. However, in the electromagnetic case
an exact form of retarded and advanced fields from a single
particle is a more subtle issue.
Considering that the field in the region ${\MOz+\MOt>0}$ may
be interpreted as the retarded effect emitted from the
charge which moves along ${\MOz>0}$, it is natural to try to
exclude advanced effects of the other particle by requiring
the field to vanish in the region ${\MOz+\MOt<0}$
(cf. Fig.~\ref{fig:Mink}). The field is then not smooth
at the null hypersurface ${\MOz=-\MOt}$. In the scalar case
such a field \emph{does} represent the pure retarded field
of the single particle, cf. Eq.~\eqref{SFraM}. However, in
the electromagnetic case the field
${\EMF_\MBorn\,\stepfc(\MOz+\MOt)}$ corresponds
to sources consisting not only of the particle but also of a
\vague{charged wall} moving along hypersurface
${\MOz+\MOt=0}$ with velocity of light
\cite{LeibovitzPeres:1963,Bondi:1981}. Nevertheless, it is possible to
obtain a pure retarded field of the only single particle by
modifying the field with a delta function valued term
localized on ${\MOz+\MOt=0}$ \cite{Boulware:1980,BondiGold:1955,Krtous:dipl}.

In de~Sitter space such a
modification is not feasible because
the advanced fields cannot be excluded. The underlying cause
is the \emph{null} character of the past conformal infinity in
Minkowski spacetime, whereas in de~Sitter spacetime both
future and past conformal infinities are \emph{spacelike}.
As a consequence, the Gauss constraint restricts the data at the spacelike past
infinity, and it can be shown that
a purely retarded field of a point-like charge cannot satisfy
this constraint \cite{BicakKrtous:2001}. The absence of purely retarded fields
is also related to a different character
of the past horizon of a particle.
Since the worldline of a particle
\vague{enters} the universe through the past spacelike infinity,
there exists the past particle horizon,
called also the \defterm{creation light cone}.
In de~Sitter space a purely retarded electromagnetic
field of a point-like charge cannot be constructed on the whole cone.
In Minkowski spacetime
the creation light cone of a particle moving asymptotically in
the past freely, coincides with the whole past null infinity,
and thus it does not belong to the physical spacetime.
Eternally accelerated particles can \vague{enter} the
Minkowski spacetime at a point of the past null infinity---as,
for example, uniformly accelerated particles do.
Like in de~Sitter case, in conformal spacetime the past horizon of such particles
forms the null cone but, in contrast to de~Sitter space,
it has one generator in common with the null infinity.
In physical spacetime this horizon thus
corresponds to a null hyperplane---for the particle $w_\ppart$ it is just the
hyperplane ${\MOz+\MOt=0}$ (cf.\ Fig.~\ref{fig:Mink})---and so
its spatial sections are not compact.
Thanks to this non-compactness the \vague{bad} behavior of
the retarded field on the horizon can be \vague{pushed out
of sight} to the infinity.
We analyzed this issue in detail in
Ref.~\cite{BicakKrtous:2001}.

\section{Many faces of de~Sitter}
\label{sc:Coor}

The fields due to various types of uniformly accelerated sources
in de~Sitter spacetime found in \cite{BicakKrtous:2001}, as well as
those described briefly in Ref.~\cite{BicakKrtous:2002},
were constructed by employing the conformal
relation between Minkowski and de~Sitter spacetimes. When analyzing the
worldlines of the sources in de~Sitter spacetime and their relation to
the corresponding worldlines in Minkowski spacetime we need to
introduce appropriate coordinate systems. Suitable coordinates will
later be used to exhibit various properties of the fields.
An extensive literature exists on various types of
coordinates in de~Sitter space (e.g. \cite{Schmidt:1993,EriksenGron:1995}),
but we want to survey some of them in this section. In particular, we relate
them to the corresponding coordinates on conformally related
Minkowski spaces since this does not appear to be given elsewhere. In
the next section, after identifying the worldlines of uniformly
accelerated particles in de~Sitter space, we shall construct new
coordinate systems tied to such particles, such as Rindler-type
\vague{accelerated} coordinates, or Robinson-Trautman-type coordinates
in which the null cones emanating from the particles have especially
simple forms. These coordinate systems will turn out to be very useful in
analyzing the fields.
Here, in the main text, however, only a brief description of relevant
coordinates will be given. More details, including both formulas
and illustrations, are relegated to the Appendix.

As it is well-known from textbooks on general relativity (for a
recent pedagogical exposition, see \cite{Rindler:book}), de~Sitter
spacetime, which is the solution of Einstein vacuum equations with a
cosmological term ${\Lambda>0}$, is best visualized as the
4-dimensional hyperboloid imbedded in flat 5-dimensional Minkowski
space. It is the homogeneous space of constant curvature equal to
$4\Lambda$. Hereafter, we use the quantity
\begin{equation}
   \DSr = \sqrt{\frac3\Lambda}
\end{equation}
(with the dimension of length) to parametrize the radius of the curvature.

\begin{figure}
\includegraphics{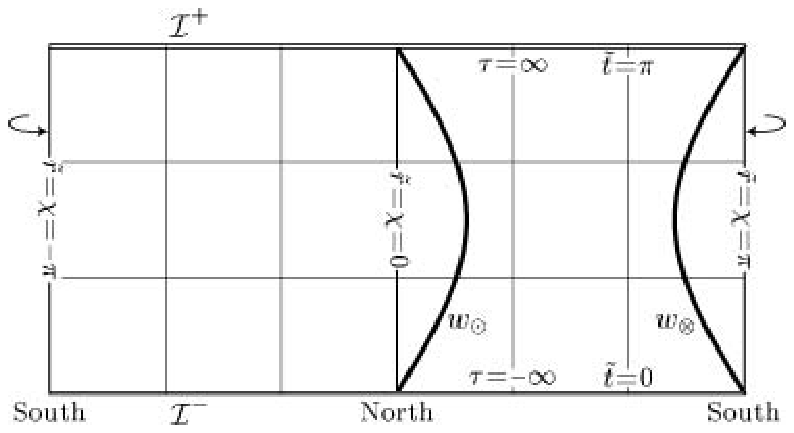}\\[24pt]
\includegraphics{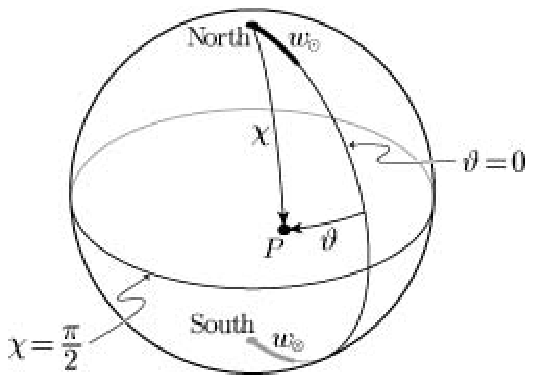}\vspace{12pt}
\caption{\label{fig:dS}%
The spherical cosmological coordinates and a pair of uniformly
accelerated particles ${w_\npart}$ and ${w_\spart}$ in de~Sitter universe:
the conformal diagram (above) and projection on the spacelike cut ${\spt=\text{constant}}$
in the standard cosmological spherical coordinates (angle $\ph$ suppressed).
The whole de~Sitter spacetime could be represented by just the \vague{right half}
of the conformal diagram. For convenience, we admit negative values of
radial coordinates and identify
${\tlr=\spr=-\pi}$  and ${\tlr=\spr=\pi}$
(see the text below Eq.~\eqref{CEinsFact} and the Appendix).
}
\end{figure}

The entire de~Sitter spacetime can be covered by a single coordinate
system---which we call \defterm{standard coordinates}---%
$\spt\in\realn$, $\spr\in(0,\pi)$, $\tht\in(0,\pi)$, $\ph\in(-\pi,\pi)$
in which the metric reads
\begin{gather}
\mtrc_\dS = -\grad\spt\formsq +
\DSr^2 \cosh^2\!\frac\spt\DSr\;
\Bigl(\grad\spr\formsq+\sin^2\!\spr\;\sphmtrc\Bigr)\commae\label{PCSphM}\\
\sphmtrc = \grad\tht\formsq+\sin^2\tht\,\grad\ph\formsq\period
\end{gather}
Clearly, we can imagine the spacetime as the time evolution of a
3-sphere which shrinks from infinite extension at $\spt\to-\infty$ to a
radius $\DSr$, and then expands again in a time-symmetric way.
Hence, we also call ${\spt,\,\spr}$ the  \defterm{spherical cosmological
coordinates}. The coordinate lines are shown in the conformal diagram,
Fig.~\ref{fig:dS}.

\begin{figure}
\includegraphics{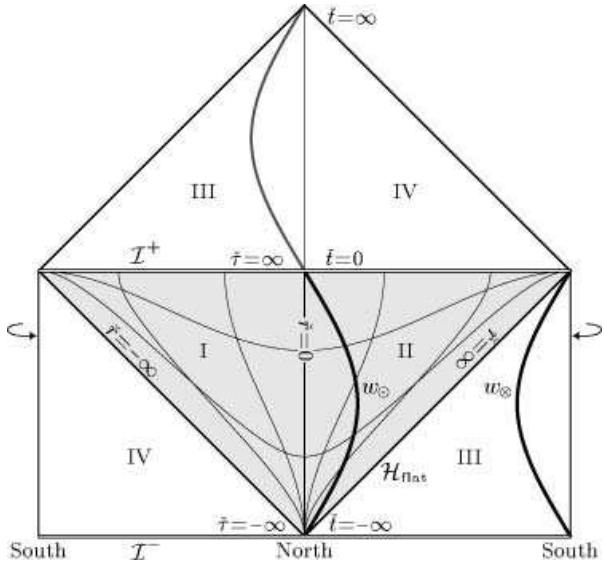}
\caption{\label{fig:dSMinkV}%
The flat cosmological coordinates and particles
${w_\npart}$, ${w_\spart}$ in de~Sitter space and
in conformally related Minkowski space.
The flat cosmological coordinates
cover shaded region. Its boundary,
${\MVr=\pm\infty}$, represents the horizon
for observers at rest in these coordinates.
}
\end{figure}

\begin{figure}
\includegraphics{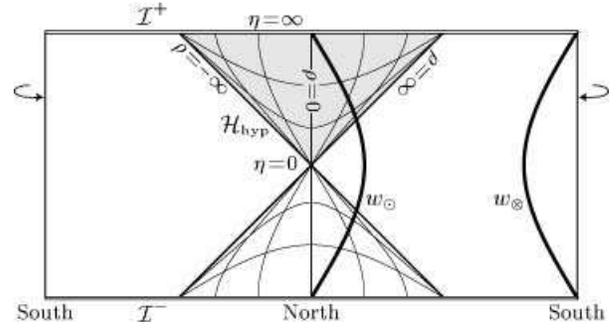}
\caption{\label{fig:dShyp}%
The hyperbolic cosmological coordinates. They cover only the shaded region and,
therefore, only a part of the worldline ${w_\npart}$. The horizon ${\hor_\cfch}$
arises for the observers who are at rest in the hyperbolic cosmological coordinates.
}
\end{figure}

In cosmology the most popular \vague{flat} de~Sitter universe is
obtained by considering only a half of de~Sitter hyperboloid foliated
by flat 3-dimensional spacelike hypersurfaces labeled by timelike
coordinate ${\MVet\in\realn}$, cf.\ Fig.~\ref{fig:dSMinkV}.
Together with appropriate radial coordinate ${\MVr\in\realn^+}$,
the new coordinates, which we call \defterm{flat cosmological
coordinates}, are given in terms of ${\spt,\,\spr}$ by
\begin{equation}\label{PCFlatCoor}
\begin{aligned}
 \MVet&=\DSr\log\Bigl(\sinh\frac\spt\DSr+\cosh\frac\spt\DSr\,\cos\spr\Bigr)\commae\\
 \MVr&=\DSr\frac{\sin\spr}{\cos\spr+\tanh(\spt/\DSr)}\commae
\end{aligned}
\end{equation}
implying the well-known \vague{inflationary} metric
\begin{equation}\label{PCFlatM}
    \mtrc_\dS=-\grad\MVet\formsq
    +\exp{\frac{2\MVet}{\DSr}}\;
    \Bigl(\grad\MVr\formsq+\MVr^2\,\sphmtrc\Bigr)\period
\end{equation}
These coordinates cover only \vague{one-half} of de~Sitter space as indicated
by shading in Fig.~\ref{fig:dSMinkV}.

de~Sitter introduced his model in what we call \defterm{hyperbolic cosmological
coordinates} ${\cht\in\realn}$, ${\chr\in\realn^+}$ (see Fig.~\ref{fig:dShyp})
related to ${\spt,\,\spr}$ by
\begin{equation}\label{PCHypCoor}
\begin{aligned}
 \cosh\frac\cht\DSr&= \cosh\frac\spt\DSr\,\cos\spr  \commae\\
 \tanh\frac\chr\DSr&= \coth\frac\spt\DSr\,\sin\spr  \period
\end{aligned}
\end{equation}
The metric
\begin{equation}\label{PCHypM}
   \mtrc_\dS=-\grad\cht\formsq+\sinh^2\!\frac\cht\DSr\,
   \Bigl(\grad\chr\formsq+\DSr^2\,{\sinh^2\!\frac\chr\DSr}\;\sphmtrc\Bigr)
\end{equation}
shows that the time slices ${\cht=\text{constant}}$ have the geometry
of constant negative curvature, i.e., as the standard time slices in an open FRW universe.

The last commonly used coordinates in de~Sitter spacetime are
\defterm{static coordinates} ${\stt\in\realn}$, ${\str\in(0,\DSr)}$:
\begin{equation}\label{PStCoor}
\begin{aligned}
 \stt&= \frac\DSr2\,\log
 \abs{\frac{\cos\spr+\tanh(\spt/\DSr)}{\cos\spr-\tanh(\spt/\DSr)}}  \commae\\
 \str&= \DSr\cosh\frac\spt\DSr\;\sin\spr \commae
\end{aligned}
\end{equation}
covering also only a part of the universe.
The metric in these coordinates reads
\begin{equation}\label{PStM}
    \mtrc_\dS=-\Bigl(1-\frac{\str^2}{\DSr^2}\Bigr)\,\grad\stt\formsq
          +\Bigl(1-\frac{\str^2}{\DSr^2}\Bigr)^{\!\!-1}\!\grad\str\formsq
          +\str^2\,\sphmtrc\commae
\end{equation}
revealing that $\cvil{\stt}$ is a timelike Killing vector in the region
${0<\str<\DSr}$.

\begin{figure}
\includegraphics{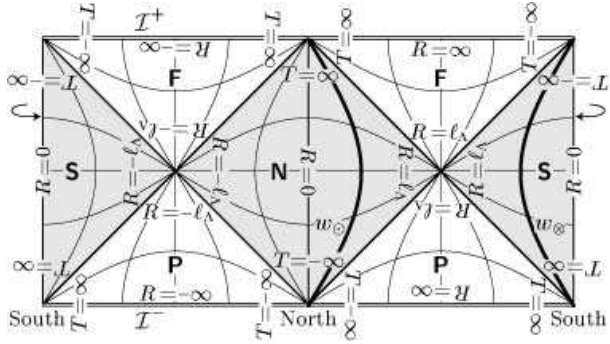}
\caption{\label{fig:dSstat}%
The static coordinates and the wordlines of particles ${w_\npart}$ and
${w_\spart}$. These coordinates can be defined in the whole spacetime,
however several coordinate patches,
in diagram indicated by shaded and nonshaded regions,
have to be used
(cf.\ Appendix \ref{asc:coorStatT} and \ref{asc:coorStatS}).
These regions are separated by the cosmological horizons at ${\str=\DSr}$,
where ${\stt=\pm\infty}$. The vector $\cvil{\stt}$ is a Killing vector
of de~Sitter spacetime. It is timelike in the domains ${\Ndom}$ and ${\Sdom}$
(shaded regions) and spacelike in the domains ${\Fdom}$ and ${\Pdom}$.
The histories of both particles ${w_\npart}$ and ${w_\spart}$ belong
to the domains ${\Ndom}$ and ${\Sdom}$.%
}
\end{figure}

Among the coordinates introduced until now only the standard
coordinates ${\spt,\,\spr,\,\tht,\,\ph}$ cover the whole de~Sitter
spacetime globally. One can easily extend flat cosmological coordinates
to cover (though not smoothly) the whole de~Sitter hyperboloid,
which will be useful in discussion of the
conformally related Minkowski spacetime, cf. Eq.~\eqref{CFlatCoor}.
We shall also use extensions of the static coordinates into
the whole spacetime, using definitions \eqref{PStCoor}, but allowing
$\str\in\realn^+$. In regions where $\str>\DSr$ coordinates $\stt$ and
$\str$ interchange their character, $\cvil{\stt}$ becomes a  spacelike
Killing vector (analogously to $\cvil{t}$ inside a Schwarzschild black
hole). However, the static coordinates ${\stt,\,\str}$ are not globally smooth
and uniquely valued. Namely, ${\stt\to\infty}$ at the cosmological horizons
$\str=\DSr$. The static coordinates, extended to the whole de~Sitter space, are
illustrated in Fig.~\ref{fig:dSstat}. Here we also indicate the regions in
which ${\cvil{\stt}}$ is spacelike by bold ${\Fdom}$ (\vague{future}) and
${\Pdom}$ (\vague{past}), whereas the regions in which it is timelike are
denoted by ${\Ndom}$ (containing the \vague{north pole} ${\spr=0}$) and
${\Sdom}$ (containing the \vague{south pole} ${\spr=\pi}$). Hereafter, this
notation will be used repeatedly.

The conformal structure of Minkowski and de~Sitter spacetimes, their conformal
relation, and their conformal relation to various regions of the Einstein
static universe have been discussed extensively  in literature (see, e.g.,
\cite{Penrose:1968,PenroseRindler:book,HawkingEllis:book,Wald:book1984}).
The complete compactified picture of these spacetimes,
in particular the 3-dimensional diagram of the compactified
Minkowski and de~Sitter spaces $M^{\#}$ as parts of the Einstein universe
represented by a \emph{solid} cylinder can be found in \cite{BicakKrtous:2001}. We refer the reader
especially to Section III of \cite{BicakKrtous:2001} where we explain and
illustrate the compactification in detail. In the present paper we shall confine
ourselves to the 2-dimensional Penrose diagrams.

The basic \defterm{standard rescaled coordinates} covering globally
de~Sitter spacetime including the conformal infinity
are simply related to the standard coordinates as follows:
\begin{equation}\label{CEinsCoor}
  \tan\frac\tlt2 = \exp\frac\spt\DSr\comma
  \tlr=\spr\commae
\end{equation}
${\tlt\in(0,\pi)}$, ${\tlr\in(0,\pi)}$. The metric \eqref{PCSphM} becomes
\begin{equation}\label{CEinsM}
    \mtrc_\dS=\DSr^2\,\sin^{\!-2}\!\tlt\;
    \bigl(-\grad\tlt\formsq+\grad\tlr\formsq+\sin^2\tlr\,\sphmtrc\bigr)\commae
\end{equation}
demonstrating explicitly the conformal relations of de~Sitter spacetime to
the Einstein universe:
\begin{equation}\label{CEinsFact}
  \mtrc_\Eins=\Omega_\dS^2\,\mtrc_\dS\comma\Omega_\dS = \sin\tlt\period
\end{equation}
Therefore, we also call coordinates ${\tlt,\,\tlr}$ the
\defterm{conformally Einstein coordinates}.
The conformal diagram of de~Sitter spacetime is illustrated in
Fig.~\ref{fig:dS}. The past and future infinities, ${\tlt=0}$ and ${\tlt=\pi}$
are spacelike, the worldlines of the north and south poles (given by the choice
of the origin of the coordinates) are described by ${\tlr=\spr=0}$ and
${\tlr=\spr=\pi}$.

The whole de~Sitter spacetime could be represented by just the
\vague{right half} of Fig.~\ref{fig:dS}. Indeed, it is customary to draw
this half only and to consider any point in the figure as a 2-sphere, except for the
poles ${\tlr=0,\pi}$. As we shall see, the formulas relating
coordinates on the conformally related de~Sitter and Minkowski spacetimes have
simpler forms if we admit negative values of the radial coordinate
${\tlr\in(-\pi,0)}$ covering the left half of the diagram.
We shall thus consider the \mbox{2-dimensional} diagrams as in Fig.~\ref{fig:dS} to represent the cuts of
de~Sitter spacetime along the axis going through the origins (through north and
south poles---analogously to the cuts along the $\MOz$~axis in $E^3$).
The axis, i.e., the main circle of the spatial spherical section
of de~Sitter spacetime, is typically chosen as ${\tht=0,\pi}$. Thus, in the
diagram the point with ${\tlr=-\tlr_\oix<0}$, ${\tht=\tht_\oix}$, ${\ph=\ph_\oix}$
is identical to that with ${\tlr=\tlr_\oix}$, ${\tht=\pi-\tht_\oix}$, and
${\ph=\ph_\oix+\pi}$.
We use the same convention also for other radial coordinates appearing later,
as explicitly stated in the Appendix (cf.\ also Appendix in \cite{BicakKrtous:2001}).
We admit negative radial coordinates only when
describing various relations between the coordinate systems. In the expressions
for the fields in the following sections only positive radial coordinates are
considered.

As mentioned above, in \cite{BicakKrtous:2001} we constructed fields on
de~Sitter spacetime by conformally transforming the fields from Minkowski
spacetime. Now  \vague{different Minkowski spaces} can be used in the conformal
relation to de~Sitter space, depending on which region of a Minkowski space is
mapped onto which region of de~Sitter space. Consider, for example, Minkowski
space with metric $\mtrc_\Mink$ given in spherical coordinates
${\MVt,\,\MVr,\,\tht,\,\ph}$. Identify it with de~Sitter
space by relations
\begin{equation}\label{CFlatCoor}
  \MVt=\frac{\DSr\sin\tlt}{\cos\tlt-\cos\tlr}\comma
  \MVr=\frac{\DSr\sin\tlr}{\cos\tlr-\cos\tlt}\commae
\end{equation}
the inverse relation \eqref{apx:tltr=MAVtr} is given in the Appendix.
In the coordinates ${\MVt,\,\MVr,\,\tht,\,\ph}$
the de~Sitter metric \eqref{CEinsM} becomes
\begin{equation}\label{CFlatM}
    \mtrc_\dS=\frac{\DSr^2}{\MVt^2}\,
    \Bigl(-\grad\MVt\formsq+\grad\MVr\formsq+\MVr^2\,\sphmtrc\Bigr)\commae
\end{equation}
so that
\begin{equation}\label{CFlatFact}
  \mtrc_\dS=\Omega_\MinkV^2\mtrc_\MinkV\comma
  \Omega_\MinkV=\frac\DSr\MVt\period
\end{equation}
The coordinates ${\MVt,\,\MVr,\,\tht,\,\ph}$ can, of course, be used in both
de~Sitter and Minkowski spaces. Fig.~\ref{fig:dSMinkV} illustrates the coordinate lines.
It also shows how four regions I, II, III, and IV of Minkowski space are
mapped onto four regions of de~Sitter space by relations \eqref{CFlatCoor}.
We call ${\MVt,\,\MVr}$ \defterm{rescaled flat cosmological coordinates}
since their radial coordinate $\MVr$ coincides with that of the
flat cosmological coordinates \eqref{PCFlatCoor} and
the time coordinate is simply related to $\MVet$ as
\begin{equation}\label{MVtRel}
  \MVt=-\DSr\,\exp(-\MVet/\DSr)\period
\end{equation}
The caron or the check (still better \vague{h\'a\v{c}ek}) \vague{$\vee$} formed by
cosmological horizon at ${\MVt=\pm\infty}$ in de~Sitter space (cf.\
Fig.~\ref{fig:dSMinkV})
inspired our notation of these coordinates. It is possible to introduce
analogously the coordinates ${\MAt,\,\MAr}$ given in the Appendix,
Eqs.~\eqref{apx:MAtr=sptr}, \eqref{apx:MAtr=tltr},
that cover nicely the past conformal infinity but are
not smooth at the cosmological horizon ${\MAt=\pm\infty}$; in this case
they form the hat \vague{$\wedge$} in the conformal diagram (see
Fig.~\ref{fig:coorCMA} in the Appendix).

From relations \eqref{CFlatCoor} it is explicitly seen why, when writing down
mappings between de~Sitter and Minkowski spaces and drawing the corresponding
2-dimensional conformal diagrams, it is advantageous to admit negative radial
coordinates. If we would restrict all radial coordinates to be non-negative, we
would have to consider the second relation in Eq.~\eqref{CFlatCoor} with
different signs for regions III and II in de~Sitter space:
in III ${\MVr= {\DSr\sin\tlr}/{(\cos\tlr-\cos\tlt)}}$, but
in III we would have ${\MVr=-{\DSr\sin\tlr}/{(\cos\tlr-\cos\tlt)}}$.

\begin{figure}
\includegraphics{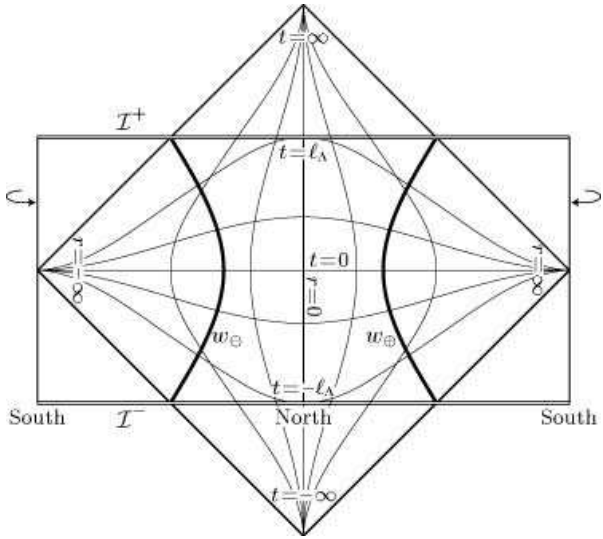}
\caption{\label{fig:dSMinkO}%
The conformally Minkowski coordinates. They cover the whole conformally related Minkowski
space but only a part of corresponding de~Sitter space.
This Minkowski space is related to that in Fig.~\ref{fig:dSMinkV}
by a shift \vague{downwards} by ${\pi/2}$ in the direction  of the
conformally Einstein coordinate $\tlt$.%
}
\end{figure}

Another mapping of Minkowski on de~Sitter space will be used to advantage in the
explicit manifestation that the generalized Born solution in de~Sitter space goes over
to the classical solution \eqref{EMBornM}. Instead of the mapping
\eqref{CFlatCoor}, consider the relations
\begin{equation}\label{CMinkCoor}
  \MOt=-\frac{\DSr\cos\tlt}{\cos\tlr+\sin\tlt}\comma
  \MOr=\frac{\DSr\sin\tlr}{\cos\tlr+\sin\tlt}
\end{equation}
(see Eq.~\eqref{apx:tltr=MOtr} for the inverse mapping), which turn the metric
\eqref{CEinsM} into
\begin{equation}\label{CMinkM}
    \mtrc_\dS=\Bigl(\frac{2\,\DSr^2}{\DSr^2-\MOt^2+\MOr^2}\Bigr)^2\,
    \Bigl(-\grad\MOt\formsq+\grad\MOr\formsq+\MOr^2\,\sphmtrc\Bigr)\period
\end{equation}
We again obtain the de~Sitter metric in the form explicitly conformal to the Minkowski metric
with, however, a different conformal factor from that in Eq.~\eqref{CFlatFact}:
\begin{equation}\label{CMinkFact}
   \mtrc_\dS=\Omega_\MinkO^2\,\mtrc_\MinkO\comma
   \Omega_\MinkO = \frac{2\,\DSr^2}{\DSr^2-\MOt^2+\MOr^2}\period
\end{equation}
(For the use of the de~Sitter metric in \vague{atypical} form \eqref{CMinkM}
in the work on the domain wall spacetimes, see \cite{Cveticetal:1993}.)
The relation of Minkowski space to de~Sitter  space based on the mapping
\eqref{CMinkCoor} is illustrated in Fig.~\ref{fig:dSMinkO}. Clearly, the
Minkowski space in this figure is shifted \vague{downwards} by ${\pi/2}$
in ${\tlt}$ coordinate, as compared with Minkowski space in Fig.~\ref{fig:dSMinkV}
(Eq.~\eqref{CFlatCoor}). Indeed, replacing ${\tlt}$ by ${\tlt+\frac\pi2}$ in
Eq.~\eqref{CFlatCoor}, we get ${\MVt=\MOt}$, ${\MVr=\MOr}$ with ${\MOt,\,\MOr}$
given by Eq.~\eqref{CMinkCoor}. Since coordinates ${\MOt,\,\MOr,\,\tht,\ph}$ are not
connected directly with any cosmological model and correspond to Minkowski space
\vague{centered} on de~Sitter space (Fig.~\ref{fig:dSMinkO}), we just call them
\defterm{conformally Minkowski coordinates}.

In Ref.~\cite{BicakKrtous:2001} still another Minkowski space is related to de~Sitter
space---one which is shifted \vague{downward} in ${\tlt}$ coordinate by another
${\pi/2}$. As mentioned below Eq.~\eqref{MVtRel}, the cosmological horizon forms hat
\vague{${\wedge}$} in this case and the corresponding coordinates are accordingly
denoted as ${\MAt,\,\MAr}$. They are given explicitly in
Appendix~\ref{asc:coorCMA} and Fig.~\ref{fig:coorCMA}.

The three sets of coordinates ${\MVt,\,\MVr}$, $\;{\MOt,\,\MOr}$, and ${\MAt,\,\MAr}$
(with the same ${\tht,\,\ph}$) relating naturally \vague{three} Minkowski spaces to
de~Sitter space are suitable for different purposes. The third set describes
conveniently the past infinity of de~Sitter space---that is why it was used
extensively in \cite{BicakKrtous:2001} where we were interested in how the sources
enter (are \vague{born in}) de~Sitter universe. The second set will be needed in
Section~\ref{sc:BorndS} for exhibiting the flat-space limit of the generalized Born
solution. The first set describes nicely the future infinity and will be employed
when analyzing radiative properties of the fields.

With all the coordinates discussed above,
corresponding double null coordinates can be associated;
some of them will also be used in the following.
Their more detailed description and illustration
is presented in section~\ref{asc:coorNull} of the Appendix.

Before concluding this section let us notice that the observers which are at
rest in cosmological coordinate systems ${\spt,\,\spr}$, $\;{\MVet,\,\MVr}$, and
${\cht,\,\chr}$ move along the geodesics with proper time $\spt$, $\MVet$, and
${\cht}$ respectively. These geodesics are also the orbits of the conformal Killing vectors.
Indeed, the symmetries of Minkowski
spacetime and of the Einstein universe become conformal symmetries in conformally
related de~Sitter spacetime. In particular, we shall employ the fact that since
${\cvil{\MVt}}$ and ${\cvil{\MOt}}$ are timelike Killing vectors in Minkowski spacetime and
${\cvil{\tlt}}$ is a timelike Killing vector in the Einstein universe, the vectors
\begin{equation}\label{CKillV}
\cv{\tlt}\comma\cv{\MVt}\comma \text{and}\quad \cv{\MOt}
\end{equation}
are timelike conformal Killing vectors in de~Sitter spacetime. As mentioned below
Eq.~\eqref{PStM}, ${\cvil{\stt}}$ is a Killing vector which is timelike for
${\abs{\str}<\DSr}$.

\section{Uniformly accelerated particles in de~Sitter}
\label{sc:AccPart}

\subsection{Particles born at the poles}
\label{ssc:patpoles}

In Section~\ref{sc:BornM} we defined uniformly accelerated motion in Minkowski
spacetime. However, the formulas given there, being in covariant forms, remain valid
in de~Sitter spacetime. As explained in \cite{BicakKrtous:2001} in detail, a simple
way of obtaining a worldline of a uniformly accelerated particle in de~Sitter
spacetime is to consider a suitable
particle moving with a \emph{uniform velocity} in
Minkowski spacetime and use the conformal relation between the spaces.

\begin{figure}
\includegraphics{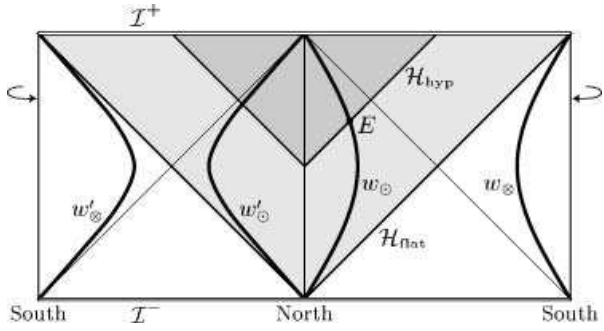}
\caption{\label{fig:wldln_p}%
The worldlines of uniformly accelerated charges.
The particles ${w_\npart}$ and ${w_\npart'}$ start and end at the \vague{north pole,}
${w_\spart}$, ${w_\spart'}$ start and end at the south pole.
Particles ${w_\npart'}$, ${w_\spart'}$ have a higher magnitude of acceleration ${a_\dS}$
than particles ${w_\npart}$, ${w_\spart}$. They are characterized by a negative
parameter ${\acp}$, whereas particles ${w_\npart}$, ${w_\spart}$ have a positive $\acp$.%
}
\end{figure}

Consider a particle moving with a constant velocity of magnitude
\begin{equation}\label{velMink}
  v_\MinkV=\tanh\acp=\text{constant}\commae
\end{equation}
such that for $\acp>0$ it moves in a negative direction along the ${\MVz}$~axis of the inertial frame in
Minkowski space ${\MinkVspc}$ with coordinates ${\MVt,\,\MVr,\,\tht,\,\ph}$ and passes
through ${\MVr=0}$ at ${\MVt=0}$:
\begin{equation}\label{wlMinkV}
\begin{aligned}
  \MVt&=\prt_\MinkV\,\cosh\acp\commae\\
  \MVr&=-\prt_\MinkV\sinh\acp\comma  \tht=0\period
\end{aligned}
\end{equation}
Substituting into transformation \eqref{apx:tltr=MAVtr}, we find
\begin{equation}\label{wlSprtM}
\begin{aligned}
  \tlt&=\arctan\Bigl(-2\DSr\frac{\prt_\MinkV\cosh\acp}{\prt_\MinkV^2-\DSr^2}\Bigr)\commae\\
  \tlr&=\arctan\Bigl(-2\DSr\frac{\prt_\MinkV\sinh\acp}{\prt_\MinkV^2+\DSr^2}\Bigr)\commae
\end{aligned}
\end{equation}
or expressing Minkowski proper time ${\prt_\MinkV}$ in terms of the proper time of
de~Sitter spacetime,
\begin{equation}\label{prtMdS}
  \prt_\MinkV=\mp\DSr\exp\bigl(\mp\prtMinkV\bigr)\commae
\end{equation}
we obtain
\begin{equation}\label{wlStl}
\begin{aligned}
 \tlt&=\arccot\Bigl(-\frac{\sinh\bigl(\prtMinkV\bigr)}{\cosh\acp}\Bigr)\commae\\
 \tlr&=\arccot\Bigl(\pm\frac{\cosh\bigl(\prtMinkV\bigr)}{\sinh\acp}\Bigr)
       \comma  \tht=0\period
\end{aligned}
\end{equation}
Here ${\prt_\dS\in\realn}$, $\arccot{}$ takes values such that
${\tlt\in(0,\pi)}$ and ${\tlr\in(0,\pi)}$ for ${\acp>0}$, or ${\tlr\in(-\pi,0)}$
for ${\acp<0}$. Upper sign is valid for the particle starting and
ending with ${\tlr=0}$ (particle ${w_\npart}$ in Fig.~\ref{fig:wldln_p}),
lower sign for the particle starting and ending at ${\tlr=\pi}$
(particle $w_\spart$ in Fig.~\ref{fig:wldln_p}).

One can make sure by direct calculations of the four-acceleration (for
its simplest form in the static coordinates, see below) that these
worldlines describe the uniformly accelerated motion as defined in
Section~\ref{sc:BornM}, the magnitude of the acceleration being
\begin{equation}\label{accdS}
   a_\dS = \sqrt{\tens{a}^\mu \tens{a}_\mu} =
   \abs{\DSr^{-1}\,\sinh\acp}\period
\end{equation}
Since de~Sitter universe represents the asymptotic state of all three
types of indefinitely expanding FRW models with $\Lambda>0$, it is of
interest to find out the form of these worldlines in the three types
of cosmological frames---spherical, flat, and hyperbolic---introduced
in Section~\ref{sc:BornM}.

In terms of cosmological spherical coordinates the worldlines are given by
\begin{equation}\label{wlcs}
\begin{aligned}
 \spt&=\DSr\,\arcsinh\Biggl(\frac{\sinh\bigl(\prtMinkV\bigr)}
       {\cosh\acp}\Biggr)\commae\\
 \spr&=\arccot\Biggl(\pm\frac
       {\cosh\bigl(\prtMinkV\bigr)}{\sinh\acp}\Biggr)\comma
       \tht=0\period
\end{aligned}
\end{equation}
In flat cosmological coordinates, which cover only half of
de~Sitter space, we obtain just particle  $w_\npart$ described by
the worldline
\begin{equation}\label{wlcf}
\begin{aligned}
  \MVet&=\prt_\dS\,\cosh\acp-\DSr\log\cosh\acp\commae\\
  \MVr&=\DSr\sinh\acp\;\,\exp\bigl(-\prtMinkV\bigr)\period
\end{aligned}
\end{equation}
Finally, in hyperbolic cosmological coordinates, which are also not
global, we obtain again one particle's worldline only given in terms
of its proper time as
\begin{equation}\label{wlch}
\begin{aligned}
  \cht&=\DSr\arccosh\frac
     {\cosh\bigl(\prtMinkV\bigr)}{\cosh\acp}\commae\\
  \chr&=\DSr\arccoth\frac
     {\sinh\bigl(\prtMinkV\bigr)}{\sinh\acp}\period
\end{aligned}
\end{equation}
These formulas have no meaning for
${\abs{{\prt_\dS}/\DSr\;\cosh\acp}<\abs{\acp}}$ where the inverse
hyperbolic functions are not defined. This corresponds to the fact
that for such ${\prt_\dS}$ the particle
occurs in the region where the hyperbolic cosmological coordinates are
not defined (cf. Fig.~\ref{fig:dShyp}).
Excluding the proper time we find the worldlines to be given by
remarkably simple formulas in the three systems of the cosmological
coordinates:
\begin{align}
\intertext{(a) spherical,}\label{wlcs1}
  \sin\spr&=\pm{\tanh\acp}\Big/{\cosh\frac\spt\DSr}\semicolone\\
\intertext{(b) flat,}\label{wlcf1}
  \frac\MVr\DSr&=\tanh\acp\Big/\exp\frac\MVet\DSr\semicolone\\
\intertext{c) hyperbolic,}\label{wlch1}
   \sinh\frac\chr\DSr&={\tanh\acp}\,\Big/\,{\sinh\frac\cht\DSr}\period
\end{align}

It is of interest to see what are the physical radial velocities which
will be observed by three types of the fundamental cosmological
observers, i.e., those with fixed $\spr$, $\MVr$, and $\chr$, respectively, whose proper
times are $\spt$, $\MVet$, and $\cht$, respectively. Such velocities can be
defined by the covariant expression
\begin{equation}\label{obsvel}
  v_\obs = \tens{u}_\alpha\,\cbv[\alpha]{1}\,
           \frac{d\prt_\dS}{d\prt_\obs}\commae
\end{equation}
where ${\tens{u}^\alpha}$ is the particle's four-velocity, ${\prt_\dS}$
its proper time, ${\cbv[\alpha]{1}}$ is the unit spacelike
vector in the direction of the radial coordinate
${x^1=\spr,\,\MVr,}$, and ${\chr}$, respectively, i.e., in directions ${\cvil{\spr}}$,
${\cvil{\MVr}}$ and ${\cvil{\chr}}$,
and ${\prt_\obs}$ is the proper time of an observer, i.e.,
${\spt,\,\MVet,}$ or ${\cht}$, respectively. Since all three cosmological
metrics are diagonal the expression \eqref{obsvel} takes on the form
\begin{equation}\label{obsvelC}
  v_\obs = \sqrt{\mtrc_{\dS\scriptscriptstyle11}}\, \frac{d x^1}{d\prt_\obs}\period
\end{equation}
The results are of interest:
\begin{align}
  v_{\obs(\spr)} &= \mp\frac
     {\sign\spt\,\sinh\acp}{\sqrt{\sinh^2\acp+\coth^2(\spt/\DSr)}}
  \commae\label{obsvelcs}\\
  v_{\obs(\MVr)} &= -\tanh\acp
  \commae\label{obsvelcf}\\
  v_{\obs(\chr)} &= -\frac
     {\sinh\acp}{\sqrt{\sinh^2\acp+\tanh^2(\cht/\DSr)}}
  \period\label{obsvelch}
\end{align}

Consider first the picture in spherical cosmological coordinates,
Eqs.~\eqref{wlcs} and \eqref{wlcs1}. Only in this frame both particles
are present. They start asymptotically at antipodes  of the spatial
section of de~Sitter space at ${\scri^-}$
(${\spt\to-\infty}$) and move one towards the other until ${\spt=0}$,
the moment of maximal contraction of de~Sitter space (\vague{the neck}
of de~Sitter hyperboloid), when they stop, ${v_{\obs(\spr)}=0}$. Then
they move, in a time-symmetric manner, apart from each other until they
reach future infinity asymptotically at the antipodes from which they
started. In contrast to the flat space case, the particles do not
approach the velocity of light in this global spherical cosmological
coordinate system, the asymptotical magnitude of their velocity being
equal to ${\abs{\tanh\acp}}$ (cf. Eq.~\eqref{obsvelcs}). Hence,
curiously enough, the particles approach the antipodes asymptotically
with a finite nonvanishing velocity (for an intuitive insight into this effect, see
below).

Although the particles $w_\npart$ and $w_\spart$
do not approach infinities with velocity of light, they \emph{are}
causally disconnected as the analogous pair of particles in
Minkowski space (cf. Fig.~\ref{fig:Mink} and Fig.~\ref{fig:wldln_p}).
No retarded or advanced effects from the particle $w_\npart$ can reach
the particle $w_\spart$ and vice versa.

Next, consider flat and hyperbolic observers.
As seen from Eq.~\eqref{obsvelcf}, with respect to the flat cosmological
coordinates the particle ${w_\npart}$ moves with the same velocity
${\abs{\tanh\acp}}$ all the time. And the same velocity is
asymptotically, at ${\cht\to\infty}$, reached by this particle
in the hyperbolic cosmological coordinates. The magnitude
of the asymptotic values of the velocity at ${\scri^+}$ is, in fact,
equal to the velocity \eqref{velMink} of the particle in Minkowski
space from which we constructed uniformly accelerated worldlines by a
conformal transformation. The identity of all these velocities is
understandable: the magnitude of the velocity with respect to an
observer can be determined by projecting the particle's four-velocity
on the observer's four-velocity, i.e., by the angle between these
directions. In de~Sitter space all three types of cosmological
observers reach ${\scri^+}$ with the same four-velocity; moreover,
this four-velocity is at ${\scri^+}$ identical to the four-velocity of observers
at rest in conformally related Minkowski space. But a conformal
transformation preserves the angles and thus, the velocities
with respect to the three types of cosmological observers in de~Sitter space
and the velocity in the conformally related Minkowski space must all
be equal---given by the \vague{Lorentzian} angle ${\acp}$.

It is worth noticing yet what is the \emph{initial} velocity
of the particle ${w_\npart}$ in hyperbolic cosmological coordinates.
Regarding Fig.~\ref{fig:dShyp} we have ${\cht\to-\infty}$, ${\chr\to0}$
at the \vague{starting point} of the particle at ${\scri^-}$. From
Eq.~\eqref{obsvelch} we get ${v_{\obs (\chr)}\to-\tanh\acp}$ which in
the magnitude is the same as in   spherical cosmological coordinates
but has opposite sign since the particle moves in the direction of
increasing negative ${\chr}$. More interesting is how the particle
enters the upper region of the hyperbolic coordinates. Fig.~\ref{fig:dShyp}
suggests that its velocity must approach the velocity of light since at
this boundary the fundamental observers of the hyperbolic cosmological
frame themselves approach the velocity of light.
Indeed, at this boundary ${\cht=0}$,
${\chr=\infty}$, and the expression \eqref{obsvelch} implies
${v_{\obs (\chr)}\to-1}$.

By far the simplest description of the particles
is obtained in the static coordinates ${\stt,\,\str}$. Using, for
example, the relation ${\str=\DSr\sin\tlr/\sin\tlt}$
(cf.\ Eqs.~\eqref{apx:sttrT=tltr}, \eqref{apx:sttrS=tltr}),
and substituting from Eq.~\eqref{wlStl}, we find
that the worldlines of both particles $w_\npart$ and $w_\spart$ are given by
remarkably lucid forms
\begin{equation}\label{wlst}
\begin{aligned}
  \stt&=\prt_\dS\,\cosh\acp
       =\frac{\prt_\dS}{\sqrt{1-{\Ro^2}/{\DSr^2}}}\commae\\
  \str&=\DSr\tanh\acp
       \equiv\Ro\period
\end{aligned}
\end{equation}
These expressions imply that the four-acceleration
${\tens{a}^\alpha = \tens{u}^\mu\,\covd_\mu\tens{u}^\alpha}$
is simply
\begin{equation}\label{accl}
  \tens{a}=-\frac{\Ro}{\DSr^2}\, \cv{\str}
          = -\frac1\DSr\tanh\acp\cv{\str}
      = \accl \cbv{\str}\commae
\end{equation}
where ${\cbv{\str}}$ is a unit spatial vector in the direction
${\cvil{\str}}$ of the static radial coordinate ${\str}$, and
we introduced constant
\begin{equation}\label{accdef}
   \accl =-\DSr^{-1}\,\sinh\acp
   =-\frac{\Ro/\DSr^2}{\sqrt{1-\Ro^2/\DSr^2}}
\end{equation}
which represents the \vague{oriented} value of the acceleration
of the particles.

We thus find the uniformly accelerated particles in de~Sitter spacetime to
be at rest in the static coordinates at fixed values ${\str=\Ro}$ of
the radial coordinate. Two charges moving along the
orbits of the boost Killing vector \eqref{KillVec} in Minkowski space
are at rest in the Rindler coordinate system and have a constant
distance from the spacetime origin, as measured along the slices
orthogonal to the Killing vector. Similarly, we see that the worldlines
${w_\npart}$ and ${w_\spart}$ are the orbits of the static Killing
vector ${\cvil{\stt}}$ of de~Sitter space. The particle ${w_\npart}$
(respectively, ${w_\spart}$) has, as measured at fixed ${\stt}$, a
constant proper distance from the origin ${\tlt=\pi/2}$ (${\spt=0}$),
${\tlr=\spr=0}$ (respectively, ${\tlr=\spr=\pi}$). As with Rindler
coordinates in Minkowski  space, the static coordinates cover only a
\vague{half} of de~Sitter space. In the other half the Killing vector
becomes spacelike. Owing to \vague{cosmic repulsion} caused by the
presence of ${\Lambda}$, fundamental cosmological observers moving
along geodesics ${\spr,\,\tht,\,\ph}$ constant are \vague{repelled}
one from the others. Their initial implosion starting at
${\spt\to-\infty}$ is stopped at ${\spt=0}$ and changes into
expansion. Clearly, a particle with constant ${\str=\Ro}$---hence a
constant proper distance from the particle at ${\str=0=\spr}$---must be
accelerated towards that \vague{central} particle.

In Eq.~\eqref{accdef} we have denoted the radial tetrad component of the
acceleration in the static coordinates by ${\accl}$; notice that, in
contrast to the magnitude of the acceleration ${a_\dS=\abs{\accl}}$
(cf. Eq.~\eqref{accdS}), ${\accl}$ can be negative as, in fact, it is
the case with \emph{both} particles ${w_\npart}$ and ${w_\spart}$,
assuming that the static radial coordinate of the particles is positive,
${\str=\Ro>0}$.
Geometrically, the four-vectors of the acceleration of the particles
point in opposite directions---towards ${\spr=0}$, the other towards ${\spr=\pi}$.
Since, however, one needs two sets of the static coordinates to cover both
particles, and the radial coordinate ${\str}$ increases from both
${\spr=0}$ and ${\spr=\pi}$ worldlines (cf.\ Fig.~\ref{fig:dSstat}), the
accelerations of both particles point in the direction of decreasing
${\str}$'s and is thus negative. All the particles we are
considering perform 1-dimensional motion only, hence we use for
the description of their worldlines the same convention as for the
2-dimensional diagrams with time and radial coordinates---we allow the
radial coordinate to take negative values. Thus, for example, consider a
particle with worldline ${w'_\npart}$ which is a \vague{reflection} of
the worldline ${w_\npart}$ with respect to
${\tlr=\spr=0}$ (see Fig.~\ref{fig:wldln_p}). The particle ${w'_\npart}$ moves in the
region of negative ${\tlr}$, respectively ${\str}$, it has an
acceleration positive, ${\accl=-\DSr^{-1}\sinh\acp>0}$ (i.e.,
${\acp<0}$), and its four-acceleration vector is pointing in the direction of
increasing ${\str}$. With our convention, the particle ${w'_\npart}$ is just that which moves from
${\spr=0}$ along the ${\tht=\pi}$ direction. This convention will be particularly useful
when we shall construct worldlines of uniformly accelerated particles
which start and end at the equator. Those which move in the region
${\spr>\pi/2}$ will have negative ${\accl}$, those moving with
${\spr<\pi/2}$ will have positive ${\accl}$---see
Section~\ref{ssc:patequat}.

An intuitive geometrical understanding of the worldlines of uniformly
accelerated particles in de~Sitter spacetime can be gained by
considering de~Sitter  space as a 4-dimensional hyperboloid
${-\hc_0{}^2+\hc_1{}^2+\hc_2{}^2+\hc_3{}^2+\hc_4{}^2=\DSr^2}$
in 5-dimensional Minkowski space. The spherical cosmological
coordinates ${\spt,\,\spr,\,\tht,\,\ph}$ are then identical to
the hyperspherical coordinates on this hyperboloid.
The worldlines of the north and south poles, ${\spr=0,\pi}$, can be obtained by
cutting the hyperboloid by a timelike 2-plane ${\mathcal{T}_2}$,
given by ${\hc_2=\hc_3=\hc_4=0}$. The worldlines of our uniformly
accelerated particles ${w_\npart}$ and ${w_\spart}$ then arise when
the hyperboloid is cut by a timelike 2-plane ${\mathcal{T}_2^*}$ parallel to
${\mathcal{T}_2}$ at a distance ${\Ro=\DSr\tanh(\acp/\DSr)}$ from the origin
\cite{Rindler:book}. ${\mathcal{T}_2^*}$ is thus given by
${\hc_2=\Ro}$, ${\hc_3=\hc_4=0}$.
From the definition of the hyperspherical
coordinates it follows ${\tht=0,\pi}$ and
${\hc_2=\DSr\cosh(\spt/\DSr)\sin\spr\cos\tht=\Ro}$, i.e.,
${\sin\spr=\pm\tanh\acp\big/\cosh(\spt/\DSr)}$, which is just
Eq.~\eqref{wlcs1} describing ${w_\npart}$ and ${w_\spart}$.

From this construction, the curious result mentioned above---that ${w_\npart}$
and ${w_\spart}$ approach antipodes ${\spr=0}$ and ${\spr=\pi}$ asymptotically
with a fixed speed ${\abs{\tanh\acp}}$
in spherical cosmological coordinates---is not so surprising:
thanks to the expansion of de~Sitter spacetime all fundamental
cosmological observers with arbitrarily small
${\spr=\text{constant}>0}$ will, in the limit
${\spt\to\infty}$, eventually cross
the plane ${\mathcal{T}_2^*}$, and thus the particle ${w_\npart}$;
however at any finite but arbitrarily large ${\spt}$
there will be observers with ${\spr=\text{constant}}$ which are still moving
towards the particle ${w_\npart}$. The same, of course, is true with
the symmetrically located particle ${w_\spart}$ and corresponding
observers close to ${\spr=\pi}$.

\subsection{Particles born at the equator}
\label{ssc:patequat}

\begin{figure}
\includegraphics{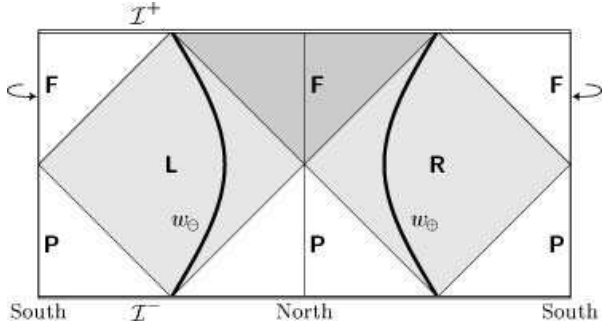}
\caption{\label{fig:wldln_e}%
The worldlines of uniformly accelerated charges located symmetrically
with respect to the origin (north pole) of the standard spherical
coordinates in de~Sitter space. The particles \vague{start} and
\vague{end} at the equator. They are causally disconnected as
a corresponding pair in Minkowski space (cf.\ Fig.~\ref{fig:Mink}).
The \vague{oriented} value $\accl$ of the acceleration of these particles is positive
(cf.\ the \vague{rotated} version of Eq.~\eqref{accdef}).
}
\end{figure}
\begin{figure}
\includegraphics{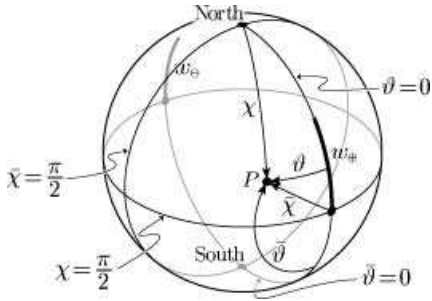}
\caption{\label{fig:rotation}%
The rotated spherical coordinates ${\rtr,\,\rtth}$ on 3-sphere
(the cut ${\ph=\text{constant}}$). The relation between
the coordinates is given in Eq.~\eqref{rotation}.
}
\end{figure}

In the classical Born solutions both charges are, at all times,
located symmetrically with respect to the origin of the Minkowski
coordinates (see Fig.~\ref{fig:Mink}). In order to demonstrate explicitly
that a limiting procedure exists in which our generalized Born's solution
goes over to its classical counterpart, we shall now construct the pair of
uniformly accelerated particles which are, at all times, symmetrically located
with respect to the origin of the standard spherical coordinates in de~Sitter space,
i.e., with respect to the \vague{north pole} at ${\spr=0}$. Asymptotically at
${\spt\to-\infty}$ these two particles both start (\vague{are born}) with the same speed at the equator,
${\spr=\pi/2}$, at the antipodal points ${\tht=0}$ and ${\tht=\pi}$.
As the universe contracts, they both move symmetrically along the axis
${\tht=0,\pi}$, reach some limiting value $\spr_\oix$ at the moment of time symmetry,
and accelerate back towards the equator, reaching the initial positions
asymptotically at ${\spt\to+\infty}$. These two particles are
illustrated in Fig.~\ref{fig:wldln_e}, with their worldlines denoted by
$w_\ppart$ and $w_\mpart$.
In Fig.~\ref{fig:rotation}, a snapshot at ${\spt=\text{constant}}$ is depicted.
Comparing Fig.~\ref{fig:wldln_e} with  Fig.~\ref{fig:wldln_p},
it is evident that the particles $w_\ppart$ and $w_\mpart$ are located with respect to the point
${\spr=\pi/2}$, ${\tht=0}$ in exactly the same manner as the particles $w_\npart$ and $w_\spart$
are located with respect to the pole ${\spr=0}$ (or, rather, as the particles
$w_\npart'$, $w_\spart'$, since we chose $w_\ppart$, $w_\mpart$ to have positive
$\accl$ in Fig.~\ref{fig:wldln_e}).

Owing to the global homogeneity of de~Sitter space and the spherical
geometry of its slices ${\spt=\text{constant}}$,
the worldlines of the particles $w_\ppart$ and $w_\mpart$
can be constructed by a suitable
rotation of the worldlines of the particles $w_\npart$ and $w_\spart$.
In Section~\ref{sc:BorndS} the same rotation will be applied
to obtain the fields of these particles \vague{born at the equator.}
We rotate the coordinates ${\spr,\,\tht,\,\ph}$ into new
coordinates ${\rtr,\,\rtth,\,\rtph}$ which, as a pole, have the point
${\spr=\pi/2}$, ${\tht=0}$ (see Fig.~\ref{fig:rotation}).
The relations between these coordinates follow from the
spherical geometry:
\begin{equation}\label{rotation}
\cos\rtr=\sin\spr\cos\tht\comma
\tan\rtth=-\tan\spr\sin\tht\comma
\rtph=\ph\period
\end{equation}
The new worldlines, $w_\ppart$ and $w_\mpart$,
will then be given by Eqs.~\eqref{wlcs} in which ${\spr,\,\tht,\,\ph}$ are replaced
by rotated coordinates ${\rtr,\,\rtth,\,\rtph}$. Substituting for these
by using relations~\eqref{rotation}, we find the worldlines $w_\ppart$, $w_\mpart$
in the original coordinates to be described by the expressions:
\begin{equation}\label{wlecs}
\begin{aligned}
 \spt&=\DSr\,\arcsinh\Biggl(\frac{\sinh\bigl(\prtMinkV\bigr)}
       {\cosh\acp}\Biggr)\commae\\
 \spr&=\pm\arctan\Biggl(-\frac
       {\cosh\bigl(\prtMinkV\bigr)}{\sinh\acp}\Biggr)\commae\\
 \tht&=0\commae
\end{aligned}
\end{equation}
with the values of $\arctan$ from ${(0,\pi)}$
and upper (lower) sign corresponding to the particle starting
at the positive (negative) value of $\spr$, i.e., to the particle
$w_\ppart$ (or $w_\mpart$, respectively).

Excluding the proper time $\prt_\dS$, we arrive at simple result (cf.\ Eq.~\eqref{wlcs1})
\begin{equation}\label{wlecselim}
\cos\spr=-\frac{\tanh\acp}{\cosh(\spt/\DSr)}\period
\end{equation}
As ${\spt\to\pm\infty}$, then indeed ${\abs{\spr}\to\pi/2}$; at ${\spt=0}$,
${\abs{\spr}=\arccos(-\tanh\acp)=\arccos(-\Ro/\DSr)}$, in agreement with the
\vague{deviation} of the \vague{original} particles $w_\npart$, $w_\spart$
from ${\spr=0}$ at ${\spt=0}$.
In the spherical rescaled coordinates, Eqs.~\eqref{wlecs} read
\begin{equation}\label{wleStl}
\begin{aligned}
 \tlt&=\arccot\Bigl(-\frac{\sinh\bigl(\prtMinkV\bigr)}{\cosh\acp}\Bigr)\commae\\
 \tlr&=\pm\arctan\Bigl(-\frac{\cosh\bigl(\prtMinkV\bigr)}{\sinh\acp}\Bigr)
       \commae\\
 \tht&=0\commae
\end{aligned}
\end{equation}
again with the values of $\arctan$ and $\arccot$ from ${(0,\pi)}$. Eq.~\eqref{wlecselim}
becomes
\begin{equation}\label{wletlelim}
\cos\tlr=-{\tanh\acp}{\sin\tlt}\period
\end{equation}

Although the flat (rescaled) cosmological coordinates cover only parts of the worldlines
$w_\ppart$, $w_\mpart$ (see Figs.~\ref{fig:wldln_e} and \ref{fig:dSMinkV}),
we transcribe the equations above also into these frames
in which the particles \vague{emerge} at ${\MVet,\,\MVt\to-\infty}$
at the cosmological horizon at ${\MVr=\pm\infty}$.
We find
\begin{align}
\frac\MVet\DSr&=-\log\biggl(\frac{-\cosh\acp}{-\sinh\bigl(\prtMinkV\bigr)+\sinh\acp}\biggr)\commae\notag\\
\frac\MVt\DSr&=\frac{\cosh\acp}{-\sinh\bigl(\prtMinkV\bigr)+\sinh\acp}\commae\\
\frac\MVr\DSr&=\mp\frac{\cosh\bigl(\prtMinkV\bigr)}{-\sinh\bigl(\prtMinkV\bigr)+\sinh\acp}\commae\notag
\end{align}
so that Eq.~\eqref{wletlelim} translates into the relations
\begin{equation}
\begin{gathered}
\MVr=\pm\sqrt{\DSr^2+\MVt^2-2\,\DSr\MVt\tanh\acp}\commae\\
\MVr=\pm\DSr\sqrt{1+2\tanh\acp\,\exp(-\MVet/\DSr)+\exp(-2\MVet/\DSr)}\period
\end{gathered}
\end{equation}
As ${\MVet\to+\infty}$, we have ${\MVr\to\pm\DSr}$, as it corresponds to ${\spr\to\pm\pi/2}$;
at ${\MVet\to-\infty}$, we get ${\MVr\to\pm\infty}$---here the particles
enter flat cosmological frame at the horizon (cf.\ Fig.~\ref{fig:wldln_e}).

The worldlines $w_\ppart$, $w_\mpart$ are situated outside the regions covered by our choice of the
hyperbolic cosmological coordinates. Similarly, we get only finite parts of $w_\ppart$, $w_\mpart$
in our static coordinates. Of course, we could rotate the static coordinates to cover
both particles but then we arrive at exactly the same picture as with the particles $w_\npart$, $w_\spart$
considered above.

Our primary reason to discuss the pair $w_\ppart$, $w_\mpart$ is to demonstrate explicitly
how our fields go over into the classical Born solution in the limit of vanishing $\Lambda$.
For this purpose, it will be important to have available also the description of the
worldlines $w_\ppart$, $w_\mpart$ in the Minkowski coordinates introduced in
Eqs.~\eqref{CMinkCoor}. As it is obvious from Fig.~\ref{fig:dSMinkO}, these
coordinates cover both worldlines $w_\ppart$ and $w_\mpart$ completely. Using the relations inverse
to Eqs.~\eqref{CMinkCoor} given in the Appendix, Eq.~\eqref{apx:tltr=MOtr}, we find
Eqs.~\eqref{wleStl} to imply
\begin{equation}\label{wleMO}
\MOt=\Bo\sinh\frac{\prt_\MinkO}{\Bo}\comma
\MOr=\pm\Bo\cosh\frac{\prt_\MinkO}{\Bo}\comma\tht=0\commae
\end{equation}
where
\begin{equation}\label{acclconst}
  \frac{\Bo}\DSr =\exp\acp=\sqrt{1+\accl^2\DSr^{-2}}-\accl\DSr=\sqrt{\frac{\DSr+\Ro}{\DSr-\Ro}}\commae
\end{equation}
and $\prt_\MinkO$ is the proper time measured in Minkowski space $\MinkOspc$
related to de~Sitter space by conformal mapping \eqref{CMinkM}, \eqref{CMinkFact}:
\begin{equation}
  \prt_\MinkO = \exp\acp\,\cosh\acp\;\prt_\dS\period
\end{equation}
Consequently,
\begin{equation}
  \MOr =\pm\sqrt{\MOt^2+\Bo^2}\comma \tht=0\commae
\end{equation}
which is the simplest form of the hyperbolic motion with the uniform
acceleration $1/\Bo$ as measured in Minkowski space (cf.\
Eqs.~\eqref{wlMink}).

\section{Frames centered on accelerated particles}
\label{sc:AccCoor}

For the investigation of the radiative properties and other physical aspects of the fields,
the use of (physically equivalent) particles $w_\npart$, $w_\spart$, i.e., those
\vague{born at the poles} of spherical coordinates is technically more advantageous.
We shall now return back and construct frames with the origins located directly on these particles.
In such frames, various properties of the fields will become more
transparent than in the coordinates introduced so far.

As we have seen in the preceding section, the uniformly accelerated
particles ${w_\npart}$ and ${w_\spart}$ are at rest in static coordinates ${\stt,\,\str}$ at given
${\str=\Ro=-\accl\DSr^2/\sqrt{1+\accl^2\DSr^2}}$, where
${\abs{\accl}}$ is the magnitude of the acceleration. In order to
investigate the properties of the fields, in particular, in order to see what is the
structure of the field along the null cones with vertices at the
particle's position, i.e., what is the field \vague{emitted} by the
particle at a given time, it is useful to construct coordinate
frames centered on the accelerated particles. Such systems of
coordinates are used to describe accelerating black holes in general
relativity (like C-metrics, known also for ${\Lambda\neq0}$, cf.\
\cite{KrtousPodolsky:2003,PodolskyOrtaggioKrtous:2003}),
so that their properties on de~Sitter background may
indicate what is their meaning in more general cases---in
situations when they are centered on gravitating objects rather than
on test particles.

We shall now describe three coordinate systems of this type:
the \defterm{accelerated coordinates}, the \defterm{C-metric-like coordinates},
and the \defterm{Robinson-Trautman coordinates}, all centered on the
worldlines ${w_\npart}$ and ${w_\spart}$. Instead of writing down just
the transformation formulas, we wish to indicate some steps how these
coordinates can be obtained naturally.
We list only the main transformation relations here, many other
formulas and forms of the metrics can be found in
the Appendix. Let us also note that in this section
we assume $\Ro,\,\acp>0$, i.e., ${\accl<0}$, and we use
only static radial coordinate with positive values, i.e., ${\str>0}$.

\vspace*{-12pt}
\subsection{Accelerated coordinates}
\label{ssc:Acoor}

We begin with the construction of \defterm{accelerated coordinates ${\At,\,\Ar,\,\Ath,\,\ph}$}.
This type of coordinates was recently introduced \cite{PodolskyGriffiths:2001}
by another method in the context of the \mbox{C-metric} with ${\Lambda>0}$.
In the preceding section we obtained the worldlines ${w_\npart}$,
${w_\spart}$ of uniformly accelerated particles in de~Sitter space by
starting from a particle moving with a uniform velocity
${v_\MinkV=\tanh\acp}$ in a negative direction of the ${\MVz}$~axis in the
inertial frame ${\MVt,\,\MVr,\,\tht,\,\ph}$ in Minkowski space
${\MinkVspc}$ which passes through ${\MVr=0}$ at ${\MVt=0}$ (see
Eqs.~\eqref{velMink}, \eqref{wlMinkV}); and we used then the conformal
relation between Minkowski and de~Sitter  spaces to find ${w_\npart}$,
${w_\spart}$. Therefore, let us first construct a frame centered on the
uniformly moving particle in ${\MinkVspc}$. Using spherical
coordinates again, this boosted frame
denoted by primes is related to the original one simply by
\begin{equation}\label{sphbst}
\begin{aligned}
  &\AVt=\MVt\cosh\acp+\MVr\cos\tht\sinh\acp\commae
\\
  &\AVr\cos\Ath=\MVt\sinh\acp+\MVr\cos\tht\cosh\acp\commae
\\
  &\AVr\sin\Ath=\MVr\sin\tht\commae
\end{aligned}
\end{equation}
\pagebreak
\begin{figure}[h]
\vspace*{20pt}
\includegraphics{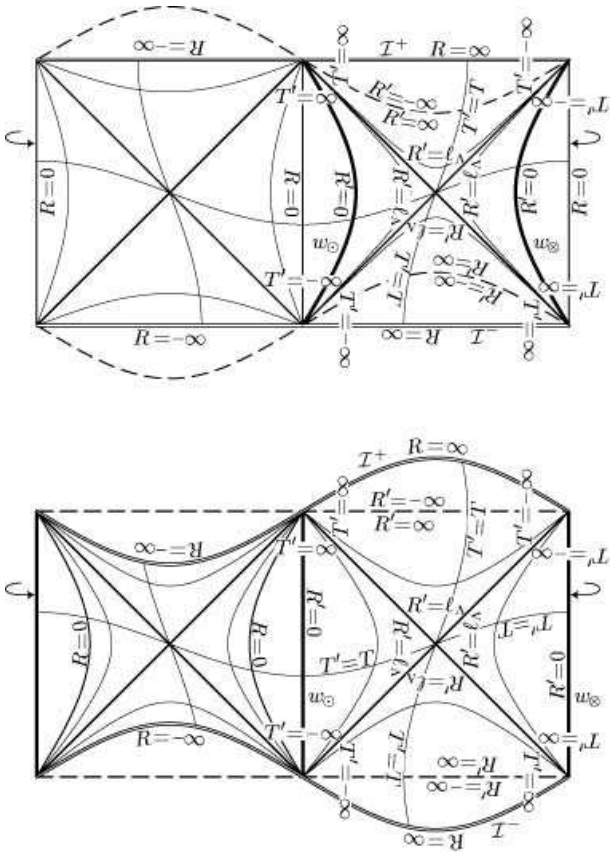}
\vspace*{20pt}
\caption{\label{fig:Acc}
  The 2-dimensional conformal diagrams of de~Sitter space based on
  the static, non-accelerated coordinates (upper diagram),
  and on the accelerated coordinates (lower diagram).
  Starting from static coordinates ${\stt,\,\str,\,\tht,\,\ph}$, one can draw
  the conformal diagram of the axis ${\tht=0,\pi}$ in which the conformal past and
  future infinities, ${\scri^\pm}$ (${\str=\pm\infty}$),
  are horizontal (double) lines. In addition to static coordinates ${\stt,\,\str}$,
  also accelerated coordinates ${\At,\,\Ar}$ are indicated in both diagrams. These
  have a coordinate singularity for ${\Ar=\infty}$ (drawn as a dashed line).
  The origins of the accelerated coordinates, ${\Ar=0}$ (thick lines), are worldlines of
  uniformly accelerated particles.
  In the conformal diagram of the axis ${\Ath=0,\pi}$
  based on accelerated coordinates, the origins ${\Ar=0}$ and
  the coordinate singularity ${\Ar=\infty}$ of the accelerated frame are
  straight lines; the true infinities $\scri^\pm$ have a \vague{bulge} upwards or
  \vague{downwards}, depending on the angle $\Ath$.
  The hypersurface ${\Ar=\infty}$ corresponds
  to the boosted hyperplane $\AVt=0$, whereas the conformal infinity corresponds to
  $\MVt=0$ (the relation of both hyperplanes can be well understood in the diagram
  of the conformally related Minkowski space $\MinkVspc$). The diagrams in which the
  conformal infinities $\scri^\pm$ are \emph{not} straight naturally arise in
  the studies of the C-metric with ${\Lambda>0}$ (de~Sitter space being a
  special case of this class of the metrics)---see \cite{KrtousPodolsky:2003}.
  In general, outside the axis ${\tht=\Ath=0,\pi}$,
  the transformations between the static and accelerated coordinates
  mix radial and angular coordinates ${\str,\tht}$ and ${\Ar,\Ath}$,
  as is seen also in the following Fig.~\ref{fig:AccRth}.
  The sections ${\Ath=\text{constant}}$ (for some general ${\Ath}$)
  are also shown in Fig.~\ref{fig:coorAcc} in the Appendix.
\vspace*{-80pt}
}
\end{figure}
\pagebreak
\begin{figure}
\includegraphics{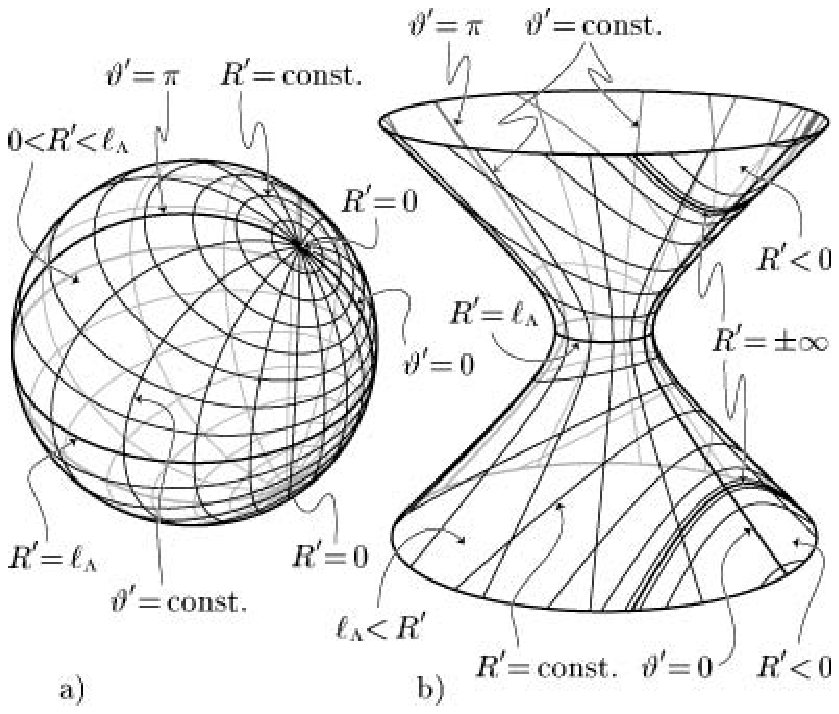}
\caption{\label{fig:AccRth}
The accelerated coordinates ${\Ar,\,\Ath}$ on the sections ${\At=\text{constant}}$
of de~Sitter space (coordinate ${\ph}$ suppressed).
In the region where ${\cvil{\At}}$ is timelike
(${0<\Ar<\DSr}$), the cut ${\At=\text{constant}}$ is a spacelike sphere (diagram a).
In the region where ${\cvil{\At}}$ is spacelike (${\DSr<\Ar}$ and ${\Ar<0}$),
it is a timelike hyperboloid (diagram b).
The diagrams are not in the same scale---the radius of
the sphere and of the neck of the hyperboloid should be the same.
The axis ${\Ath=0,\pi}$ corresponds to the lines ${\At=\text{constant}}$ of Fig.~\ref{fig:Acc}.
The coordinate singularity ${\Ar=\pm\infty}$ is also indicated.
For more details see the text.
}
\end{figure}
the ${\ph}$-coordinate does not change and will be suppressed in the
following. From here
\begin{equation}\label{sphbstrel}
\begin{gathered}
  -\AVt^2+\AVr^2=-\MVt^2+\MVr^2\commae\\
  \tan\Ath=\frac{\sin\tht}{(\MVt/\MVr)\sinh\acp+\cos\tht\cosh\acp}\period
\end{gathered}
\end{equation}
The original frame ${\MVt,\,\MVr,\tht}$ in Minkowski space ${\MinkVspc}$ is related to
the static coordinates ${\stt,\,\str,\tht}$ in de~Sitter space by
(cf.\ Eqs.~\eqref{apx:sttrT=MAVtr}, \eqref{apx:sttrS=MAVtr})
\begin{equation}\label{stCoorDef}
  \stt=-\frac\DSr2\log\Bigabs{\frac{\MVt^2-\MVr^2}{\DSr^2}}\comma
  \str=-\DSr\frac\MVr\MVt\comma \tht=\tht\period
\end{equation}
The metrics of the two spaces are related by ${g_\dS=(\DSr^2/\MVt^2)\mtrc_\MinkV}$,
${\mtrc_\dS}$ being given by Eq.~\eqref{PStM}---cf.\
Eq.~\eqref{CFlatM}. Now, let us introduce
coordinates ${\At,\,\Ar,\,\Ath}$ given in terms of
${\AVt,\,\AVr,\,\Ath}$ by exactly the same formulas as
coordinates ${\stt,\,\str,\,\tht}$ are given in terms of ${\MVt,\,\MVr,\,\tht}$ in
Eq.~\eqref{stCoorDef}. In this way we obtain
${\mtrc_{\dS'}=(\DSr^2/\AVt^2)\mtrc_{\MinkV}}$. Combining the last
relation with ${g_\dS=(\DSr^2/\MVt^2)\mtrc_\MinkV}$, we find the
metric of the original de~Sitter space in the new coordinates
${\At,\,\Ar,\,\Ath}$ in the form
\begin{equation}\label{dSdS}
  \mtrc_\dS=\frac{\AVt^2}{\MVt^2}\,\mtrc_{\dS'}\commae
\end{equation}
${\mtrc_{\dS'}}$ is given by the \vague{primed} version of
Eq.~\eqref{PStM}. Expressing then the factor ${(\AVt/\MVt)^2}$
by using Eqs.~\eqref{sphbst}, \eqref{sphbstrel}, and \vague{primed}
relations \eqref{stCoorDef},
we arrive at the de~Sitter metric in the accelerated
coordinates in the form
\begin{equation}\label{accM}
\begin{split}\raisetag{42pt}
   \mtrc_\dS=\,&\frac{1-{\Ro^2}/{\DSr^2}}{\bigl[1+({\Ar\Ro}/{\DSr^{2}})\,\cos\Ath\bigr]^2}
     \Biggl(-\Bigl(1-\frac{\Ar^2}{\DSr^2}\Bigr)\,\grad\At\formsq
     \\&\;
       +\Bigl(1-\frac{\Ar^2}{\DSr^2}\Bigr)^{\!\!-1}\!\grad\Ar\formsq
       +\Ar^2\,\bigl(\grad\Ath\formsq+\sin^2\Ath\,\grad\ph\formsq\bigr)\Biggr)\period\\
\end{split}
\end{equation}
Here the accelerated coordinates ${\At,\,\Ar,\,\Ath,\,\ph}$ are given in
terms of static coordinates by the relation obtained by the procedure
described above as follows:\pagebreak[1]
\begin{equation}\label{stacc}
\begin{gathered}
  \Ar=\DSr\sqrt{1
  -\frac{\bigl(1-{\str^2}/{\DSr^2}\bigr)\bigl(1-{\Ro^2}/{\DSr^2}\bigr)}
  {\bigl[1-({\str\Ro}/{\DSr^2})\,\cos\tht\bigr]^2}}\commae\\
  \At=\stt\comma
  \tan\Ath=\frac{\sqrt{1-{\Ro^2}/{\DSr^2}}\;\str\sin\tht}{\str\cos\tht-\Ro}
  \period
\end{gathered}
\end{equation}

Notice that the time coordinate of static and accelerated frames
coincide. Technically, this is easy to see from the first relation in
Eqs.~\eqref{sphbstrel} and Eq.~\eqref{stCoorDef}. Intuitively, this is
evident since the uniformly accelerated particles are at rest in the
static coordinates, as well as in the accelerated coordinates, the
only difference being that they are located \emph{at the origin} of the
accelerated frame. Setting ${\Ro=0}$ in Eq.~\eqref{stacc}, we get
${\Ar=\str}$, ${\Ath=\tht}$, as expected. The static coordinates are
centered on the poles ${\spr=0,\,\pi}$, hence, on the
\emph{un}accelerated worldlines. The name \emph{accelerated}
coordinates is thus inspired by the fact that their origin \emph{is}
accelerated, and the value of this acceleration enters the
form of the metric \eqref{accM} explicitly through the quantity ${\Ro}$.

The 2-dimensional conformal diagram of de~Sitter space with coordinate lines
${\At=\text{constant}}$, ${\Ar=\text{constant}}$ of the accelerated frame is
given in Fig.~\ref{fig:Acc}. For details, see the figure caption. Here let
us just notice that the cosmological horizons are still described by
${\Ar^2=\DSr^2}$.
Infinite values of ${\Ar}$ can, however, be encountered
\vague{before} the conformal infinities
$\scri^\pm$ are reached. This depends on the angle $\Ath$.
Indeed, $\Ar=\infty$ corresponds to $\AVt=0$, whereas
$\scri^\pm$ is given by $\MVt=0$, i.e.,
\begin{equation}
\Ar=-\frac{\DSr^2}{\Ro}\frac1{\cos\Ath}\commae
\end{equation}
(cf.\ metric \eqref{accM}). Relation of these two surfaces is best viewed
in Minkowski space ${\MinkVspc}$. We see that for ${\tht,\Ath<\pi/2}$,
the conformal infinity $\scri^+$ ($\scri^-$) lies \vague{above} (\vague{below})
the surface ${\Ar=\pm\infty}$. Thus the infinity
${\Ar=\pm\infty}$ is just a coordinate singularity, which can be removed using,
for example, the C-metric-like coordinate $\Cy$ introduced below.

Fig.~\ref{fig:AccRth}a shows the cut ${\stt=\At=\text{constant}}$ located
in the region of de~Sitter space where the Killing vector $\cvil{\stt}=\cvil{\At}$
is timelike (${\str,\Ar<\DSr}$); $\ph$-direction is suppressed.
The cut is a \emph{spacelike} sphere~$S^3$ with homogeneous spherical metric.
The coordinate lines ${\Ar=\text{constant}}$ and
${\Ath=\text{constant}}$ are plotted, with two origins ${\Ar=0}$ indicated:
here the accelerated particles occur. The coordinate $\Ar$ grows from
$\Ar=0$ at the origins to the equator where ${\Ar=\DSr}$.
In Fig.~\ref{fig:AccRth}b the cut ${\stt=\At=\text{constant}}$
located in the regions where $\cvil{\stt}=\cvil{\At}$ is spacelike (${\str,\Ar>\DSr}$) is illustrated,
again with $\ph$-direction suppressed.
Here the cut is \emph{timelike} with the geometry of three-dimensional de~Sitter space.
The coordinate lines ${\Ar=\text{constant}}$ and
${\Ath=\text{constant}}$ are also shown.

As we have just seen, the points with ${\Ar=\infty}$ can be \vague{nice} points in
de~Sitter manifold. It may thus be convenient to introduce
the inverse of $\Ar$ as a new coordinate. Also, we consider ${-\cos\Ath}$
as a new coordinate, and make the time
coordinate dimensionless. We thus arrive at the \defterm{C-metric-like coordinates
${\Ct,\,\Cy,\,\Cx,\,\ph}$}:
\begin{equation}\label{CmtrcCoor}
  \Ct=\frac\At\DSr\comma\Cy=\frac\DSr\Ar\comma\Cx=-\cos\Ath\period
\end{equation}
The metric \eqref{accM} becomes
\begin{equation}\label{Cmtrc}
\begin{split}
  \mtrc_\dS=\RTr^2\,&\Bigl(-(\Cy^2-1)\,\grad\Ct\formsq+\frac1{\Cy^2-1}\,\grad\Cy\formsq\\
          &\qquad\quad+\frac1{1-\Cx^2}\,\grad\Cx\formsq+(1-\Cx^2)\,\grad\ph\formsq\Bigr)\commae
\end{split}
\end{equation}
with the conformal factor $\RTr$ given by
\begin{equation}\label{RTr}
  \RTr=\frac{\DSr}{\Cy\cosh\acp-\Cx\sinh\acp}\period
\end{equation}
This is de~Sitter spacetime in the \vague{C-metric form}:
setting the mass and charge parameters, ${m}$ and ${e}$, equal to zero
in the the C-metric with a positive cosmological constant
(written in the form (2.8) of Ref.~\cite{KrtousPodolsky:2003}),
and choosing the acceleration parameter equal to
${A=\DSr^{-1}\abs{\sinh\acp}=\abs{\accl}}$,
we obtain the metric \eqref{Cmtrc}.

\subsection{Robinson-Trautman coordinates}
\label{ssc:RTcoor}

In order to arrive naturally to the Robinson-Trautman form of the metric,
notice that the coefficients in the metric \eqref{Cmtrc}
become singular at ${\Cy\to\pm1}$, similarly as they do on the horizon of the Schwarzschild spacetime in
the standard Schwarzschild coordinates. Analogously to that case, we choose a
\vague{tortoise-type} coordinate $\CTy$ by
\begin{equation}\label{CTy}
  \CTy=\frac12\log\abs{\frac{1-\Cy}{1+\Cy}}\period
\end{equation}
Similarly to the Schwarzschild case again, we introduce a suitable \emph{null}
coordinate $\RTu$ in terms of the radial and time coordinates $\Ct$ and $\CTy$ as follows:
\begin{equation}\label{RTu}
  \RTu=(\DSr\tanh\acp)\,(\Ct+\CTy)\period
\end{equation}
Together with the conformal factor $\RTr$
defined in Eq.~\eqref{RTr}, we arrive at the de~Sitter metric in coordinates
${\RTu,\,\RTr,\,\Cx,\,\ph}$ (cf.\ Eq.~\eqref{apx:RTthpmtrc}) which is very near to being in the
Robinson-Trautman form. However, there is a non-vanishing mixed metric
coefficient at ${\grad\RTu\stp\grad\Cx}$ which is absent in the
Robinson-Trautman metric. Such a term can be made to vanish by introducing a new
angular coordinate $\RTp$ by
\begin{equation}\label{RTp}
  \RTp=\arctanh\Cx+\frac\RTu\DSr\,\sinh\acp\period
\end{equation}
The de~Sitter metric then becomes
\begin{equation}\label{RTM}
    \mtrc_\dS=-\RTH\,\grad\RTu\formsq-\grad\RTu\stp\grad\RTr
          +\frac{\RTr^2}{\RTP^2}\,
      \bigl(\grad\RTp\formsq+\grad\ph\formsq\bigr)\commae
\end{equation}
where
\begin{gather}
   \RTH=-\frac{\RTr^2}{\DSr^2}
     +2\,\frac\RTr\DSr\,\sinh\acp\,
     \tanh\Bigl(\RTp-\frac\RTu\DSr\,\sinh\acp\Bigr)+1\commae
   \notag\\\label{RTMfc}
   \RTP=\cosh\Bigl(\RTp-\frac\RTu\DSr\,\sinh\acp\Bigr)\period
\end{gather}
This is precisely the form of the Robinson-Trautman metric---see,
e.g., \cite{Stephanietal:book}.
Tracking back the transformations leading to the metric \eqref{RTM}, the
connection between the Robinson-Trautman coordinates and the static coordinates
${\stt,\,\str,\,\tht,\,\ph}$ turns out to be not as complicated as our
procedure might have indicated, in particular, for the radial coordinate. We
find a nice formula for ${\RTr}$,
\begin{equation}\label{RTrst}
\RTr=\frac{\DSr}{\sqrt{1-\frac{\Ro^2}{\DSr^2}}}\,
     \Biggl(\Bigl(1-\frac{\str\Ro}{\DSr^2}\cos\tht\Bigr)^2
     -\Bigl(1-\frac{\str^2}{\DSr^2}\Bigr)
      \Bigl(1-\frac{\Ro^2}{\DSr^2}\Bigr)\Biggr)^{\frac12}\!\!\!\commae
\end{equation}
whereas the other two coordinates are simply expressed only in terms of
accelerated coordinates ${\At=\stt,\,\Ar,\,\Ath,\,\ph}$:
\begin{align}
\RTu&=\sqrt{1-\frac{\Ro^2}{\DSr^2}}\,\Biggl(\At+\frac\DSr2\log\abs{\frac{\Ar-\DSr}{\Ar+\DSr}}\Biggr)
\commae\\
\RTp&=\frac\Ro\DSr\,\Biggl(\frac\At\DSr+\frac12\log\abs{\frac{\Ar-\DSr}{\Ar+\DSr}}\Biggr)
      +\log\abs{\tan\frac\Ath2}\period\notag
\end{align}
Coordinates ${\Ar,\,\Ath}$ can then be obtained in terms of the original
static coordinates by using Eqs.~\eqref{stacc}.

The Robinson-Trautman coordinates with metric \eqref{RTM} are centered on the
accelerated particles. As with static or accelerated frames, we need two sets
of such coordinates to cover both $w_\npart$ and $w_\spart$. The relations
to the static coordinates become, of course, much simpler if the particles are
not accelerated, ${\Ro=0}$, and when both the Robinson-Trautman and static coordinates are
centered on the pole ${\spr=0}$:
\begin{equation}\label{RTstzeroa}
\begin{gathered}
  \RTr=\str\comma
  \RTp=\log\tan\frac\tht2\commae\\
  \RTu=\stt+\frac\DSr2\,\log\abs{\frac{\str-\DSr}{\str+\DSr}}\period
\end{gathered}
\end{equation}
However, even \vague{accelerated} Robinson-Trautman coordinates possess some
very convenient features. The radial coordinate $\RTr$ is an affine parameter
along null rays ${\RTu,\,\RTp,\,\ph=\text{constants}}$, normalized at
the particle's worldline by the condition
\begin{equation}\label{RTrnorm}
\cv[\mu]{\RTr}\, \mtrc_{\dS\mu\nu}\, \tens{u}^\nu=-1\commae
\end{equation}
where ${\tens{u}}$ is the particle's four-velocity. These null rays form a diverging but
nonshearing and nonrotating congruence of geodesics. The null vector
${\cvil{\RTr}}$, tangent to the rays, is parallelly propagated along them.
Its divergence is given by ${\covd_\mu\cv[\mu]{\RTr}=2/\RTr}$ so that $\RTr$ is both the
affine parameter and the luminosity distance (see, e.g., \cite{BondiBurgMetzner:1962}).
With Robinson-Trautman coordinates, one can also associate
a null tetrad (explicitly written down in the Appendix, Eq.~\eqref{apx:RTnulltetr})
which is parallelly transported along
the null rays from the particle (${\RTr=0}$) up to infinity (${\RTr=\infty}$).

\begin{figure}
\includegraphics{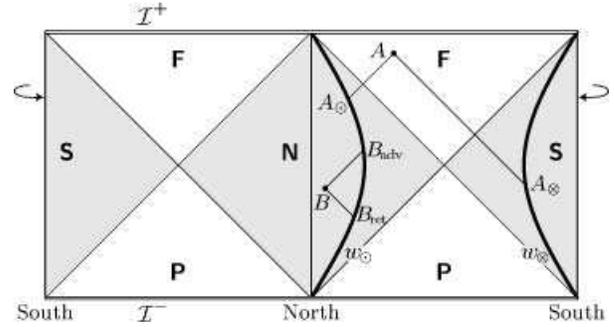}
\caption{\label{fig:affpar}%
The field at an event ${A}$ can be interpreted as ${1/2}$ of the sum
of the retarded fields produced by particle ${w_\npart}$ at ${A_\npart}$
and particle ${w_\spart}$ at ${A_\spart}$. The field at ${B}$ can be interpreted
as ${1/2}$ of the sum of the retarded and advanced effects from particle ${w_\npart}$.
The affine parameter distances ${BB_\ret}$ and ${BB_\adv}$ are equal,
the same being true for the distances ${AA_\npart}$ and ${AA_\spart}$.
}
\end{figure}

Owing to the boost symmetry of both the worldlines and de~Sitter
space, an interesting feature arises, which is analogous to the situation in
Minkowski space. Consider a point ${B}$ in region ${\Ndom}$
(Fig.~\ref{fig:affpar}). There are two generators of the null cone with the
origin at ${B}$ which cross the worldline ${w_\npart}$ at two points, ${B_\ret}$
and ${B_\adv}$. Then the affine parameter distance ${B B_\ret}$ is
the same as  ${B B_\adv}$. (In order to go towards the past from
${B_\adv}$ to ${B}$, the \vague{advanced} Robinson-Trautman coordinates built
on the past null cones with origins on ${w_\npart}$ can easily be introduced.)
This is evident because ${B}$ lies on one orbit of the boost Killing vector ${\cvil{\stt}}$
and a boost can be applied which leaves the worldline ${w_\npart}$ invariant
but moves ${B}$ into event ${B'}$ on the slice of time symmetry, ${\spt=0}$ (also ${\stt=\At=0}$), where
the particle is at rest. Then ${B_\adv}$ and ${B_\ret}$ move to the new points ${B'_\adv}$, ${B'_\ret}$,
which are located symmetrically with respect to ${\spt=0}$. The equality of
the affine parameter distances then follows from the symmetry immediately.
Similarly, for an event ${A}$ in region ${\Fdom}$ one can show that the affine
parameter distance ${A A_\npart}$ is equal to the distance ${AA_\spart}$ (see
Fig.~\ref{fig:affpar}). The point ${A}$  lies on a boost orbit (which now has
a spatial character) along which it can be brought, by an appropriate boost, to
the point located symmetrically between the worldlines ${w_\npart}$ and ${w_\spart}$
(lying so on the equator, ${\spr=\pi/2}$). The same consideration can, of
course, be applied to an event in the \vague{past} region ${\Pdom}$---showing
that the affine distances along future-oriented null rays from an event to the
particles are equal.

Although the symmetries just described are common to the worldlines of
uniformly accelerated particles in Minkowski and de~Sitter spacetimes, an
important difference exists. In Minkowski space, the affine parameter
distance along the null ray from an event on particle's worldline, such as
${A_\npart}$, to an \vague{observation} point ${A}$
is equal to the proper distance between ${A_\perp}$ and ${A_\npart}$
where ${A_\perp}$ is the \emph{orthogonal} projection of ${A}$ onto the
spacelike slice ${\stt=\stt(A_\npart)}$. This is not the case in de~Sitter
space if, as it appears natural, under an `orthogonal projection' we understand
the projection of the observation point ${A}$ onto the spacelike slice
${\stt=\text{constant}}$ containing ${A_\npart}$
performed along a timelike geodesic orthogonal to such a slice.
Nevertheless, the proper distance $s$ between ${A_\npart}$ and ${A_\perp}$
is still related to the affine parameter distance $\RTr$ by a simple expression:
\begin{equation}\label{RTrprojdist}
\frac\RTr\DSr=\tan\frac{s}{\DSr}\period
\end{equation}
This relation can be derived as follows. Consider, without loss of generality,
${A_\npart}$ located at the turning point of the particle ${w_\npart}$ at ${\stt=0}$.
The condition that the events $A$ and ${A_\npart}$ are connected by a null ray
implies that the distance $s$ between ${A_\npart}$ and ${A_\perp}$
is the same as the time interval between ${A_\perp}$ and $A$ as measured by
the metric \eqref{CEinsFact} of the conformally related static Einstein universe.
Since $A$ occurs at some time $\tlt$ whereas $A_\npart$ and $A_\perp$
at ${\tlt=\pi/2}$ (i.e., at static time ${\stt=0}$), this time interval is
equal to ${\DSr(\tlt-\pi/2)}$, cf.\ Eq.~\eqref{CEinsM}.
The  static radial coordinate $\str$ of $A$ thus reads (cf.\ Eqs.~\eqref{apx:sttrT=tltr},
\eqref{apx:sttrS=tltr})
\begin{equation}
\str=\frac{\sin\tlr}{\sin\tlt}=\frac{\sin\tlr}{\cos(s/\DSr)}\period
\end{equation}
The slice ${\stt=0}$ has a geometry of the 3-sphere of radius $\DSr$.
Using the standard law of cosines in spherical trigonometry for the sides of the triangle
spanned by $A_\npart$, $A_\perp$, and the north pole,
we can eliminate $\tlr$.
Finally, employing Eq.~\eqref{RTrst}, we obtain the result \eqref{RTrprojdist}.
Clearly, near the particle ${w_\npart}$ we have ${s\ll\DSr}$, and
Eq.~\eqref{RTrprojdist} then gives ${\RTr\approx s}$, as in Minkowski space.

In the following section we shall explore the character of the fields of the
particles ${w_\npart}$ and ${w_\spart}$. We shall see that the affine
parameter distance ${\RTr}$ will play most important role, simplifying their
description enormously. Namely, as we will see in
Section~\ref{ssc:Fcentonpart}, Eq.~\eqref{KfactRT},
the affine parameter ${\RTr}$ is identical to the
factor ${\Kfact}$ which will be introduced in the
following and will appear in all expressions for the fields.

\section{Fields of uniformly accelerated sources and their many faces}
\label{sc:Fields}

In this section we wish to construct the scalar and electromagnetic
fields of uniformly accelerated (scalar and electric) charges in
de~Sitter universe. A general procedure, suitable in case of any---not
necessarily uniform---acceleration would be to seek for appropriate
Green's functions. Alternatively, in particular for sources
moving along uniformly accelerated worldlines, we
can make use of the conformal relations between Minkowski and de~Sitter spaces,
and of the properties of scalar and electromagnetic fields under conformal
mappings. This method is advantageous not only for finding the fields in de~Sitter
spacetime, but also for understanding their relationships to the known fields
of corresponding sources in special relativity. The only delicate issue is the
fact that there are no conformal mappings between Minkowski and de~Sitter space
which are globally smooth. We discussed, in Section~\ref{sc:Coor}, how various
regions of one space can be mapped onto the regions of the other space. In
Ref.~\cite{BicakKrtous:2001} we carefully treated the fields at the
hypersurfaces where the conformal transformation fails to be regular. In order
to obtain well-behaved fields, one must continue analytically across such a
hypersuface the field obtained in one region into the whole de~Sitter space.
In Section~II in Ref.~\cite{BicakKrtous:2001}, we also analyzed in detail the
behavior of the scalar field wave equation with sources and of Maxwell's
equations with sources under (general) conformal transformations.

In Ref.~\cite{BicakKrtous:2001} we primarily concentrated on the absence of
purely retarded fields at the past infinity $\scri^-$ of de~Sitter
spacetime---in fact, in any spacetime in which $\scri^-$ is spacelike. In order to
analyze this problem we also considered, in addition to monopole charges, more
complicated sources like rigid and geodesic dipoles; and we constructed some
retarded solutions to show their patological features.
However, we confined
ourselves to the sources the worldlines of which start and end at the poles; we
did not employ coordinates best suited for exhibiting the properties of the
fields at future infinity $\scri^+$, and the frames corresponding to
cosmological models like flat ($k=0$) or hyperbolic  ($k=-1$) cosmological
coordinates; and we did not give the physical components of the fields. In the
following we shall find the fields and discuss their properties in various
physically important coordinate systems, in particular those significant at
$\scri^+$ or in a cosmological context. In the next section, we also obtain the
fields due to the uniformly accelerated scalar and electric charges starting at $\scri^-$ at
${\spr=\pi/2}$ (\vague{born at the equator}). This, among others, will be
important when we wish to regain the classical Born fields in the limit
${\Lambda\to0}$.

We start by using the analysis of the conformal behavior of the fields and
sources given in Section~II in \cite{BicakKrtous:2001}, and we also take over
from \cite{BicakKrtous:2001} the resulting forms of the fields due to the
sources starting and ending at the poles of de~Sitter space, as described in
standard coordinates.

\subsection{Fields in coordinates centered on the poles}
\label{ssc:Fcentsonpoles}

Consider two uniformly accelerated point sources starting at $\scri^-$ (i.e., at
${\spt\to-\infty}$, ${\tlt\to0}$) at the poles
${\spr=\tlr=0}$ and ${\spr=\tlr=\pi}$ (Fig.~\ref{fig:wldln_p}).
Their worldlines $w_\npart$, $w_\spart$ are given by Eqs.~\eqref{wlcs} (or
\eqref{wleStl}) in these standard (rescaled) coordinates, by Eqs.~\eqref{wlcf}
and \eqref{wlch} in the flat and hyperbolic cosmological coordinates, and by
Eqs.~\eqref{wlst} in the static coordinates.
Their simplest description is, of
course, given by ${\Ar=0}$ and ${\RTr=0}$ in the accelerated and Robinson-Trautman
coordinates since these frames are centered exactly on their worldlines.
In Section~\ref{sc:AccPart} we discussed physical velocities and
other properties of these particles.

Now, as noticed at the beginning of Section~\ref{sc:AccPart}, these \emph{two}
worldlines can be obtained by conformally mapping the worldline of \emph{one}
uniformly moving particle in Minkowski space into de~Sitter space. The fields
of \emph{uniformly moving} sources in Minkowski space are just boosted Coulomb
fields. Under a conformal rescaling of the metric,
${\mtrc_{\alpha\beta}\to\hat\mtrc_{\alpha\beta}=\Omega^2\mtrc_{\alpha\beta}}$,
the fields behave as follows: ${\SF\to\hat\SF=\Omega^{-1}\SF}$,
${\EMF_{\alpha\beta}\to\hat{\EMF}_{\alpha\beta}=\EMF_{\alpha\beta}}$ (see
\cite{BicakKrtous:2001}, Section~II, where the behavior of the source
terms is also analyzed). Hence, the fields due to two \emph{uniformly accelerated}
sources in de~Sitter spacetime can be obtained by conformally transforming the
boosted Coulomb fields in Minkowski spacetime. Employing the conformal mapping
\eqref{CFlatCoor}--\eqref{CFlatFact}, we arrive at the following results
(note \cite{nt:dictASDS}). The scalar field is given by the expression
\begin{equation}
\label{SFsym}
  \SF = \frac{\SFcharge}{4\pi}\frac1\Kfact  \commae
\end{equation}
where
\begin{equation}\label{Kfactsp}
\begin{split}
  \Kfact &= \DSr
  \Bigl[\Bigl(\sqrt{1\!+\!\accl^{2}\DSr^2}
  + \accl\DSr\cosh\frac\spt\DSr\sin\spr\cos\tht\Bigr)^{2}\\
  &\mspace{120mu}- \Bigl(1 - \cosh^2\frac\spt\DSr\,\sin^2\spr\Bigr)\Bigr]^{\frac12}
  \commae
\end{split}
\end{equation}
or, written in the standard rescaled coordinates,
\begin{equation}\label{Kfacttl}
  \Kfact= \DSr
  \Bigl[\!\Bigl(\sqrt{1\!+\!\accl^{2}\DSr^2}
  + \accl\DSr\frac{\sin\tlr}{\sin\tlt}\cos\tht\Bigr)^{\!2}
  \!- 1 + \frac{\sin^2\tlr}{\sin^2\tlt}\Bigr]^{\frac12}
  \!\period
\end{equation}
This field is produced by \emph{two} identical charges of magnitude $\SFcharge$ moving
along worldlines $w_\npart$ and $w_\spart$. It is smooth everywhere outside
the charges and it can be written as a symmetric combination of retarded and
advanced effects from both charges (cf.\ Eq.~(6.6) in \cite{BicakKrtous:2001}).

Similarly to the scalar-field case, by using conformal technique the
electromagnetic field produced by two uniformly accelerated charges moving
along $w_\npart$ and $w_\spart$ can be obtained in the form
\begin{equation}\label{EMFsymcs}
\begin{split}\raisetag{18pt}
\EMF =& \frac{\EMcharge}{4\pi}\,
\frac{\DSr^2}{\Kfact^3}\cosh\frac\spt\DSr\,
\Bigl[\,\accl\DSr \sin\spr\cos\spr\sin\tht\,
\grad\spt\wedge\grad\tht\\
&\;-\!\Bigl(\!\sqrt{1\!+\accl^{\,2}\DSr^2}\cosh\frac\spt\DSr\sin\spr\! + \accl\DSr \cos\tht\Bigr)\,
\grad\spt\wedge\grad\spr\\
&\;+\accl\DSr^2 \sinh\frac\spt\DSr\cosh\frac\spt\DSr\sin^2\spr\sin\tht\,
\grad\spr\wedge\grad\tht\,\Bigr]\commae
\end{split}
\end{equation}
where $\Kfact$ is again given by Eq.~\eqref{Kfactsp}. As in the
scalar-field case, the field is smooth, non-vanishing in the whole de~Sitter
spacetime and involving thus both retarded and advanced effects (cf.\
Section~VIIA in \cite{BicakKrtous:2001}). However, an important difference
between the scalar and electromagnetic case exists: the magnitude of the
scalar charges is the same, whereas the electromagnetic charges producing the
fields \eqref{EMFsymcs} have \emph{opposite} signs. This is analogous to the
situation in Minkowski spacetime described in Section~\ref{sc:BornM} (see the
discussion below Eq.~\eqref{EMBornM}). At the root of this fact appears to be
CPT theorem---cf.\ \cite{Bicak:1968} for the analogous gravitational case where
the masses of the particles uniformly accelerated in the opposite direction are the same.
In de~Sitter spacetime, as in any spacetime with compact spacelike sections, a simpler argument
exists: the total charge in a compact space must vanish as a consequence of the
Gauss theorem \cite{BicakKrtous:2001}.

To gain a better physical insight into the electromagnetic fields, we
shall introduce the orthonormal tetrad ${\{\cbv{\mu}\}}$ and
the dual tetrad ${\{\cbf{\mu}}\}$ tied to
each coordinate frame used, and we shall
decompose the electromagnetic field $\EMF$ into the electric and magnetic
parts. Such a decomposition, of course, depends on the choice of the tetrad.
For example, in the standard spherical coordinates
${\spt,\,\spr,\,\tht,\,\ph}$ the electromagnetic field (2-form) $\EMF$ can be
written as
\begin{equation}\label{FtoEBdecomp}
\begin{split}
\EMF&=\EME^\spr\,\cbf{\spr}\wedge\cbf{\spt}+\EME^\tht\,\cbf{\tht}\wedge\cbf{\spt}+\EME^\ph\,\cbf{\ph}\wedge\cbf{\spt}\\
&+\EMB^\spr\,\cbf{\tht}\wedge\cbf{\ph}+\EMB^\tht\,\cbf{\ph}\wedge\cbf{\spr}+\EMB^\ph\,\cbf{\spr}\wedge\cbf{\tht}\commae
\end{split}
\end{equation}
and the electric and magnetic field spatial vectors are given in terms of their
\emph{frame} components as follows:
\begin{equation}\label{EBcomp}
\begin{split}
\EME&=\EME^\spr\,\cbv{\spr}+\EME^\tht\,\cbv{\tht}+\EME^\ph\,\cbv{\ph}\commae\\
\EMB&=\EMB^\spr\,\cbv{\spr}+\EMB^\tht\,\cbv{\tht}+\EMB^\ph\,\cbv{\ph}\period
\end{split}
\end{equation}
In the present case of the standard spherical coordinates, using the explicit
forms of the tetrad given in the Appendix (Eqs.~\eqref{apx:sphtetr}), we find
\begin{align}
&\EME_{\cfsp} = \frac{\EMcharge}{4\pi}
\frac{\DSr}{\Kfact^3}
\Bigl[-\accl\DSr \cos\spr\sin\tht\;\cbv{\tht}\notag\\
&\mspace{40mu}+\Bigl(\sqrt{1+\accl^{\,2}\DSr^2}\,\cosh\frac\spt\DSr\,\sin\spr
+ \accl\DSr \cos\tht\Bigr)\,\cbv{\spr}\Bigr]\commae\notag\\
&\EMB_{\cfsp} = \frac{\EMcharge}{4\pi}\,
\frac{\accl\DSr^2}{\Kfact^3}\; \sinh\frac\spt\DSr\,\sin\spr\sin\tht\;\cbv{\ph}
\period\label{EMEBsymsp}
\end{align}

In the Appendix the orthonormal tetrads tied to the coordinate systems considered
in this paper are all listed explicitly. The only exception is the
Robinson-Trautman coordinate system with one coordinate null and thus with a
nondiagonal metric; in this case the null tetrad is given in which the
Newman-Penrose-type components are more telling.

The tetrad components of the electric intensity and the magnetic induction vectors
are physically meaningful objects:
they can be measured by observers who move with the four-velocities given by the
timelike vector of the tetrad (as, e.g., $\cbv{\spt}$ for spherical cosmological observers),
and are equipped with an orthonormal triad of the spacelike vectors (e.g.,
${\cbv{\spr},\,\cbv{\tht},\,\cbv{\ph}}$).

We first list the resulting electromagnetic field tensor and its electric and
magnetic parts in the coordinate systems
centered on the poles ${\spr=0,\pi}$. The scalar field is always given by
expression \eqref{SFsym}, the explicit form of the scalar factor $\Kfact$ changes
according to the coordinates used. Since this factor enters all the
electromagnetic quantities as well, we always give $\Kfact$ first and then write the
electromagnetic field quantities.

In the flat cosmological coordinates (see Eqs.~\eqref{PCFlatCoor},
\eqref{PCFlatM}) we find:
\begin{equation}\label{Kfactcf}
\begin{split}
  \Kfact &=
  \DSr\Bigl[\Bigl(\cosh\acp
  -\sinh\acp\, \frac\MVr\DSr\, \exp\frac\MVet\DSr\, \cos\tht\Bigr)^2\\
  &\mspace{140mu}- \Bigl(1-\frac{\MVr^2}{\DSr^2}\,\exp\Bigl(2\frac\MVet\DSr\Bigr)\Bigr)\Bigr]^{\frac12}
  \commae
\end{split}
\end{equation}
\begin{equation}\label{EMFsymcf}
\begin{split}\raisetag{18pt}
\EMF &= -\frac{\EMcharge}{4\pi}\,
\frac1{\Kfact^3}\,
\exp\frac\MVet\DSr\,
\Bigl[\,\DSr\MVr\,\sinh\acp\, \sin\tht\,
\grad\MVet\wedge\grad\tht\\
&\quad+\Bigl(\MVr\,\cosh\acp\,\exp\frac\MVet\DSr -\DSr\sinh\acp\, \cos\tht\Bigr)\,
\grad\MVet\wedge\grad\MVr\\
&\quad+ \MVr^2\,\sinh\acp\, \exp\Bigl(2\frac\MVet\DSr\Bigr)\sin\tht\,
\grad\MVr\wedge\grad\tht\,\Bigr]\commae
\end{split}
\end{equation}
\begin{equation}\label{EMEBsymcf}
\begin{aligned}
\EME_{\cfcf} &= \frac{\EMcharge}{4\pi}
\frac{\DSr}{\Kfact^3}
\Bigl[\,\sinh\acp\; \sin\tht\;\cbv{\tht}\\
&\mspace{15mu}-\!\Bigl(\cosh\acp\frac{\MVr}{\DSr}
\exp\frac\MVet\DSr-\sinh\acp \cos\tht\Bigr)\,\cbv{\MVr}
\Bigr]\commae\\
\EMB_{\cfcf} &= -\frac{\EMcharge}{4\pi}\,
\frac{\DSr\sinh\acp}{\Kfact^3}\,\frac{\MVr}{\DSr}\,\exp\frac\MVet\DSr\, \sin\tht\;\cbv{\ph}
\period
\end{aligned}
\end{equation}
In the hyperbolic cosmological coordinates (see Eqs.~\eqref{PCHypCoor},
\eqref{PCHypM}), the results are slightly lengthier:
\begin{equation}\label{Kfactch}
\begin{split}\raisetag{18pt}
  \Kfact &= \DSr
  \Bigl[\Bigl(\cosh\acp-\sinh\acp\,\sinh\frac\cht\DSr\,\sinh\frac\chr\DSr\,\cos\tht\Bigr)^{2}\\
  &\mspace{100mu}- \Bigl(1 - \sinh^2\frac\cht\DSr\,\sinh^2\frac\chr\DSr\Bigr)\Bigr]^{\frac12}
  \commae
\end{split}
\end{equation}
\begin{equation}\label{EMFsymch}
\begin{split}
\EMF &= -\frac{\EMcharge}{4\pi}\,
\frac\DSr{\Kfact^3}\,\Bigl[\;\sinh\frac\cht\DSr\\
&\;\times\!\!\bigl(\cosh\acp\,\sinh\frac\cht\DSr\,\sinh\frac\chr\DSr -\sinh\acp\,\cos\tht\bigr)
\,\grad\cht\wedge\grad\chr\\
&\;+\sinh\acp\,\sinh\frac\cht\DSr\,\sinh\frac\chr\DSr\,\cosh\frac\chr\DSr\,\sin\tht\,
\DSr\grad\cht\wedge\grad\tht\\
&\;+\sinh\acp\,\sinh^2\frac\cht\DSr\,\cosh\frac\cht\DSr\,\sinh^2\frac\chr\DSr\,\sin\tht\,
\DSr\grad\chr\wedge\grad\tht
\Bigr]\!\commae
\end{split}
\end{equation}
\begin{align}
&\EME_{\cfch} = \frac{\EMcharge}{4\pi}
\frac{\DSr}{\Kfact^3}
\Bigl[\,\sinh\acp\,\cosh\frac\chr\DSr\sin\tht\,\cbv{\tht}\notag\\
&\qquad\quad+\bigl(\cosh\acp\,\sinh\frac\cht\DSr\,\sinh\frac\chr\DSr
-\sinh\acp\,\cos\tht\bigr)\,\cbv{\chr}\Bigr]\commae\notag\\
&\EMB_{\cfch} = -\frac{\EMcharge}{4\pi}
\frac{\DSr}{\Kfact^3} \sinh\acp\cosh\frac\cht\DSr\sinh\frac\chr\DSr\sin\tht\;\cbv{\ph}
\period\label{EMEBsymch}
\end{align}
Much simpler expressions for the fields arise in the static coordinates (see
Eqs.~\eqref{PStCoor}, \eqref{PStM}). We obtain
\begin{equation}\label{KfactTR}
  \Kfact^2  = \frac{(\DSr^2-\str \Ro \cos\tht)^2}{(\DSr^2-\Ro^2)}-(\DSr^2-\str^2)\commae
\end{equation}
\begin{equation}\label{EMFsymTR}
\begin{split}\raisetag{20pt}
\EMF = & -\frac{\EMcharge}{4\pi}\,
\frac{\DSr}{\sqrt{\DSr^2-\Ro^2}}\,
\frac1{\Kfact^3}\;
\biggl[(\str-\Ro\cos\vartheta)\; \grad \stt \wedge \grad \str\\
&\mspace{50mu}+ \Bigl(1-\frac{\str^2}{\DSr^2}\Bigr) \str \Ro \sin\tht \;
\grad \stt \wedge \grad\tht\biggr]\commae
\end{split}
\end{equation}
\begin{align}
&\EME_{\cfst} = \frac{\EMcharge}{4\pi}\,\frac1{\Kfact^3}
\biggl[\frac{\DSr(\str-\Ro\cos\vartheta)}{\sqrt{\DSr^2-\Ro^2}}\;\cbv{\str}
+ \Ro \sin\tht \;\cbv{\tht}\biggr]\commae\notag\\
&\EMB_{\cfst} = 0\period\label{EMEBsymTR}
\end{align}

Since for practical calculations and for an understanding of the conformal
relations between Minkowski  and de~Sitter spaces the rescaled coordinates are
very useful, we also give the fields in these coordinates.
The rescaled coordinates are tied with the same orthonormal tetrad as non-rescaled
ones, and they define the same splitting into electric and magnetic parts
($\EME$ and $\EMB$ are the same spatial vectors); the functional dependence on
the coordinates, however, is different.
In the standard rescaled (conformally Einstein) coordinates
(see Eqs.~\eqref{CEinsCoor}--\eqref{CEinsFact}),
which cover the whole de~Sitter spacetime including its
conformal infinities globally, we get Eq.~\eqref{Kfacttl} for ${\Kfact}$ and
\begin{equation}\label{EMFsymtl}
\begin{split}\raisetag{18pt}
\EMF=& -\frac{\EMcharge}{4\pi}\,
\frac1{\Kfact^3}
\frac{\DSr^3}{\sin^3\tlt}
\Bigl[\,\accl\DSr \cos\tlt\sin^2\tlr\sin\tht\,
\grad\tlr\wedge\grad\tht\\
&+\Bigl(\sqrt{1+\accl^{\,2}\DSr^2}\sin\tlr + \accl\DSr \sin\tlt\cos\tht\Bigr)\,
\grad\tlt\wedge\grad\tlr\\
&-\accl\DSr \sin\tlt\cos\tlr\sin\tlr\sin\tht\,
\grad\tlt\wedge\grad\tht\,\Bigr]\commae
\end{split}
\end{equation}
\begin{equation}\label{EMEBsymtl}
\begin{aligned}
\EME_{\cfsp} &= \frac{\EMcharge}{4\pi}
\frac{\DSr}{\Kfact^3}
\Bigl[-\accl\DSr \cos\tlr\sin\tht\;\cbv{\tht}\\
&\qquad+
\Bigl(\!\sqrt{1+\accl^{\,2}\DSr^2}\;\frac{\sin\tlr}{\sin\tlt}
+ \accl\DSr \cos\tht\Bigr)\,\cbv{\spr}\Bigr]\commae\\
\EMB_{\cfsp} &= -\frac{\EMcharge}{4\pi}\,
\frac{\accl\DSr^2}{\Kfact^3}\; \cot\tlt\sin\tlr\sin\tht\;\cbv{\ph}
\commae
\end{aligned}
\end{equation}
whereas in the flat rescaled cosmological coordinates
\eqref{CFlatCoor}--\eqref{CFlatFact}, which cover globally the conformally
related Minkowski space (see also Fig.~\ref{fig:dSMinkV}), we arrive at
\begin{equation}\label{KfactMV}
  \Kfact =
  \DSr\biggl[\Bigl(\cosh\acp
  +\sinh\acp \frac\MVr\MVt \cos\tht\Bigr)^2
  \!- \Bigl(1-\frac{\MVr^2}{\MVt^2}\Bigr)\biggr]^{\frac12}
  \commae
\end{equation}
\begin{equation}\label{EMFsymMV}
\begin{split}
\EMF =\,& \frac{\EMcharge}{4\pi}\,
\frac1{\Kfact^3}
\frac{\DSr^3}{\MVt^3}
\Bigl[\,\sinh\acp\; \MVr^2\sin\tht\,
\grad\MVr\wedge\grad\tht\\
&\;\;+\bigl(\cosh\acp\;\MVr +\sinh\acp\; \MVt\cos\tht\bigr)\,
\grad\MVt\wedge\grad\MVr\\
&\;\;-\sinh\acp\; \MVt\,\MVr\sin\tht\,
\grad\MVt\wedge\grad\tht\,\Bigr]\commae
\end{split}
\end{equation}\pagebreak[2]\vspace*{-6ex}
\begin{equation}\label{EMEBsymMV}
\begin{aligned}
\EME_{\cfcf} &= \frac{\EMcharge}{4\pi}
\frac{\DSr}{\Kfact^3}
\biggl[\,\sinh\acp\; \sin\tht\;\cbv{\tht}\\
&\qquad-\Bigl(\!\cosh\acp\,\frac{\MVr}{\MVt}
+\sinh\acp \cos\tht\Bigr)\,\cbv{\MVr}\biggr]\commae\\
\EMB_{\cfcf} &= \frac{\EMcharge}{4\pi}\,
\frac{\DSr\sinh\acp}{\Kfact^3}\,\frac{\MVr}{\MVt}\, \sin\tht\;\cbv{\ph}
\period
\end{aligned}
\end{equation}

In various contexts the electromagnetic field four-potential form $\EMA$, implying
the field ${\EMF=\grad\EMA}$, may be needed. In the standard rescaled (conformally
Einstein) coordinates the potential reads
\begin{equation}\label{EMA}
\begin{split}
\mspace{-10mu}
\EMA=-\frac{\EMcharge}{4\pi} \frac{1}{\Kfact} &\frac{\DSr}{\sin\tlt}\,
\frac{\sqrt{1+\accl^{\,2}\DSr^2}\sin\tlt + \accl\DSr \sin\tlr\cos\tht}{\sin^2\tlt-\sin^2\tlr}\\
&\times\Bigl(\sin\tlt\cos\tlr\,\grad\tlt-\cos\tlt\sin\tlr\,\grad\tlr\Bigr)
\period
\end{split}\raisetag{24pt}
\end{equation}
From this expression the frame components can easily be obtained and the
four-potential form can be transformed directly to any coordinate system of interest.
The four-potential acquires a particularly simple form in static coordinates:
\begin{equation}
\EMA=-\frac{\EMcharge}{4\pi} \,\frac{1}{\Kfact}\;
\frac{\DSr^2-\str\Ro\cos\tht}{\DSr\sqrt{\DSr^2-\Ro^2}}\;\grad\stt
\period
\end{equation}

Inspecting now the expressions \eqref{EMFsymcs}--\eqref{EMEBsymMV}, we first
notice few basic features of the fields. As a consequence of the axisymmetry,
the azimuthal $\ph$ component of the electric field vanishes. On the other
hand, only the azimuthal $\ph$ component of the magnetic field is non-zero. At
the axis of symmetry, ${\tht=0,\pi}$, the latitudinal $\tht$ component of the
electric field and magnetic field vanish as ${\sim\sin\tht}$. The electric
field points along the axis.

In the classical Born solution in Minkowski space, both charges are, at any time,
located symmetrically with respect to the equatorial plane ${\tht=\pi/2}$.
Consequently, the radial part of the electric field vanishes for ${\tht=\pi/2}$
(cf.\ Eq.~\eqref{EMBornM}). In de~Sitter spacetime the charges outgoing from
the poles are, at all times, symmetrically located with respect to the sphere
${\spr=\pi/2}$ (illustrated as the circle in Fig.~\ref{fig:dS}). We thus expect
the $\tht$ component of the electric field to vanish for ${\spr=\pi/2}$.
This, indeed, follows from Eq.~\eqref{EMEBsymsp}. This symmetry can be seen only in the
standard spherical coordinates since the sphere ${\spr=\pi/2}$ is not covered
by the hyperbolic cosmological coordinates and in the flat cosmological
coordinates only one particle occurs.

Another typical feature of the Born solution in Minkowski space is its time
symmetry. As a consequence, the magnetic field vanishes at ${\MOt=0}$ (cf.\ Eq.
\eqref{EMBornM}). In the past, it was this fact which led some investigators,
W.~Pauli \cite{Pauli:book} among them, to the conclusion that
there is ``no formation of a wave zone nor any corresponding
radiation'' since ${\EMB=0}$ at ${\MOt=0}$.
However, it is not at a spacelike hypersurface
${\MOt=\text{constant}}$ but at ${\scri^+}$, which is reached by taking
${\MOu=\MOt-\MOr}$ constant, ${\MOt,\,\MOr\to\infty}$, where the Born field has
typical radiative features, i.e., ${\abs{\EME}=\abs{\EMB}\sim\MOr^{-1}}$ (see
\cite{FultonRohrlich:1960}, \cite{Bicak:1968}, \cite{BicakKrtous:2002}). In our
generalized Born solution, the time symmetry of the fields is clearly
demonstrated in the global standard coordinates: under inversion
${\spt\to-\spt}$ the electric field in Eq.~\eqref{EMEBsymsp} is invariant,
whereas the magnetic field changes the sign; ${\EMB_\ph=0}$ at ${\spt=0}$. The
field also exhibits radiative character when we approach ${\scri^+}$ in an
appropriate way, as it is briefly indicated in \cite{BicakKrtous:2002}. As mentioned
in the Introduction, a detailed analysis of the radiative properties of the
generalized Born field will be given elsewhere.

The fields take the simplest form in the static coordinates,
Eq.~\eqref{EMFsymTR}. In these coordinates the particles are at rest, and they both have
a constant distance from the poles; their world lines are the orbits of the
\vague{static} Killing vector ${\cvil{\stt}}$ of de~Sitter space. The electric
field is time independent, the magnetic field vanishes. This is fully analogous
to the Born field in Minkowski spacetime: it is static, and
purely electric in the Rindler coordinates, the time coordinate of which is
aligned along the orbit of the boost Killing vectors (see, e.g.,
\cite{Boulware:1980}). However, as we discussed in Section~\ref{sc:Coor},
the static coordinates cover only a \vague{half} of de~Sitter space. In
the other half, the Killing vector ${\cvil{\stt}}$ becomes spacelike. It is in
this non-static domain (regions ${\Fdom}$ and ${\Pdom}$ in
Fig.~\ref{fig:affpar}) where we expect, in analogy with the results in
Minkowski spacetime, to find fields which have radiative properties.
${\cvil{\stt}}$ is the Killing vector also in the non-static regions,
however, it is spacelike here, as it is typical for a boost Killing vector in
Minkowski space. The fields of uniformly accelerated charges in de~Sitter
spacetime are invariant under the boosts along ${\cvil{\stt}}$ everywhere. They
are thus boost-rotation symmetric as the Born fields in Minkowski spacetime.

In the cosmological coordinates, respectively, in their rescaled versions, the
fields are, of course, time dependent. Here we expect the effects of the
expansion/contraction of de~Sitter universe to be manifested. Indeed, considering
in any of the cosmological frames the spatial coordinates fixed, and examining
the fields along the timelike geodesics, we discover that the fields
exponentially decay at large times, i.e., as ${\scri^+}$ is approached. More
specifically, with the spherical coordinates ${\spr,\,\tht,\,\ph}$ fixed,
the factor ${\Kfact}$ behaves as ${\exp(\spt/\DSr)}$ at large times ${\spt}$,
and hence, we obtain ${\EME_{\cfsp}\approx
c_1\exp(-2\spt/\DSr)\cbv{\spr}}{+c_2\exp(-3\spt/\DSr)\cbv{\tht}}$,
${\EMB_{\cfsp}\approx b_1\exp(-2\spt/\DSr)\cbv{\ph}}$, ${c_1,\,c_2,\,b_1}$ being
constants. The electric field thus becomes radial at large ${\spt}$. Similarly,
in flat cosmological coordinates we find
${\EME_{\cfcf}\approx c_1\exp(-2\MVet/\DSr)\cbv{\MVr}}$,
${\EMB_{\cfcf}\approx b_1\exp(-2\MVet/\DSr)\cbv{\ph}}$.
In the hyperbolic cosmological coordinates the proper time ${\cht}$
appears instead of ${\MVet}$. The rapid decay of the fields along timelike
worldlines at large times is caused by the
exponential expansion (at large times) of the spatial slices ${\spt=\text{constant}}$
(respectively ${\MVet,\,\cht=\text{constant}}$). Although our fields are
just test fields, their exponential decay is another manifestation of the
\vague{cosmic no-hair phenomenon}: geodesic observers in spacetimes with
${\Lambda>0}$ see at large times these spacetimes to approach the de~Sitter
universe exponentially fast---the universe becomes \vague{bald} (see, e.g.,
\cite{Maeda:1989,Rendall:2004}). Clearly, as one approaches \emph{past} infinity ${\scri^-}$
(${\spt\to-\infty}$), the fields also decay exponentially.

It is interesting to notice the character of the field as it would be seen by
the observers at rest with respect to the hyperbolic cosmological coordinates
in the limit at which the particles \vague{enter} the region covered by these observers
across the horizon ${\tlt=\tlr}$ (cf.\ Fig.~\ref{fig:dShyp}),
given in the hyperbolic coordinates by ${\cht\to0}$, ${\chr\to\infty}$.
As discussed in Section~\ref{sc:AccPart}, the observed
velocity \eqref{obsvelch} of the charges at this boundary is (in the limit)
equal to the velocity of light.  Employing the transformation formulas
\eqref{apx:chtr=tltr}, it is easy to see that at this boundary
${\abs{\sinh(\cht/\DSr)\,\sinh(\chr/\DSr)}\to1}$.
Hence, the factor ${\Kfact}$ is finite here (as it is evident from its scalar
character and its finiteness in the global standard coordinates). Also, the radial
part of the electric field remains finite.
However, ${\EME^\tht}$ diverges as ${\exp(\chr/\DSr)}$ here, indicating that the
field has a character of an impulse, in fact, rather of an impulsive wave:
indeed, Eq.~\eqref{EMEBsymcf} implies ${\abs{\EME^\tht}=\abs{\EMB^\ph}}$. The
situation appears to be analogous to the field of a static charge viewed from
an inertial frame boosted to the velocity of light in Minkowski spacetime
(see, e.g., \cite{Jackson:book}).

\subsection{Fields in coordinates centered on the particles}
\label{ssc:Fcentonpart}

As expected, a remarkable simplification occurs when the fields are evaluated
in the coordinates at the origin of which the charges are situated at all times.
Since the accelerated coordinates ${\At,\,\Ar,\,\Ath}$ and the
C-metric-like coordinates are simply related by  Eqs.~\eqref{CmtrcCoor},
the discussion of the field properties is the same in both these frames.
Namely, notice that both coordinate systems are
tied with the same orthonormal tetrad, and they thus define the same splitting of the field
into the electric and magnetic parts.
In these coordinates, we find the factor ${\Kfact}$ to read
\begin{equation}\label{SFsymA}
\begin{split}
  \Kfact &= \cosh\acp\,\frac1\Ar\,+\sinh\acp\,\frac1\DSr\cos\Ath\\
  &=\frac1\DSr \bigl(\Cy\cosh\acp-\Cx\sinh\acp\bigr)\period
\end{split}
\end{equation}
The scalar field is again given by ${\SF=(\SFcharge/4\pi)\,\Kfact^{-1}}$, and
the electromagnetic field also acquires now an extremely simple form:
\begin{equation}\label{EMFsymA}
\EMF = \frac{\EMcharge}{4\pi}\,\frac1{\Ar^2}\,\grad\Ar\wedge\grad\At
= \frac{\EMcharge}{4\pi}\,\grad\Ct\wedge\grad\Cy\commae
\end{equation}
\begin{equation}\label{EMEBsymA}
\EME_{\cfA} = \frac{\EMcharge}{4\pi}\,
\frac1{\Kfact^2}\,\cbv{\Ar}\comma
\EMB_{\cfA} = 0\period
\end{equation}
The magnetic field vanishes in the frame tied to the accelerated
and C-metric coordinates, the electric field has precisely the Coulomb
form, with the factor ${\Kfact}$ playing the role of a distance.

As signalized above already, the factor ${\Kfact}$ turns out to be the
Robinson-Trautman radial coordinate (see Eq.~\eqref{KfactRT} below), i.e., the
affine parameter distance allong null geodesics.
The geometrical role of ${\Kfact}$ was elucidated in Section~\ref{ssc:RTcoor}.
Considering a fixed point in de~Sitter universe and a light cone emanating from
this point, three typical situations can arise as illustrated in
Fig.~\ref{fig:affpar}.
For a point ${B}$ from the regions ${\Ndom}$ or ${\Sdom}$,
there are two null geodesics, one past-pointing, the other
future-pointing, each of which crosses the worldline of \emph{the same} particle, say ${w_\npart}$
(in case of ${B}$ from ${\Ndom}$), at points ${B_\ret}$ and ${B_\adv}$ (see
Fig.~\ref{fig:affpar}).
Since ${\Kfact}$ is equal to the (specific) affine
parameter distance which is the same from ${B_\ret}$ as from ${B_\adv}$ (see
Section~\ref{ssc:RTcoor}), we can interpret the field \eqref{EMEBsymA} as arising from
purely retarded, respectively, advanced effects from ${B_\ret}$, respectively
${B_\adv}$;
or, equivalently, as a combination of retarded and advanced effects from these points.
In the second situation, when the fixed point, say ${A}$, is located \vague{above the
roof} (in the region ${\Fdom}$), there are two past-oriented
null geodesics emanating from it which cross now \emph{both} particles ${w_\npart}$ and ${w_\spart}$
at points ${A_\npart}$ and ${A_\spart}$ (see Fig.~\ref{fig:affpar}). The field can be
interpreted as arising from retarded effects only, either as a combination from
both particles ${w_\npart}$ and ${w_\spart}$, or as the retarded field from just one
of them. Finally, for a point from the region ${\Pdom}$ the field can analogously be interpreted
in terms of advanced effects.

As we discussed in Section~\ref{ssc:Acoor} and illustrated in detail in the Appendix,
the accelerated coordinates (similarly as the static coordinates to which they go
over for a vanishing acceleration) are \emph{static}, i.e., the
vector ${\cvil{\At}}$ tangent to the orbits of the Killing vector is timelike, only in
the regions ${\Ndom}$ and ${\Sdom}$ (cf. Figs.~\ref{fig:Acc}, \ref{fig:affpar}).
Observers following the orbits of the Killing vector are thus confined to the regions ${\Ndom}$ and
${\Sdom}$, and they cannot detect the fields in the region ${\Fdom}$ (respectively
${\Pdom}$). Nevertheless, notice that although the time coordinate ${\At}$
diverges at the horizon ${\str=\DSr}$, the radial coordinate ${\Ar}$ is perfectly finite
there, ${\Ar=\DSr}$ (cf.\ Eq.~\eqref{stacc} with ${\str=\DSr}$), and the field
\eqref{EMFsymA} is meaningful in the region ${\Fdom}$ (or ${\Pdom}$) as
well. Since here the roles of the coordinates ${\Ar}$ and ${\At}$ are interchanged, ${\Ar}$
becoming a \emph{time coordinate}, the field becomes \emph{time-dependent}. As mentioned
above, we do not expect to find radiative properties in the regions ${\Ndom}$ and
${\Sdom}$. Indeed, in accelerated coordinates the field \eqref{EMEBsymA}
is static Coulomb field, with ${\Kfact}$ playing the role of a distance.
However, the radiative properties of the whole field in the wave zone in the region ${\Fdom}$
are not evident from the time-dependent, purely electric field in the
accelerated coordinates with ${\Ar}$ as a time coordinate.

It is worthwhile to recall that with finite sources in Minkowski spacetime the
field at \emph{any} event is of a \emph{general} algebraic type; only
asymptotically, at large distances, its features approach those of a
\defterm{null field} (${\EME^2-\EMB^2=0}$, ${\EME\scp\EMB=0}$), if there is a
radiation (see, e.g., \cite{GoldbergKerr:1964,PenroseRindler:book}). In case of a non-null field, one can
always introduce a frame in which the
electric and magnetic fields are collinear, or, in the
language of the Newman-Penrose formalism,
to choose such a null tetrad ${\kG,\,\lG,\,\mG,\,\bG}$,
corresponding to the orthonormal tetrad, that the only non-vanishing null-tetrad
component is ${\EMP{1}=\frac1{2\sqrt2}\,(\EME-i\EMB)\scp(\kG-\lG)}$ (see
Eqs.~\eqref{apx:RTnulltetr} for the explicit expressions of the null tetrad and
Eqs.~\eqref{PhiDef} for the null-tetrad components of the electromagnetic field).
Such a situation arises precisely for the null
tetrad associated with the accelerated coordinates:
the null-tetrad components are simply
\begin{equation}\label{EMPSymA}
\EMP{1}^{\cfA} = -\frac12\,\frac{\EMcharge}{4\pi}\,\frac1{\Kfact^2}\comma
\EMP{0}^{\cfA} = \EMP{2}^{\cfA} = 0\period
\end{equation}
The vanishing of the other
two null-tetrad components, ${\EMP{0}^{\cfA}}$ and ${\EMP{2}^{\cfA}}$, has a deeper
algebraic explanation: the null tetrad tied to the accelerated
coordinates is special in the sense that it contains \emph{both}
principal null directions of the electromagnetic field.
Inspecting the form of the null tetrad constructed from the orthonormal
tetrad \eqref{apx:AccTetrad}, we observe that both these principal null directions
are tangent to the \vague{radial} surfaces ${\Ath,\,\ph=\text{constant}}$
in the accelerated coordinates.

The radiative properties are well exhibited in the Robinson-Trautman
coordinates. As we discussed in Section~\ref{ssc:RTcoor}, these
coordinates are tied to the future null cones centered on the worldline of a
particle. We consider the null cones with vertices on the particle ${w_\npart}$.
Let us recall that the radial coordinate ${\RTr}$ is the affine parameter along
the generators of the null cones, each of which is given by ${\RTu,\,\RTp,\,\ph}$
fixed. Now, as mentioned above, it turns out that the \emph{factor
${\Kfact}$ is precisely equal to this affine parameter ${\RTr}$}:
\begin{equation}\label{KfactRT}
\Kfact=\RTr\period
\end{equation}
The scalar field is then simply given by
\begin{equation}\label{SFsymRT}
  \SF = \frac{\SFcharge}{4\pi}\,\frac1\RTr\period
\end{equation}
A remarkably nice form also acquires the electromagnetic field:
\begin{equation}\label{EMFsymRT}
\begin{split}
\EMF
&= \frac{\EMcharge}{4\pi}\,
\Bigl(\frac1{\RTr^2}\,\grad\RTu\wedge\grad\RTr
+\accl\sin^2\Ath\,\grad\RTu\wedge\grad\Ath\Bigr)\\
&= \frac{\EMcharge}{4\pi}\,
\Bigl(\frac1{\RTr^2}\,\grad\RTu\wedge\grad\RTr
+\accl\sin^2\Ath\,\grad\RTu\wedge\grad\RTp\Bigr)\period
\end{split}
\end{equation}
The Newman-Penrose scalars are defined in terms of the null tetrad \eqref{apx:RTnulltetr}, which is
parallelly propagated from the source to the \vague{observation point} along the
rays ${\RTu,\,\RTp,\,\ph=\text{constant}}$.
They look as follows:
\begin{equation}\label{EMPhisymRT}
\begin{aligned}
\EMP{0}^{\cfRT} &= 0\commae\\
\EMP{1}^{\cfRT} &= -\frac12\,\frac{\EMcharge}{4\pi}\,\frac1{\RTr^2}\commae\\
\EMP{2}^{\cfRT} &=
\frac1{\sqrt2}\,\frac{\EMcharge}{4\pi}\,\frac1{\RTr}\,\accl\sin\Ath\period
\end{aligned}
\end{equation}
Now the radiative character of the field is transparent: the first term
entering the peeling behavior, the scalar ${\EMP{2}}$, decays indeed as
${\RTr^{-1}}$, and it is non-vanishing for a non-zero acceleration ${\accl}$. In the
expressions \eqref{EMFsymRT} and \eqref{EMPhisymRT}, the de~Sitter background is
completely \vague{hidden}. The same form of the fields are obtained in case of
uniformly accelerated charges in Minkowski space if the coordinates built on
the null cones emanating from the particles are employed. A difference between
both cases reveals itself only in the explicit dependence of the affine
parameter ${\RTr}$ on the coordinates of spacetime points.

\section{Born in de~Sitter}
\label{sc:BorndS}

Finally, we turn to the fields from the particles symmetrically located with
respect to the origin  ${\spr=0}$ (the \vague{north pole}) of the standard spherical
coordinates. The particles are thus \vague{born} asymptotically at
the equator, ${\spr=\pi/2}$, at ${\spt\to-\infty}$, and return back at ${\spt\to\infty}$
with the opposite speeds (Fig.~\ref{fig:wldln_e}). Their fields, of course, are
intrinsically the same as those considered in the preceding section but only now
they represent the \emph{direct} generalization of the classical Born solutions
due to uniformly accelerated charges symmetrically located with respect to the
origin of Minkowski space.

We shall find the generalized Born fields easily by using the transformation
\eqref{rotation} which we applied to obtain the worldlines of the particles born
at the equator from those born at the poles. The scalar field
due to two equal scalar charges ${\SFcharge}$ moving along the worldlines
${w_\ppart}$and ${w_\mpart}$ reads
\begin{equation}\label{SFBdS}
  \SF = \frac{\SFcharge}{4\pi}\,\Omega_\MinkO^{-1}\,\frac1\Rfact
  \commae
\end{equation}
where the factor ${\Rfact}$ is determined by
\begin{equation}\label{Rfactsp}
\begin{split}
  \Rfact &= \frac{\DSr}{1+\cosh\frac\spt\DSr\cos\spr}
  \Bigl[\,\cosh^2\!\frac\spt\DSr\,\sin^2\!\spr\,\sin^2\!\tht\\
  &\quad+\Bigl(\sqrt{1+\accl^{\!2}\DSr^2}\cosh\!\frac\spt\DSr\,\cos\spr
  -\accl\DSr\Bigr)^2\Bigr]^{\frac12}\commae
\end{split}
\end{equation}
and the conformal factor ${\Omega_\MinkO}$ is given by (cf.\
Eq.~\eqref{CMinkFact})
\begin{equation}
  \Omega_\MinkO
  = 1+\cosh\frac\spt\DSr\cos\spr\period
\end{equation}
This factor is left in the explicit form here, in contrast to the preceding section,
since it explicitly exhibits conformal relation of the scalar field under conformal
mappings \eqref{CMinkFact} between de~Sitter space and Minkowski space ${\MinkOspc}$.
This relation will be used in the following
to perform the limit from the Born field in de~Sitter to the Born field in
Minkowski spacetime.

The electromagnetic field produced by charge ${\EMcharge}$ moving along the
worldline ${w_\ppart}$ and by symmetrically located charge ${-\EMcharge}$
moving along ${w_\mpart}$ has the following form:
\begin{equation}\label{EMFBdSsp}
\begin{split}\raisetag{72pt}
  &\EMF_{\dSBorn} = \frac{\EMcharge}{4\pi}
  \frac{\DSr^2}{\Rfact^3}
  \frac{\cosh\frac\spt\DSr\,\sin\tht}{(1+\cosh\frac\spt\DSr\cos\spr)^3}\\
  &\quad\times\biggl[
  \accl\DSr^2 \sinh\frac\spt\DSr\,\cosh\frac\spt\DSr\,\sin^2\!\spr\,
  \grad\spr\wedge\grad\tht
  \\&\quad
  + \Bigl(\sqrt{1+\accl^2\DSr^2} \cosh\frac\spt\DSr\,\cos\spr - \accl\DSr\Bigr)\,
  \cot\tht\,\grad\spt\wedge\grad\spr
  \\&\quad
  - \Bigl(\sqrt{1+\accl^{2}\DSr^{2}}\cosh\frac\spt\DSr - \accl\DSr\cos\spr\Bigr)\,
  \sin\spr\,\grad\spt\wedge\grad\tht
  \biggr]\commae
\end{split}
\end{equation}
with factor ${\Rfact}$ given by \eqref{Rfactsp}. In the tetrad tied to the
standard spherical coordinates the electric and magnetic fields become
\begin{align}
  &\EME_{\cfsp}^{\dSBorn}=-\frac{\EMcharge}{4\pi}
  \frac{\DSr}{\Rfact^3}
  \frac{1}{(1+\cosh\frac\spt\DSr\cos\spr)^3}\notag
  \\&\mspace{47mu}\times
  \Bigl[
  \Bigl(\sqrt{1+\accl^2\DSr^2} \cosh\frac\spt\DSr\,\cos\spr - \accl\DSr\Bigr)
  \cot\tht\;\cbv{\spr}\notag
  \\&\mspace{57mu}
  - \Bigl(\sqrt{1+\accl^{2}\DSr^{2}}\cosh\frac\spt\DSr - \accl\DSr\cos\spr\Bigr)
  \sin\spr\;\cbv{\tht}\,
  \Bigr]\commae\notag\\
  &\EMB_{\cfsp}^{\dSBorn}=\frac{\EMcharge}{4\pi}
  \frac{\accl\DSr^2}{\Rfact^3}\,
  \frac{ \sinh\frac\spt\DSr\, \sin\spr\,\sin\tht}{(1+\cosh\frac\spt\DSr\cos\spr)^3}
  \;\cbv{\ph}\period\label{EMEBBdSsp}
\end{align}

In the standard rescaled (conformally Einstein) coordinates the expressions
\eqref{EMFBdSsp}--\eqref{EMEBBdSsp} slightly simplify:
\begin{gather}
  \frac{\Rfact}{\DSr} = \frac{
  \bigl[(\accl\DSr\sin\tlt -
  \sqrt{1+\accl^2 \DSr^2}\cos\tlr)^2 +
  \sin^2\tlr\sin^2\tht\bigr]^{\frac12}}
  {\sin\tlt+\cos\tlr}\commae\label{Rfacttl}\\
  \Omega_\MinkO
  = \frac{\cos\tlr+\sin\tlt}{\sin\tlt}\commae
\end{gather}
\begin{equation}\label{EMFBdStl}
\begin{split}
  \EMF_{\dSBorn} =& -\frac{\EMcharge}{4\pi}
  \frac{\DSr^3}{\Rfact^3}
  \frac{\sin\tht}{(\sin\tlt+\cos\tlr)^3}\Bigl[
  \accl\DSr\sin^2\tlr\cos\tlt\,
  \grad\tlr\wedge\grad\tht
  \\&
  - \bigl(\sqrt{1+\accl^2\DSr^2} \cos\tlr - \accl\DSr\sin\tlt\mspace{2mu}\bigr)\,
  \cot\tht\,\grad\tlt\wedge\grad\tlr
  \\&
  + \bigl(\sqrt{1+\accl^{2}\DSr^{2}} - \accl\DSr\cos\tlr\sin\tlt\mspace{2mu}\bigr)\,
  \sin\tlr\,\grad\tlt\wedge\grad\tht
  \Bigr]\commae
\end{split}
\end{equation}
\begin{equation}\label{EMEBBdStl}
\begin{aligned}
  \EME_{\cfCE}^{\dSBorn}&=\frac{\EMcharge}{4\pi}
  \frac{\DSr\sin^2\tlt}{\Rfact^3(\sin\tlt+\cos\tlr)^3}
  \\&\quad\times
  \Bigl[
  -\bigl(\sqrt{1+\accl^2\DSr^2}\cos\tlr
  -\accl\DSr\sin\tlt\bigr)\cos\tht\,\cbv{\tlr}
  \\&\qquad\;
  +\bigl(\sqrt{1+\accl^2\DSr^2}
  -\accl\DSr\sin\tlt\cos\tlr\bigr)\sin\tht\,\cbv{\tht}
  \Bigr]\commae\\
  \EMB_{\cfCE}^{\dSBorn}&=-\frac{\EMcharge}{4\pi}
  \frac{\accl\DSr^2\sin^2\tlt}
  {\Rfact^3(\sin\tlt+\cos\tlr)^3}
  \cos\tlt\sin\tlr\sin\tht\,\cbv{\ph}\period
\end{aligned}
\end{equation}

The character of these fields was discussed in the preceding section for
the particles ${w_\npart}$ and ${w_\spart}$. One must only rotate all the structures by
${\pi/2}$ in the ${\spr}$ direction; hence, for example, the sphere of symmetry changes from
${\spr=\pi/2}$ to ${\tht=\pi/2}$.

There is some interest in having the fields available also in the hyperbolic
cosmological coordinates. They cover only those regions of the fields in
which we assume the radiative properties will be manifested. The sources
producing the fields are not covered by these coordinates (cf.\
Fig.~\ref{fig:wldln_e}). The fields in the hyperbolic cosmological coordinates
look as follows:
\begin{gather}
\begin{split}
  \Rfact &= \frac{1}{2 \Bo}
  \Bigl[ \Bigl(\Bo^2+\DSr^2\tanh^2\frac\cht{2\DSr}\Bigr)^2 \\
  &\qquad\qquad\;  +4\, \Bo^2\,\tanh^2\frac\cht{2\DSr}\sinh^2\frac\chr\DSr \sin^2\tht\Bigr]^{\frac12}\commae
  \end{split}\label{Rfactch}\\
  \Omega_\MinkO
  = 1+\cosh\frac\cht\DSr=2\cosh^2\frac\cht{2\DSr}\commae
\end{gather}
\vspace{-1ex}
\begin{equation}\label{EMFBdSch}
\begin{split}
  &\EMF_\dSBorn = \frac{\EMcharge}{4\pi}\,
  \frac{\DSr^3}{\Rfact^3}\,\frac{1}{2 \Bo\Omega_\MinkO^2}\,
  \biggl[\Bigl(\frac{\Bo^2}{\DSr^2}+\tanh^2\!\frac\cht{2\DSr}\Bigr)\sinh\frac\cht\DSr\\
  &\quad\;\times\!\Bigl(\frac1\DSr\cos\tht\,\grad\cht\wedge\grad\chr
  -\sinh\frac\chr\DSr\cosh\frac\chr\DSr\sin\tht\,\grad\cht\wedge\grad\tht\Bigr)\\
  &\quad\;-\!\Bigl(\frac{\Bo^2}{\DSr^2}-\tanh^2\!\frac\cht{2\DSr}\Bigr)\sinh^2\!\frac\cht\DSr\,\sinh^2\!\frac\chr\DSr
  \sin\tht\,\grad\chr\wedge\grad\tht
  \biggr]\!\commae
\end{split}
\end{equation}
\begin{align}
  &\EME_{\cfch}^\dSBorn =
  \frac{\EMcharge}{4\pi}\,
  \frac{\DSr^2}{\Rfact^3}\,
  \frac{1}{2\Bo\Omega_\MinkO^2}
  \Bigl(\frac{\Bo^2}{\DSr^2}+\tanh^2\!\frac\cht{2\DSr}\Bigr)\notag\\
  &\mspace{95mu}\times\!\Bigl(-\cos\tht\;\cbv{\chr}
  +\cosh\frac\chr\DSr\sin\tht\;\cbv{\tht}\Bigr)\commae\label{EMEBBdSch}\\
  &\EMB_{\cfch}^\dSBorn =
  -\frac{\EMcharge}{4\pi}\,
  \frac{\DSr^2}{\Rfact^3}\,
  \frac{\sinh\frac\chr\DSr}{2\Bo\Omega_\MinkO^2}\,
  \Bigl(\frac{\Bo^2}{\DSr^2}-\tanh^2\!\frac\cht{2\DSr}\Bigr)
  \sin\tht\;\cbv{\ph}
  \period\notag
\end{align}
We shall use these results in the forthcoming paper on the radiative properties of
the generalized Born solution.

Here, finally, we wish to describe the limiting procedure which leads from the
generalized Born solutions directly to their counterparts in Minkowski spacetime. For
this purpose it is natural to employ the conformally Minkowski coordinates
${\MOt,\,\MOr,\,\tht,\,\ph}$ introduced in Eq.~\eqref{CMinkCoor}, with the
inverse transformation given in the Appendix, Eq.~\eqref{apx:tltr=MOtr}. Transforming the
fields of the particles ${w_\ppart}$, ${w_\mpart}$ from the conformally
Einstein coordinates to the conformally Minkowski coordinates, we arrive at the
following intriguing forms.
The scalar field is given by Eq.~\eqref{SFBdS} where now the factors
${\Rfact}$ and ${\Omega_\MinkO}$ are determined by
\begin{gather}
  \Rfact = \frac{1}{2 \Bo}
  \sqrt{ (\Bo^2+\MOt^2-\MOr^2)^2 +
  4\, \Bo^2\,\MOr^2 \sin^2\tht}\commae\\
  \Omega_\MinkO
  = \frac{2\,\DSr^2}{\DSr^2-\MOt^2+\MOr^2}\period
\end{gather}
Notice that factor ${\Rfact}$ coincides with the expression \eqref{Rfact} in Minkowski
space. The electromagnetic field reads
\begin{equation}\label{EMFBdSMO}
\begin{split}
  \EMF_\dSBorn &= -\frac{\EMcharge}{4\pi}
  \frac{1}{2 \Bo}\,\frac{1}{\Rfact^3}\;
  \Bigl[
  - 2\, \MOt\,\MOr^2\sin\tht\,
  \grad\MOr\wedge\grad\tht
  \\&\qquad
  - (\Bo^2+\MOt^2-\MOr^2)\cos\tht\,
  \grad\MOt\wedge\grad\MOr
  \\&\qquad
  + \MOr\, (\Bo^2+\MOt^2+\MOr^2)\sin\tht\,
  \grad\MOt\wedge\grad\tht
  \Bigr]\commae
\end{split}
\end{equation}
and the electric and magnetic parts of the field turn out to be
\begin{equation}\label{EMEBBdSMO}
\begin{aligned}
  \EME_{\cfMO}^\dSBorn &=
  \frac{\EMcharge}{4\pi}\,
  \frac{1}{\Rfact^3}\,
  \frac{1}{2\Bo\Omega_\MinkO^2}\Bigl[
  (\Bo^2+\MOt^2-\MOr^2)\cos\tht\,\cbv{\MOr}\\
  &\mspace{120mu}-(\Bo^2+\MOt^2+\MOr^2)\sin\tht\,\cbv{\tht}
  \Bigr]  \commae\\
  \EMB_{\cfMO}^\dSBorn &=
  \frac{\EMcharge}{4\pi}\,
  \frac{1}{\Rfact^3}\,
  \frac{1}{\Bo\Omega_\MinkO^2}\,
  \MOt\,\MOr\sin\tht\,\cbv{\ph}
  \period
\end{aligned}
\end{equation}

To connect these fields with their counterparts in flat space, note
that they are conformally related by the conformal transformation \eqref{CMinkFact}.
Under the conformal mapping, the field $\SF_\dSBorn$
must be multiplied by factor $\Omega_\MinkO$, which gives
${\SF_\MinkO = ({\SFcharge}/{4\pi})\,{\Rfact}^{-1}}$, and $\EMF_\dSBorn$ in \eqref{EMFBdSMO}
remains unchanged. The transformed fields then
coincide with the classical Born fields
\eqref{SFsymM}, \eqref{Rfact}, and \eqref{EMBornM}.

In order to see the limit for $\Lambda\rightarrow 0$, we
parametrize the sequence of de~Sitter spaces by $\Lambda$, and identify
them in terms of coordinates ${\MOt,\MOr,\tht,\ph}$. As
$\Lambda={3/\DSr^2\to0}$, Eq.~\eqref{CMinkFact} implies
${(\Omega_\Mink)_\Lambda\to2}$,
${(\mtrc_\dS)_\Lambda\to4\mtrc_\MinkO}$. After the trivial
rescaling of $\MOt$, $\MOr$ by factor $2$, the standard Minkowski
metric is obtained. The limit of the scalar and electromagnetic fields
\eqref{SFBdS} and \eqref{EMFBdSMO}, in which $\Bo$ is kept constant (with
$\accl={(1-\Bo^2\DSr^{-2})/(2\Bo)}$---cf.\ Eq.~\eqref{acclconst}),
leads precisely to the scalar and
electromagnetic Born fields \eqref{SFsymM} and \eqref{EMBornM}
in flat space. Because of the rescaling of
the coordinates by factor $2$, we get the physical acceleration equal to
${1/{\Bo}=2\accl}$, and the scalar field rescaled by $1/2$.
The explicit limiting procedure carrying the generalized Born fields in de~Sitter universe back into
the classical Born solution in Minkowski space has thus been demonstrated.

\section{Concluding remarks}
\label{sc:Concl}

Since 1998 the observations of high-redshift supernovae
indicate, with an increasing evidence, that we live in an
accelerating universe with a positive cosmological constant
(for most recent observations see, e.g., \cite{Riessetal:2004}).
Vacuum energy seems to dominate in the universe and it is thus of interest to
understand fundamental physics in the vacuum dominated
de~Sitter spacetime.

In the present work, we constructed the fields of uniformly accelerated
charges in this universe. They go over to the classical
Born fields in Minkowski space in the limit of a vanishing
cosmological constant. Aside from some similarities found,
the generalized fields provide the models showing how a
positive cosmological constant implies essential
differences from physics in flat spacetime. For example,
advanced effects occur inevitably due to the spacelike
character of the past infinity $\scri^-$ and its consequence---the existence of
the past particles' horizons, respectively, of the \vague{creation
light cones} of the particles' worldlines.

Since de~Sitter spacetime, according to
our present understanding, appears to be not only an appropriate basic model for studying
future cosmological epochs, but it is commonly used
also for exploring the inflationary era, various physical processes have been
investigated in de~Sitter space from the perspective of the early universe,
among them, the effects of quantum field theory. Also in quantum contexts,
however, problems arise from combining the causal structure of
the full de~Sitter spacetime with the constraint equations (see
\cite{Woodard:2004} for a recent review).
These problems are associated with the \vague{insufficiency
of purely retarded fields} in spacetimes with a spacelike $\scri^-$.
We analyzed this issue in detail for
the classical electromagnetic and scalar fields with sources in Ref.~\cite{BicakKrtous:2001}.

Another intriguing implication of
the rapid expansion of de~Sitter universe due to
a positive cosmological constant is
manifested in the exponential decay of the fields at large
times. We noticed this \vague{cosmic no-hair phenomenon}
explicitly on the late-time behaviour of the fields due
to accelerated charges.

In the present paper we wished to give all details on the
construction of the fields and on coordinate frames useful
in understanding their various aspects, including their
relation to their counterparts in flat spacetime. We did
not here analyze the radiative characteristics of the
fields. In the Introduction we indicated that radiative
properties depend on the way in which a given point of
infinity is approached. This is briefly described at
the end of our paper \cite{BicakKrtous:2002}.

In de~Sitter spacetime it is not a~priori clear, as it is
in special relativity, how to define global physical
quantities like energy or energy flux. Such issues connected
with the question of radiation from \vague{Born in
de~Sitter} will be considered in a future presentation.

\vspace{-6pt}
\section*{Acknowledgments}
\vspace{-3pt}

The authors are grateful to the Division of geometric analysis and gravitation,
Albert Einstein Institute, Golm, for
the kind hospitality. Here most of the work was done.
J.B. also thanks the Institute of Astronomy, University of Cambridge, for hospitality.
We also acknowledge support by Grants
GA\v{C}R 202/02/0735 of the Czech Republic and Research
Project No.~MSM113200004.


\appendix

\vspace{-6pt}
\section{The palette of coordinate systems in de~Sitter spacetime}\label{asc:coorIntro}
\vspace{-3pt}

Nine families of coordinate systems are here introduced, described analytically and
illustrated graphically. The corresponding forms of de~Sitter metric, orthonormal
tetrads and interrelations between the systems are given. All these systems
are suitable for exhibiting various features of de~Sitter space;
two families are directly associated with uniformly accelerated particles.
Although the majority (though not all) of these
coordinate systems
undoubtedly appeared in literature in some form already, they are scattered
and, as far as we know, not summarized as comprehensively as in the following.
In the main text we refer frequently to this Appendix, but the Appendix can be read
independently. We hope it can serve as a catalogue useful for analyzing
various aspects of physics in de~Sitter universe.

By a \defterm{family of coordinate systems} we mean the systems with the \emph{same}
coordinate lines; e.g., ${\{x^\mu\}}$ and ${\{y^\mu\}}$ where
${x^1=x^1(y^1)}$, ${x^2=x^2(y^2)}$, etc. Seven of our families
have the same spherical angular coordinates ${\tht}$, ${\ph}$,
accelerated and Robinson-Trautman coordinates mix three coordinates,
only azimuthal coordinate ${\ph}$ remains unchanged.

The homogeneous normalized metric on
two-spheres (the metric \vague{in angular direction}) is denoted by
\begin{equation}\label{twospheremtrc}
\sphmtrc=\grad\tht\formsq+\sin^2\tht\,\grad\ph\formsq\period
\end{equation}

The radial coordinates label directions pointing out from the pole and acquire only positive values.
However, transformations among coordinates  take simpler forms if we allow radial coordinates to take
on negative values as well. This causes no problems if, denoting by ${t}$ and ${r}$
the prototypes of time and radial coordinates, we adopt the convention that the
following two values of coordinates describe the same point:
\begin{equation}\label{apx:negr}
\{t,\,r,\,\tht,\,\ph\}\;
\leftrightarrow\; \{t,\,-r,\,\pi-\tht,\,\ph+\pi\}\period
\end{equation}
Hence, intuitively we may consider a point with ${-r<0}$ and ${\tht,\,\ph}$ fixed to lie
on diametrically opposite side of the pole ${r=0}$ with respect to the point ${r>0,\,\tht,\,\ph}$.

The orthonormal tetrad ${\cbv{t},\,\cbv{r},\,\cbv{\tht},\,\cbv{\ph}}$
associated with a coordinate system is tangent to the coordinate lines and
oriented (with few exceptions) in the directions of growing coordinates.
It is chosen in such a way that the external product
${\cbf{t}\wedge\cbf{r}\wedge\cbf{\tht}\wedge\cbf{\ph}}$
of \mbox{1-forms} of the dual tetrad has
always the same orientation. 
Since all forms of the metric contain the term \eqref{twospheremtrc}
the only component ${(\cbv\ph)^\ph}$ of the tetrad
vector ${\cbv{\ph}}$ in coordinate frame ${\{\cv{x^\mu}\}}$ is
related to the ${\tht}$-component of ${\cbv{\tht}}$ as
\begin{equation}
(\cbv{\ph})^\ph=\frac1{\sin\tht}\,(\cbv{\tht})^\tht\commae
\end{equation}
and we thus omit ${\cbv{\ph}}$ henceforth.

In the standard Newman-Penrose null complex tetrad ${\kG,\,\lG,\,\mG,\,\bG}$
with only nonvanishing inner products
${\kG\scp\lG = -1}$, ${\mG\scp\bG = 1}$,
the electromagnetic field ${\EMF}$ is represented by three
complex components:\pagebreak[2]
\begin{equation}\label{PhiDef}
\begin{gathered}
  \EMP{0} = \EMF_{\alpha\beta}\,
    \kG^\alpha\,\mG^\beta\comma
  \EMP{2} = \EMF_{\alpha\beta}\,
    \bG^\alpha\,\lG^\beta\commae\\
  \EMP{1} = {\textstyle\frac12}\,\EMF_{\alpha\beta}\,
    \bigl(\kG^\alpha\,\lG^\beta-\mG^\alpha\,\bG^\beta\bigr)\period
\end{gathered}
\end{equation}
The null tetrad can be specified directly (as it will be done in the case of Robinson-Trautman
coordinates in Eq.~\eqref{apx:RTnulltetr}), or it can be associated with any orthonormal tetrad,
say ${\tG,\,\qG,\,\rG,\,\sG}$, by relations
\begin{equation}\label{NormNullTetr}
\begin{aligned}
  \kG &= \textstyle{\frac1{\sqrt{2}}} (\tG+\qG)\comma&
  \lG &= \textstyle{\frac1{\sqrt{2}}} (\tG-\qG)\commae\\
  \mG &= \textstyle{\frac1{\sqrt{2}}} (\rG-i\,\sG)\comma&
  \bG &= \textstyle{\frac1{\sqrt{2}}} (\rG+i\,\sG)\period
\end{aligned}
\end{equation}
Here, ${\tG}$ and ${\qG}$ are timelike and spacelike unit vectors
respectively, typically in a direction
of \vague{time} and \vague{radial} coordinate,
and ${\rG,\,\sG}$ are spacelike unit vectors in angular directions,
${\rG=\cbv{\tht}}$,  ${\sG=\cbv{\ph}}$.

For each coordinate family we give the diagram illustrating
section ${\tht,\,\ph=\text{constant}}$
with the radial coordinate taking on \emph{both} positive and negative values.
The diagrams thus represent the history of the \emph{entire} main circle
of the spatial spherical section of de~Sitter universe. The left and right edges
of the diagrams represent the south pole and should be considered as identified;
the central vertical line describes the history of the north pole.
Recalling the meaning of the negative radial
coordinate we could eliminate the left half of each of the diagrams by transforming
it into the right one by replacements ${\{\tht,\,\ph\}\rightarrow \{\pi-\tht,\,\ph+\pi\}}$.
However, it is instructive to keep both halves for better understanding of the spatial topology
of the sections. All diagrams are compactified---they are adapted to
the standard rescaled coordinates ${\tlt,\,\tlr}$ (see below).
The past and future conformal infinities
are drawn as double lines. The ranges of time and radial
coordinates are shown, the orientation of coordinate labels indicates
the directions of the growth of corresponding coordinates.

\begin{figure}[b]
\includegraphics{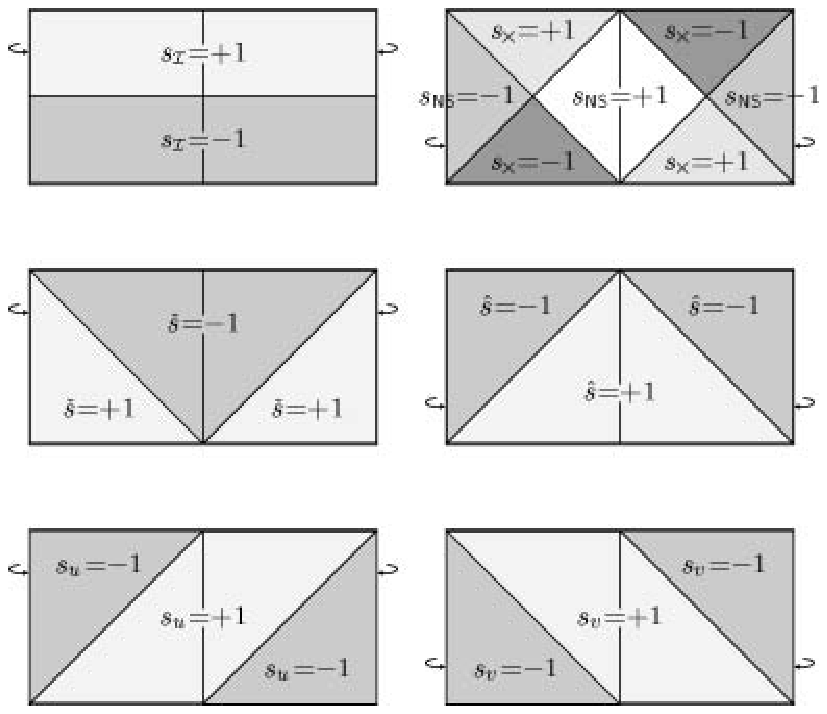}
\caption{\label{fig:sign}
The values of the factors $\signscri$, $\signNS$, $\signx$, $\signMV$, $\signMA$,
$\signu$ and $\signv$ in various regions of de~Sitter space. The factors are
defined in Eqs.~\eqref{apx:signscri}, \eqref{apx:signNS}, \eqref{apx:signx},
\eqref{apx:signMV}, \eqref{apx:signMA} and \eqref{apx:signuv},
respectively. The factor $\signx$ is used only in the expressions for static
coordinates in the region where the Killing vector is spacelike. Therefore,
we indicated the values of $\signx$ only in those regions, although
Eq.~\eqref{apx:signx} defines $\signx$ everywhere. The factors $\signx$,
$\signu$, and $\signv$ are defined only for any given section ${\tht=\text{constant}}$,
but not as unique functions on the whole spacetime (they are not
symmetric with respect to the pole). This is related to our convention using
negative radial coordinates, cf.\ the text below Eq.~\eqref{twospheremtrc}.
}
\end{figure}

We will also introduce several sign factors. The values of these factors
in different domains of spacetime are indicated in Fig.~\ref{fig:sign}.

\pagebreak
\coorsection{The spherical cosmological family}\label{asc:coorCE}
\begin{figure}[h]
\includegraphics{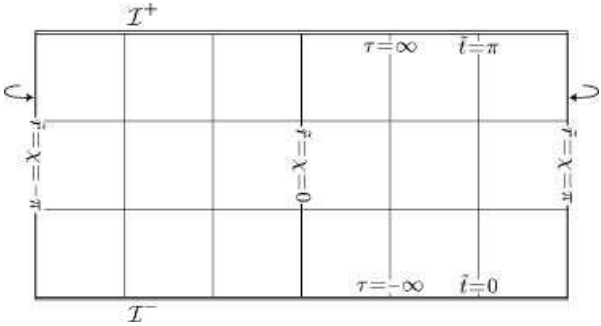}
\caption{\label{fig:coorCE}The spherical cosmological family of coordinates.}
\end{figure}

The first family consists of the \defterm{standard} or \defterm{spherical cosmological coordinates}
${\spt,\,\spr,\,\tht,\,\ph}$, and of the \defterm{standard rescaled} or \defterm{conformally Einstein coordinates}
${\tlt,\,\tlr,\,\tht,\,\ph}$ (where ${\tlr\equiv\spr}$).
These coordinates cover de~Sitter spacetime globally. They are associated with cosmological observers
with homogeneous spatial sections of positive spatial curvature. The coordinates are adjusted
to the spherical symmetry of the spatial sections: ${\spr}$, ${\tht}$, and ${\ph}$
are standard angular coordinates.
The coordinate ${\tau}$ is a proper time along the worldlines of the cosmological observers
given by ${\spr,\,\tht,\,\ph=\text{constant}}$.
The vector ${\cvil{\spt}}$ is a conformal Killing vector which is everywhere timelike.
The rescaled coordinates ${\tlt,\,\tlr,\,\tht,\,\ph}$ can also be viewed as
the standard coordinates of the conformally related Einstein universe;
they cover smoothly both conformal
infinities ${\scri^\pm}$ of de~Sitter spacetime.

\coorsubsection{Metric and relation between coordinates}
\begin{gather}
    \mtrc=-\grad\spt\formsq+\DSr^2\,\cosh^2\!({\spt}/{\DSr})\;
    \bigl(\grad\spr\formsq+\sin^2\!\spr\,\sphmtrc\bigr)
    \commae\\
    \mtrc=\DSr^2\,\sin^{\!-2}\!\tlt\;
    \bigl(-\grad\tlt\,\formsq+\grad\tlr\formsq+\sin^2\!\tlr\,\sphmtrc\bigr)
    \commae
\end{gather}
\begin{subequations}
\begin{gather}
\begin{gathered}
  \tan\frac\tlt2=\exp\frac\spt\DSr\comma
  \cot\tlt=-\sinh\frac\spt\DSr\commae\\
  \sin\tlt=\cosh^{\!-1}\!\frac\spt\DSr\comma
  \cos\tlt=-\tanh\frac\spt\DSr\commae
\end{gathered}\\
  \tlr=\spr\period
\end{gather}
\end{subequations}
The ranges of coordinates are
\begin{equation}
\begin{gathered}
\spt\in\realn\comma \spr\in(-\pi,\pi)\commae\\
\tlt\in(0,\pi)\comma \tlr\in(-\pi,\pi)\commae
\end{gathered}
\end{equation}
with negative values of radial coordinates $\spr$, $\tlr$
interpreted in accordance with Eq.~\eqref{apx:negr}.

\coorsubsection{Orthonormal tetrad}
\begin{align}
  \cbv{\spt}&=\cv{\spt}=\frac1\DSr\sin\tlt\,\cv{\tlt}\commae\notag\displaybreak[1]\\
  \cbv{\spr}&=\frac1\DSr\cosh^{\!-1}\!\frac\spt\DSr\,\cv{\spr}=\frac1\DSr\sin\tlt\,\cv{\tlr}\commae\displaybreak[1]\label{apx:sphtetr}\\
  \cbv{\tht}&=\frac1\DSr\frac1{\cosh(\spt/\DSr)\sin\spr}\cv{\tht}=\frac1\DSr\frac{\sin\tlt}{\sin\tlr}\cv{\tht}\period\notag
\end{align}

\coorsubsection{Relation to flat cosmological family}
\begin{equation}\label{apx:tltr=MAVtr}
\begin{aligned}
  \tan\tlt&=\frac{2\,\DSr\MAt}{\DSr^2-\MAt^2+\MAr^2}
           =\frac{2\,\DSr\MVt}{\DSr^2-\MVt^2+\MVr^2}\commae\\
  \tan\tlr&=\frac{2\,\DSr\MAr}{\DSr^2+\MAt^2-\MAr^2}
           =\frac{2\,\DSr\MVr}{\DSr^2+\MVt^2-\MVr^2}\period
\end{aligned}
\end{equation}

\coorsubsection{Relation to hyperbolic cosmological coordinates}
\begin{equation}
\begin{aligned}
  \cot\tlt&=
    -\sinh\frac\cht\DSr\,\cosh\frac\chr\DSr\commae\\
  \tan\tlr&=
    \spcm\tanh\frac\cht\DSr\,\sinh\frac\chr\DSr\period
\end{aligned}
\end{equation}

\coorsubsection{Relation to static family in timelike domains ${\Ndom}$, ${\Sdom}$}
\begin{equation}
\begin{aligned}
  \tan\tlt&=-\signNS\,\frac{\DSr}{\sqrt{\DSr^2-\str^2}}\;{\sinh^{\!-1}\!\frac\stt\DSr}\commae\\
  \tan\tlr&=\signNS\,\frac{\str}{\sqrt{\DSr^2-\str^2}}\;{\cosh^{\!-1}\!\frac\stt\DSr}\commae
\end{aligned}
\end{equation}
\begin{equation}
  \tan\tlt=-\signNS\,\frac{\cosh\frac\brr\DSr}{\sinh\frac\brt\DSr}\comma
  \tan\tlr=\signNS\,\frac{\sinh\frac\brr\DSr}{\cosh\frac\brt\DSr}\commae
\end{equation}
where ${\signNS=+1}$ (${-1}$) in domain $\Ndom$ ($\Sdom$), cf.\ Eq.~\eqref{apx:signNS}.

\coorsubsection{Relation to static family in spacelike domains ${\Fdom}$, ${\Pdom}$}
\begin{equation}
\begin{aligned}
  \tan\tlt&=\frac{-\signscri\,\DSr}{\sqrt{\str^2-\DSr^2}}\;{\cosh^{\!-1}\!\frac\stt\DSr}\commae\\
  \tan\tlr&=\frac{\signscri\,\str}{\sqrt{\str^2-\DSr^2}}\;{\sinh^{\!-1}\!\frac\stt\DSr}\commae
\end{aligned}
\end{equation}
\begin{equation}
  \tan\tlt=\signx\,\frac{\sinh\frac\brr\DSr}{\cosh\frac\brt\DSr}\comma
  \tan\tlr=-\signx\,\frac{\cosh\frac\brr\DSr}{\sinh\frac\brt\DSr}\commae
\end{equation}
where ${\signscri=-\sign\cos\tlt}$\; and ${\signx=-\signscri\,\sign\tlr}$,
cf.\ Eqs. \eqref{apx:signscri} and \eqref{apx:signx}.

\coorsubsection{Relation to conformally Minkowski coordinates}
\begin{equation}\label{apx:tltr=MOtr}
  \cot\tlt=\frac{2\DSr\MOt}{\MOt^2-\MOr^2-\DSr^2}\comma
  \tan\tlr=\frac{2\DSr\MOr}{\MOt^2-\MOr^2+\DSr^2}\period
\end{equation}

\pagebreak
\coorsection{The flat cosmological family, type \vague{$\vee$}}\label{asc:coorCMV}
\begin{figure}[h]
\includegraphics{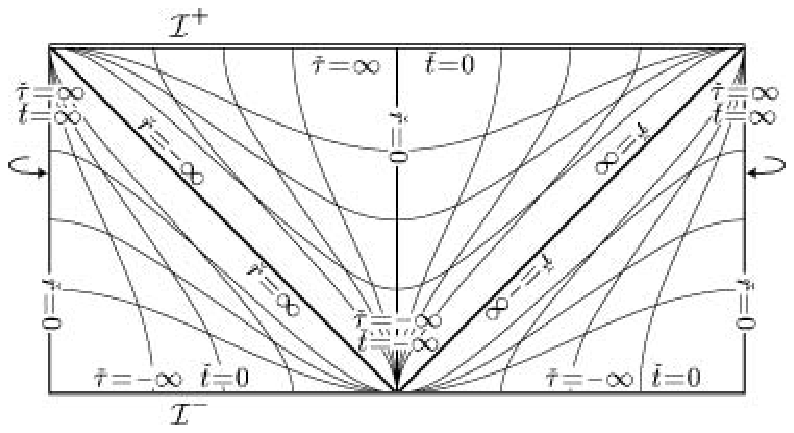}
\caption{\label{fig:coorCMV}The flat cosmological family, type~\vague{$\vee$}.}
\vspace*{-3pt}
\end{figure}

The first flat cosmological coordinate family consists
of the \defterm{flat cosmological coordinates} ${\MVet,\,\MVr,\,\tht,\,\ph}$ and
of the \defterm{rescaled flat cosmological coordinates} ${\MVt,\,\MVr,\,\tht,\,\ph}$.
Hypersurfaces ${\MVet=\text{constant}}$ are homogeneous flat spaces and coordinate
lines ${\MVr,\,\tht,\,\ph=\text{constant}}$ are worldlines of cosmological observers
orthogonal to these hypersurfaces. They are geodesic with proper time $\MVet$,
the vector $\cvil{\MVet}$ is a conformal Killing vector. The coordinates cover de~Sitter
spacetime smoothly, except for the past cosmological horizon, ${\tlr=\tlt}$, of the north pole
where ${\MVr,\,\MVt\to\pm\infty}$. The coordinates thus split into two
coordinate patches---\vague{above} and \vague{below} the horizon. The domain
above the horizon has a cosmological interpretation of an exponentially expanding flat
three-space. The rescaled coordinates can be viewed as inertial coordinates
in the conformally related Minkowski space $\MinkVspc$, cf.\ Fig.~\ref{fig:dSMinkV};
the domain above the horizon corresponds to the \vague{lower half}, ${\MVt<0}$, of
$\MinkVspc$, the domain below corresponds to the \vague{upper half}, ${\MVt>0}$.

\vspace{-3.5ex}
\coorsubsection{Metric and relation between coordinates}
\begin{gather}
    \mtrc=\frac{\DSr^2}{\MVt^2}\,
    \Bigl(-\grad\MVt\,\formsq+\grad\MVr\formsq+\MVr^2\,\sphmtrc\Bigr)\commae\\
    \mtrc=-\grad\MVet\formsq+\exp\bigl({-\signMV\,2\,{\MVet}/{\DSr}}\bigr)\,
    \bigl(\grad\MVr\formsq+\MVr^2\,\sphmtrc\bigr)\period\\[6pt]
  \MVt=\signMV\,\DSr\exp\Bigl(\signMV\frac\MVet\DSr\Bigl)\commae\\
  \signMV=\sign \MVt\period\label{apx:signMV}
\end{gather}
The ranges of coordinates are
\begin{equation}
\begin{gathered}
\MVet\in\realn\comma \MVt\in\realn^-\comma\MVr\in\realn\quad\text{above the horizon}\commae\\
\MVet\in\realn\comma \MVt\in\realn^+\comma\MVr\in\realn\quad\text{below the horizon}\commae
\end{gathered}
\end{equation}
with negative values of radial coordinate $\MVr$ interpreted as described  in Eq.~\eqref{apx:negr}.

\vspace{-3ex}
\coorsubsection{Orthonormal tetrad}
\begin{equation}
\begin{gathered}
  \cbv{\MVt}=\cv{\MVet}=\frac{\signMV\,\MVt}\DSr\,\cv{\MVt}\comma
  \cbv{\MVr}=\exp\frac{\signMV\MVet}{\DSr}\,\cv{\MVr}=\frac{\signMV\,\MVt}\DSr\,\cv{\MVr}\commae\\
  \cbv{\tht}=-\frac\signMV\MVr\exp\frac{\signMV\MVet}{\DSr}\,\cv{\tht}=-\frac1\DSr\frac\MVt\MVr\,\cv{\tht}\period
\end{gathered}
\end{equation}

\coorsubsection{Relation to spherical cosmological family}
\begin{equation}
  \MVt=\frac{-\DSr\cosh^{\!-1}\!(\spt/\DSr)}{\cos\spr+\tanh(\spt/\DSr)}\comma
  \MVr=\frac{\DSr\cosh^{\!-1}\!(\spt/\DSr)}{\cos\spr+\tanh(\spt/\DSr)}\period
\end{equation}
\begin{equation}
  \MVt=\frac{\DSr\sin\tlt}{\cos\tlt-\cos\tlr}\comma
  \MVr=\frac{\DSr\sin\tlr}{\cos\tlr-\cos\tlt}\period
\end{equation}

\coorsubsection{Relation to flat cosmological family, type \vague{${\wedge}$}}
\begin{equation}
  \MVt=-\frac{\MAt\,\DSr^2}{\MAt^2-\MAr^2}
  \comma
  \MVr=\frac{\MAr\,\DSr^2}{\MAt^2-\MAr^2}
  \commae
\end{equation}
\begin{equation}
\begin{gathered}
  \MVt\,\MAr+\MAt\,\MVr=0\comma
  \MAt\,\MVt+\MAr\,\MVr=-\DSr^2\commae\\
  \bigl(-\MAt^2+\MAr^2\bigl)\bigl(-\MVt^2+\MVr^2\bigl)=\DSr^4\commae\\
  (\MAt+\MAr)(\MVt+\MVr)=(\MAt-\MAr)(\MVt-\MVr)=-\DSr^2\period
\end{gathered}
\end{equation}

\coorsubsection{Relation to static family in timelike domains ${\Ndom}$, ${\Sdom}$}
\begin{equation}
\begin{gathered}
  \frac\MVt\DSr=-\signNS\frac\DSr{\sqrt{\DSr^2-\str^2}}\,\exp\Bigl(-\frac\stt\DSr\Bigr)
  \commae\\
  \frac\MVr\DSr=\spcm\signNS\frac\str{\sqrt{\DSr^2-\str^2}}\,\exp\Bigl(-\frac\stt\DSr\Bigr)\commae
\end{gathered}
\end{equation}
\begin{equation}
\begin{gathered}
  \MVt=-\signNS\,\DSr\exp\Bigl(-\frac\brt\DSr\Bigr)\,\cosh\frac\brr\DSr\commae\\
  \MVr=\spcm\signNS\DSr\exp\Bigl(-\frac\brt\DSr\Bigr)\,\sinh\frac\brr\DSr\commae
\end{gathered}
\end{equation}
where ${\signNS=+1}$ (${-1}$) in domain $\Ndom$ ($\Sdom$), cf.\ Eq.~\eqref{apx:signNS}.

\coorsubsection{Relation to static family in spacelike domains ${\Fdom}$, ${\Pdom}$}
\begin{equation}
\begin{gathered}
  \frac\MVt\DSr=\signx\frac{\DSr}{\sqrt{\str^2-\DSr^2}}\,\exp\Bigl(-\frac\stt\DSr\Bigr)\commae\\
  \frac\MVr\DSr=-\signx\frac{\str}{\sqrt{\str^2-\DSr^2}}\,\exp\Bigl(-\frac\stt\DSr\Bigr)\commae
\end{gathered}
\end{equation}
\begin{equation}
\begin{gathered}
  \MVt=\signx\DSr\exp\Bigl(-\frac\brt\DSr\Bigr)\,\sinh\frac\brr\DSr\commae\\
  \MVr=-\signx\DSr\exp\Bigl(-\frac\brt\DSr\Bigr)\,\cosh\frac\brr\DSr\commae
\end{gathered}
\end{equation}
where ${\signx=\sign\tlr\,\sign\cos\tlt}$,
cf.\ Eqs.~\eqref{apx:signscri} and \eqref{apx:signx}.

\coorsubsection{Relation to conformally Minkowski coordinates}
\begin{equation}
  \frac\MVt\DSr=-\frac{\DSr^2-\MOt^2+\MOr^2}{(\DSr+\MOt)^2-\MOr^2}\comma
  \frac\MVr\DSr=\frac{2\,\DSr\MOr}{(\DSr+\MOt)^2-\MOr^2} \period
\end{equation}

\pagebreak
\coorsection{The flat cosmological family, type \vague{$\wedge$}}\label{asc:coorCMA}
\begin{figure}[h]
\includegraphics{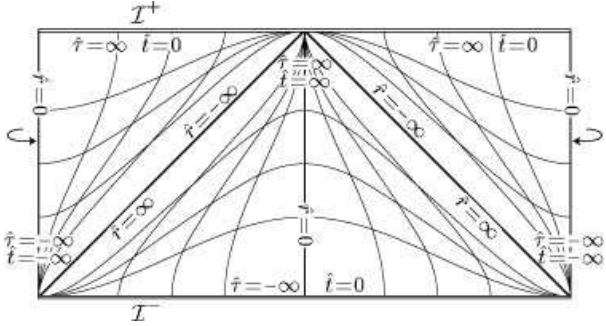}
\caption{\label{fig:coorCMA}The flat cosmological family, type \vague{$\wedge$}.}
\end{figure}

The second flat cosmological coordinate family consists
of the \defterm{flat cosmological coordinates} ${\MAet,\,\MAr,\,\tht,\,\ph}$ and
of the \defterm{rescaled flat cosmological coordinates} ${\MAt,\,\MAr,\,\tht,\,\ph}$.
They can be built analogously to the flat coordinates introduced above,
with north and south poles interchanged only. They thus have similar properties:
Hypersurfaces ${\MAt=\text{constant}}$ are homogeneous flat three-spaces,
coordinate lines ${\MAr,\,\tht,\,\ph=\text{constant}}$ are geodesics with
proper time $\MAet$, and $\cvil{\MAet}$ is a conformal Killing vector.
The coordinates cover de~Sitter spacetime everywhere except the future cosmological
horizon, ${\tlr=\pi-\tlt}$, of the north pole (i.e., the past horizon of the south pole),
and the rescaled coordinates can be viewed as inertial coordinates in
the conformally related Minkowski space $\MinkAspc$.

\coorsubsection{Metric and relation between coordinates}
\begin{gather}
    \mtrc=\frac{\DSr^2}{\MAt^2}\,
    \Bigl(-\grad\MAt\,\formsq+\grad\MAr\formsq+\MAr^2\,\sphmtrc\Bigr)\commae\\
    \mtrc=-\grad\MAet\formsq+\exp\bigl({-\signMA\,2\,{\MAet}/{\DSr}}\bigr)\,
    \bigl(\grad\MAr\formsq+\MAr^2\,\sphmtrc\bigr)\commae
\end{gather}
\begin{equation}
  \MAt=\signMA\,\DSr\exp\Bigl(\signMA\frac\MAet\DSr\Bigl)\commae
\end{equation}
where
\begin{equation}\label{apx:signMA}
  \signMA=\sign \MAt\period
\end{equation}
The ranges of coordinates are
\begin{equation}
\begin{gathered}
\MAet\in\realn\comma \MAt\in\realn^-\comma\MAr\in\realn\quad\text{above the horizon}\commae\\
\MAet\in\realn\comma \MAt\in\realn^+\comma\MAr\in\realn\quad\text{below the horizon}\commae
\end{gathered}
\end{equation}
with negative values of radial coordinate $\MAr$ interpreted as described  in Eq.~\eqref{apx:negr}.

\vspace{-1ex}
\coorsubsection{Orthonormal tetrad}
\begin{equation}
\begin{gathered}
  \cbv{\MAt}=\cv{\MAet}=\frac{\signMA\,\MAt}\DSr\,\cv{\MAt}\comma
  \cbv{\MAr}=\exp\frac{\signMA\MAet}{\DSr}\,\cv{\MAr}=\frac{\signMA\,\MAt}\DSr\,\cv{\MAr}\commae\\
  \cbv{\tht}=\frac\signMA\MAr\exp\frac{\signMA\MAet}{\DSr}\,\cv{\tht}=\frac1\DSr\frac\MAt\MAr\,\cv{\tht}\period
\end{gathered}
\end{equation}

\coorsubsection{Relation to spherical cosmological family}
\begin{equation}\label{apx:MAtr=sptr}
  \MAt=\frac{\DSr\cosh^{\!-1}\!(\spt/\DSr)}{\cos\spr-\tanh(\spt/\DSr)}\comma
  \MAr=\frac{\DSr\cosh^{\!-1}\!(\spt/\DSr)}{\cos\spr-\tanh(\spt/\DSr)}\period
\end{equation}
\begin{equation}\label{apx:MAtr=tltr}
  \MAt=\frac{\DSr\sin\tlt}{\cos\tlt+\cos\tlr}\comma
  \MAr=\frac{\DSr\sin\tlr}{\cos\tlr+\cos\tlt}\period
\end{equation}

\coorsubsection{Relation to flat cosmological family, type \vague{${\wedge}$}}
\begin{equation}
  \MAt=-\frac{\MVt\,\DSr^2}{\MVt^2-\MVr^2}\comma
  \MAr=\frac{\MVr\,\DSr^2}{\MVt^2-\MVr^2}\commae
\end{equation}
\begin{equation}
\begin{gathered}
  \MVt\,\MAr+\MAt\,\MVr=0\comma
  \MAt\,\MVt+\MAr\,\MVr=-\DSr^2\commae\\
  \bigl(-\MAt^2+\MAr^2\bigl)\bigl(-\MVt^2+\MVr^2\bigl)=\DSr^4\commae\\
  (\MAt+\MAr)(\MVt+\MVr)=(\MAt-\MAr)(\MVt-\MVr)=-\DSr^2\period
\end{gathered}
\end{equation}

\coorsubsection{Relation to static family in timelike domains ${\Ndom}$, ${\Sdom}$}
\begin{equation}
\begin{gathered}
  \frac\MAt\DSr=\signNS\,\frac\DSr{\sqrt{\DSr^2-\str^2}}\,\exp\frac\stt\DSr\commae\\
  \frac\MAr\DSr=\signNS\frac\str{\sqrt{\DSr^2-\str^2}}\,\exp\frac\stt\DSr\commae
\end{gathered}
\end{equation}
\begin{equation}
\begin{gathered}
  \MAt=\signNS\,\DSr\exp\frac\brt\DSr\,\cosh\frac\brr\DSr\commae\\
  \MAr=\signNS\,\DSr\exp\frac\brt\DSr\,\sinh\frac\brr\DSr\commae
\end{gathered}
\end{equation}
where ${\signNS=+1}$ (${-1}$) in domain $\Ndom$ ($\Sdom$), cf.\ Eq.~\eqref{apx:signNS}.

\coorsubsection{Relation to static family in spacelike domains ${\Fdom}$, ${\Pdom}$}
\begin{equation}
\begin{gathered}
  \frac\MAt\DSr=\signx\,\frac{\DSr}{\sqrt{\str^2-\DSr^2}}\,\exp\frac\stt\DSr\commae\\
  \frac\MAr\DSr=\signx\,\frac{\str}{\sqrt{\str^2-\DSr^2}}\,\exp\frac\stt\DSr\commae
\end{gathered}
\end{equation}
\begin{equation}
\begin{gathered}
  \MAt=\signx\,\DSr\exp\frac\brt\DSr\,\sinh\frac\brr\DSr\commae\\
  \MAr=\signx\,\DSr\exp\frac\brt\DSr\,\cosh\frac\brr\DSr\commae
\end{gathered}
\end{equation}
where ${\signx=\sign\tlr\,\sign\cos\tlt}$,
cf.\ Eqs.~\eqref{apx:signscri} and \eqref{apx:signx}.

\coorsubsection{Relation to conformally Minkowski coordinates}
\begin{equation}
  \frac\MAt\DSr=\frac{\DSr^2-\MOt^2+\MOr^2}{(\DSr-\MOt)^2-\MOr^2}\comma
  \frac\MAr\DSr=\frac{2\,\DSr\MOr}{(\DSr-\MOt)^2-\MOr^2}\period
\end{equation}

\pagebreak
\coorsection{The conformally Minkowski family}\label{asc:coorCMO}
\begin{figure}[h]
\includegraphics{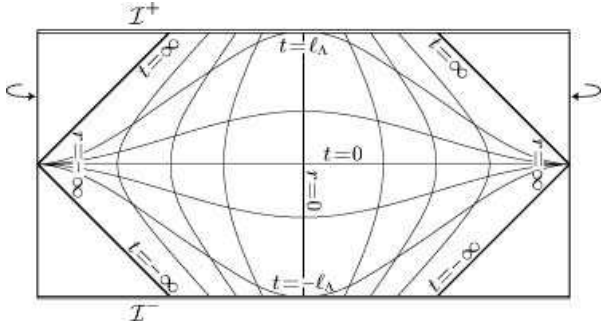}
\caption{\label{fig:coorCMO}The conformally Minkowski family of coordinates.}
\end{figure}

The \defterm{conformally Minkowski coordinates} ${\MOt,\,\MOr,\,\tht,\,\ph}$ can be
understood as spherical coordinates in the conformally related Minkowski space ${\MinkOspc}$.
The coordinates do not cover de~Sitter spacetime globally---they cover only a region
around north pole, see Fig.~\ref{fig:coorCMO}. The boundary of this region is given by the conformal infinity
of the Minkowski spacetime. These coordinates are useful for studying the
limit ${\Lambda\to 0}$.

\coorsubsection{The metric}
\begin{equation}
    \mtrc=\Bigl(\frac{2\,\DSr^2}{\DSr^2-\MOt^2+\MOr^2}\Bigr)^2\,
    \Bigl(-\grad\MOt\formsq+\grad\MOr\formsq+\MOr^2\,\sphmtrc\Bigr)\commae
\end{equation}
the ranges of coordinates
\begin{equation}
\MOt\in\realn\comma \MOr\in\realn\comma\text{such that}\quad\MOt^2-\MOr^2<\DSr^2\commae
\end{equation}
with negative values of radial coordinate $\MOr$ interpreted as described  in Eq.~\eqref{apx:negr}.

\coorsubsection{Orthonormal tetrad}
\begin{equation}
\begin{gathered}
  \cbv{\MOt}=\frac{\DSr^2-\MOt^2+\MOr^2}{2\DSr^2}\,\cv{\MOt}\commae\\
  \cbv{\MOr}=\frac{\DSr^2-\MOt^2+\MOr^2}{2\DSr^2}\,\cv{\MOr}\commae\\
  \cbv{\tht}=\frac{\DSr^2-\MOt^2+\MOr^2}{2\DSr^2}\frac1\MOr\,\cv{\MOt}\period
\end{gathered}
\end{equation}

\coorsubsection{Relation to spherical cosmological family}
\begin{equation}
\begin{gathered}
  \MOt=-\frac{\DSr\cos\tlt}{\cos\tlr+\sin\tlt}\commae\\
  \MOr=\spcm\frac{\DSr\sin\tlr}{\cos\tlr+\sin\tlt}\period
\end{gathered}
\end{equation}

\coorsubsection{Relation to flat cosmological family}
\begin{equation}
\begin{aligned}
  \frac\MOt\DSr&
  =-\frac{\DSr^2-\MAt^2+\MAr^2}{(\DSr+\MAt)^2-\MAr^2}
  =\frac{\DSr^2-\MVt^2+\MVr^2}{(\DSr-\MVt)^2-\MVr^2}
  \commae\\
  \frac\MOr\DSr&
  =\spcm\frac{2\,\DSr\MAr}{(\DSr+\MAt)^2-\MAr^2}
  =\frac{2\,\DSr\MVr}{(\DSr-\MVt)^2-\MVr^2}
  \period
\end{aligned}
\end{equation}

\coorsubsection{Relation to hyperbolic cosmological coordinates}
\begin{equation}
\begin{aligned}
  \frac\MOt\DSr&=\tanh\frac\cht{2\DSr}\,\cosh\frac\ROr\DSr\commae\\
  \frac\MOr\DSr&=\tanh\frac\cht{2\DSr}\,\sinh\frac\ROr\DSr\period
\end{aligned}
\end{equation}

\coorsubsection{Relation to static family in timelike domains ${\Ndom}$, ${\Sdom}$}
\begin{equation}
\begin{aligned}
  \frac\MOt\DSr&=\frac{\sinh\frac\brt\DSr}{\cosh\frac\brt\DSr+\signNS\cosh\frac\brr\DSr}
  \commae\\
  \frac\MOr\DSr&=\frac{\sinh\frac\brr\DSr}{\cosh\frac\brr\DSr+\signNS\cosh\frac\brt\DSr}
  \commae
\end{aligned}
\end{equation}
\begin{equation}
\begin{aligned}
  \frac\MOt\DSr&=\frac{\sqrt{\DSr^2-\str^2}\,\sinh\frac\stt\DSr}{\signNS\DSr+\sqrt{\DSr^2-\str^2}\,\cosh\frac\stt\DSr}
  \commae\\
  \frac\MOr\DSr&=\frac{\str}{\DSr+\signNS\sqrt{\DSr^2-\str^2}\,\cosh\frac\stt\DSr}
  \commae
\end{aligned}
\end{equation}
where ${\signNS=+1}$ (${-1}$) in domain $\Ndom$ ($\Sdom$), cf.\ Eq.~\eqref{apx:signNS}.

\coorsubsection{Relation to static family in spacelike domains ${\Fdom}$, ${\Pdom}$}
\begin{equation}
\begin{aligned}
  \frac\MOt\DSr&=\frac{\cosh\frac\brt\DSr}{\sinh\frac\brt\DSr-\signx\sinh\frac\brr\DSr}
  \commae\\
  \frac\MOr\DSr&=\frac{\cosh\frac\brr\DSr}{\sinh\frac\brr\DSr-\signx\cosh\frac\brt\DSr}
  \commae
\end{aligned}
\end{equation}
\begin{equation}
\begin{aligned}
  \frac\MOt\DSr&=\frac{\sqrt{\str^2-\DSr^2}\,\cosh\frac\stt\DSr}
  {-\signx\DSr+\sqrt{\str^2-\DSr^2}\,\sinh\frac\stt\DSr}\commae\\
  \frac\MOr\DSr&=\frac{\str}{\DSr-\signx\sqrt{\str^2-\DSr^2}\,\sinh\frac\stt\DSr}\commae
\end{aligned}
\end{equation}
with ${\signx=\sign\tlr\,\sign\cos\tlt}$,
cf.\ Eqs.~\eqref{apx:signscri} and \eqref{apx:signx}.

\pagebreak
\coorsection{The static family in timelike domains $\Ndom$~and~$\Sdom$}\label{asc:coorStatT}
\begin{figure}[h]
\includegraphics{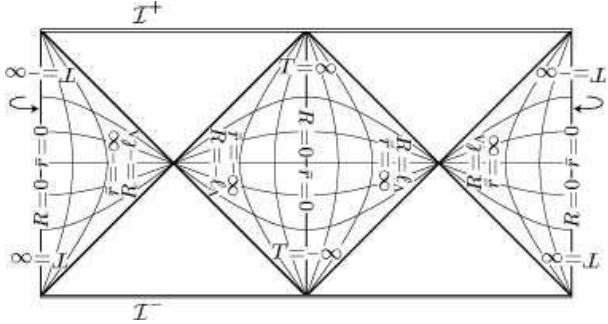}
\caption{\label{fig:coorStatT}The static family of coordinates, timelike domains.}
\end{figure}

This family consists of the \defterm{static coordinates} $\stt$, $\str$, $\tht$, $\ph$
and the \defterm{\vague{tortoidal} static coordinates $\brt,\,\brr,\,\tht,\,\ph$}.
The metric does not depend on time coordinate ${\stt=\brt}$---the coordinates
are associated with a Killing vector. Since the
Killing vector changes its character, the coordinates do not cover the spacetime smoothly.
We first describe the static coordinates in domains ${\Ndom}$ and~${\Sdom}$,
where the Killing vector is timelike.
In domain~${\Ndom}$~the orbits of the Killing vector (corresponding to the worldlines of static
observers) start and end at the north pole, in domain ${\Sdom}$---at the south pole.
They are orthogonal to slices ${\stt=\text{constant}}$, each of which consists of two hemispheres
(one in domain ${\Ndom}$, the other in ${\Sdom}$)
with homogeneous spherical 3-metric. The distances between static observers
(measured within these slices) do not change. Since
the static observers must overcome first the cosmological contraction and then the expansion,
they move with a (uniform) acceleration.

\vspace*{-3pt}
\coorsubsection{Metric and relation between coordinates}
\begin{gather}
    \mtrc=\cosh^{\!-2}\!\frac\brr\DSr\,
    \Bigl(-\grad\brt\formsq+\grad\brr\formsq
          +\DSr^2\sinh^2\frac\brr\DSr\,\sphmtrc\Bigr)\commae\\
    \mtrc=\!-\Bigl(1-\frac{\str^2}{\DSr^2}\Bigr)\,\grad\stt\formsq
          +\Bigl(1-\frac{\str^2}{\DSr^2}\Bigr)^{\!\!-1}\!\grad\str\formsq
          +\str^2\,\sphmtrc\!\!\commae\!\!
\end{gather}
\begin{subequations}
\vspace*{-6pt}
\begin{equation}
  \stt=\brt\commae
\end{equation}
\vspace*{-9pt}
\begin{equation}
\begin{aligned}
  &\exp\frac\brr\DSr=\sqrt{\frac{\DSr+\str}{\DSr-\str}}
  \commae&
  \sinh\frac\brr\DSr&=\frac{\str}{\sqrt{\DSr^2-\str^2}}
  \commae\\
  &\tanh\frac\brr\DSr=\frac{\str}{\DSr}\comma&
  \cosh\frac\brr\DSr&=\frac{\DSr}{\sqrt{\DSr^2-\str^2}}\commae
\end{aligned}
\end{equation}
\end{subequations}
\begin{equation}\label{apx:signNS}
\signNS =
\begin{cases}
+1\quad\text{in domain ${\Ndom}$}\commae\\[0.5ex]
-1\quad\text{in domain ${\Sdom}$}\period
\end{cases}
\end{equation}
The ranges  of coordinates are
\begin{equation}
\begin{gathered}
\stt\in\realn\comma\str\in(-\DSr,\DSr)\commae\\
\brt\in\realn\comma\brr\in\realn\commae
\end{gathered}
\end{equation}\pagebreak[1]%
with negative values of radial coordinate $\str$ and $\brr$ interpreted as described  in Eq.~\eqref{apx:negr}.

\coorsubsection{Orthonormal tetrad}
\begin{equation}
\begin{gathered}
  \cbv{\stt}=\Bigl(1-\frac{\str^2}{\DSr^2}\Bigr)^{\!\!-1/2}\cv{\stt}=\cosh\frac\brr\DSr\,\cv{\brt}\commae\\
  \cbv{\str}=\Bigl(1-\frac{\str^2}{\DSr^2}\Bigr)^{1/2}\cv{\str}=\cosh^{\!-1}\!\frac\brr\DSr\,\cv{\brr}\commae\\
  \cbv{\tht}=\frac1\str\,\cv{\tht}=\frac1\DSr\coth\frac\brr\DSr\,\cv{\tht}\period
\end{gathered}
\end{equation}

\coorsubsection{Relation to spherical cosmological family}
\begin{equation}\label{apx:sttrT=tltr}
  \stt=\frac\DSr2\,\log\frac{\cos\tlr-\cos\tlt}{\cos\tlr+\cos\tlt}\comma
  \str=\DSr\,\frac{\sin\tlr}{\sin\tlt}\commae
\end{equation}
\begin{subequations}
\begin{gather}
  \brt
  =\frac\DSr2\,\log\Bigl(\tan{\frac{\tlt+\tlr}2}\;\tan{\frac{\tlt-\tlr}2}\Bigr)
  \commae\\
\begin{aligned}
  &\exp\frac\brt\DSr=\sqrt{\frac{\cos\tlr-\cos\tlt}{\cos\tlr+\cos\tlt}}
  \commae\!\!&
  \sinh\frac\brt\DSr&=\frac{-\signNS\cos\tlt}{\sqrt{\cos^2\tlr-\cos^2\tlt}}
  \commae\\
  &\tanh\frac\brt\DSr=-\frac{\cos\tlt}{\cos\tlr}
  \commae&
  \cosh\frac\brt\DSr&=\frac{\signNS\cos\tlr}{\sqrt{\cos^2\tlr-\cos^2\tlt}}
  \commae
\end{aligned}\notag\\[9pt]
  \brr
  =\frac\DSr2\,\log\Bigl(\tan{\frac{\tlt+\tlr}2}\;\cot{\frac{\tlt-\tlr}2}\Bigr)
  \commae\\
\begin{aligned}
  &\exp\frac\brr\DSr=\sqrt{\frac{\sin\tlt+\sin\tlr}{\sin\tlt-\sin\tlr}}
  \commae\!\!&
  \sinh\frac\brr\DSr&=\frac{\sin\tlr}{\sqrt{\sin^2\tlt-\sin^2\tlr}}
  \commae\\
  &\tanh\frac\brr\DSr=\frac{\sin\tlr}{\sin\tlt}
  \commae&
  \cosh\frac\brr\DSr&=\frac{\sin\tlt}{\sqrt{\sin^2\tlt-\sin^2\tlr}}
  \period
\end{aligned}\notag
\end{gather}
\end{subequations}

\coorsubsection{Relation to flat cosmological family}
\begin{equation}
\begin{gathered}
  \brt
  =\frac\DSr2\log\frac{\MAt^2-\MAr^2}{\DSr^2}
  =-\frac\DSr2\log\frac{\MVt^2-\MVr^2}{\DSr^2}
  \commae\\
  \brr
  =\frac\DSr2\log\frac{\MAt+\MAr}{\MAt-\MAr}
  =\frac\DSr2\log\frac{\MVt-\MVr}{\MVt+\MVr}
  \commae
\end{gathered}
\end{equation}
\begin{equation}\label{apx:sttrT=MAVtr}
\begin{gathered}
  \frac\stt\DSr
  =\frac12\log\frac{\MAt^2-\MAr^2}{\DSr^2}
  =-\frac12\log\frac{\MVt^2-\MVr^2}{\DSr^2}
  \commae\\
  \frac\str\DSr
  =\frac\MAr\MAt
  =-\frac\MVr\MVt
  \period
\end{gathered}
\end{equation}

\coorsubsection{Relation to conformally Minkowski coordinates}
\begin{gather}
  \tanh\frac\stt\DSr=\frac{2\,\DSr\MOt}{\DSr^2+\MOt^2-\MOr^2}
  \comma
  \frac\str\DSr=\frac{2\,\DSr\MOr}{\DSr^2+\MOr^2-\MOt^2}
  \commae\\
  \brt=\frac\DSr2\log\frac{(\DSr+\MOt)^2-\MOr^2}{(\DSr-\MOt)^2-\MOr^2}
  \comma
  \brr=\frac\DSr2\log\frac{(\DSr+\MOr)^2-\MOt^2}{(\DSr-\MOr)^2-\MOt^2}
  \period
\end{gather}

\pagebreak
\coorsection{The static family in spacelike domains $\Fdom$~and~$\Pdom$}\label{asc:coorStatS}
\begin{figure}[h]
\includegraphics{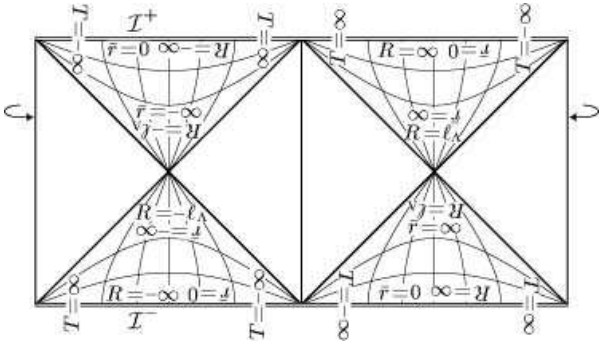}
\caption{\label{fig:coorStatS}\!The static family of coordinates, spacelike domains.}
\end{figure}

Here we describe the \defterm{static coordinates ${\stt,\,\str,\,\tht,\,\ph}$}
and the \defterm{\vague{tortoidal} static coordinates ${\brt,\,\brr,\,\tht,\,\ph}$}
from the preceding section in domains ${\Fdom}$ and ${\Sdom}$ where
the Killing vector is spacelike. These \vague{non-static} domains
extend up to infinity, namely, domain ${\Fdom}$ up to ${\scri^+}$,
domain ${\Pdom}$ up to ${\scri^-}$.
The orbits of the Killing vector start at the south pole and end at the north
pole in ${\Fdom}$, and they point in opposite direction in ${\Pdom}$.
The motion along them could thus be characterized
as a \vague{translation} from one pole to the other.
The Lorentzian hypersurfaces ${\stt=\text{constant}}$ are
homogeneous spaces with positive curvature, i.e., 3-dimensional
de~Sitter spacetimes.

\vspace*{-9pt}
\coorsubsection{Metric and relation between coordinates}
\begin{gather}
    \mtrc=\sinh^{\!-2}\!\frac\brr\DSr\,
    \Bigl(-\grad\brr\formsq+\grad\brt\formsq
          +\DSr^2\cosh^2\frac\brr\DSr\,\sphmtrc\Bigr)\commae\\
    \mtrc=\!-\Bigl(1-\frac{\str^2}{\DSr^2}\Bigr)\,\grad\stt\formsq
          +\!\Bigl(1-\frac{\str^2}{\DSr^2}\Bigr)^{\!\!-1}\!\grad\str\formsq
          +\str^2\,\sphmtrc\!\!\commae\!\!
\end{gather}
\begin{subequations}
\begin{equation}
  \stt=\brt\commae
\end{equation}\\[-6ex]
\begin{equation}
\begin{aligned}
  &\exp\frac\brr\DSr=\sqrt{\frac{\str+\DSr}{\str-\DSr}}
  \commae\!\!&
  &\abs{\sinh\frac\brr\DSr}=\frac{\DSr}{\sqrt{\str^2-\DSr^2}}
  \commae\\
  &\tanh\frac\brr\DSr=\frac{\DSr}{\str}
  \commae\!\!&
  &\cosh\frac\brr\DSr=\frac{\abs{\str}}{\sqrt{\str^2-\DSr^2}}\period
\end{aligned}
\end{equation}
\end{subequations}
The signature factors $\signscri$ and $\signx$ are defined as
\begin{equation}\label{apx:signscri}
\signscri =
\begin{cases}
+1\quad\text{in domain ${\Fdom}$}\commae\\[0.5ex]
-1\quad\text{in domain ${\Pdom}$}\commae
\end{cases}
\end{equation}
and\\[-4ex]
\begin{equation}\label{apx:signx}
\signx = -\signscri\,\sign\tlr\period
\end{equation}
The coordinates ranges are
\begin{equation}
\begin{gathered}
\stt\in\realn\comma\abs{\str}\in(\DSr,\infty)\commae\\
\brt\in\realn\comma\brr\in\realn\commae
\end{gathered}
\end{equation}
with negative values of radial coordinate $\str$ and $\brr$ interpreted as described  in Eq.~\eqref{apx:negr}.

\coorsubsection{Orthonormal tetrad}
\begin{equation}
\begin{gathered}
  \cbv{\stt}=\Bigl(\frac{\str^2}{\DSr^2}-1\Bigr)^{\!\!-1/2}\cv{\stt}=\abs{\,\sinh\frac\brr\DSr}\,\cv{\brt}\commae\\
  \cbv{\str}=\Bigl(\frac{\str^2}{\DSr^2}-1\Bigr)^{1/2}\cv{\str}=-\!\abs{\,\sinh^{\!-1}\!\frac\brr\DSr}\,\cv{\brr}\commae\\
  \cbv{\tht}=\frac1\str\,\cv{\tht}=\frac1\DSr\abs{\,\tanh\frac\brr\DSr}\,\cv{\tht}\period
\end{gathered}
\end{equation}

\coorsubsection{Relation to spherical cosmological family}
\begin{equation}\label{apx:sttrS=tltr}
  \stt=\frac\DSr2\,\log\frac{\cos\tlt-\cos\tlr}{\cos\tlt+\cos\tlr}\comma
  \str=\DSr\,\frac{\sin\tlr}{\sin\tlt}\commae
\end{equation}
\begin{subequations}
\begin{gather}
  \brt
  = \frac\DSr2\,\log\Bigl(-\tan{\frac{\tlt+\tlr}2}\;\tan{\frac{\tlt-\tlr}2}\Bigr)
  \commae\\
\begin{aligned}
  &\exp\frac\brt\DSr=\sqrt{\frac{\cos\tlt-\cos\tlr}{\cos\tlt+\cos\tlr}}
  \commae\!\!&
  \sinh\frac\brt\DSr&=\frac{\signscri\cos\tlr}{\sqrt{\cos^2\tlt-\cos^2\tlr}}
  \commae\\
  &\tanh\frac\brt\DSr=-\frac{\cos\tlr}{\cos\tlt}
  \commae&
  \cosh\frac\brt\DSr&=\frac{-\signscri\cos\tlt}{\sqrt{\cos^2\tlt-\cos^2\tlr}}
  \commae
\end{aligned}\notag\\[9pt]
  \brr
  =\frac\DSr2\,\log\Bigl(-\tan{\frac{\tlt+\tlr}2}\;\cot{\frac{\tlt-\tlr}2}\Bigr)
  \commae\\
\begin{aligned}
  &\exp\frac\brr\DSr=\sqrt{\frac{\sin\tlr+\sin\tlt}{\sin\tlr-\sin\tlt}}
  \commae\!\!&
  \abs{\sinh\frac\brr\DSr}&=\frac{\sin\tlt}{\sqrt{\sin^2\tlr-\sin^2\tlt}}
  \commae\\
  &\tanh\frac\brr\DSr=\spcm\frac{\sin\tlt}{\sin\tlr}
  \commae&
  \cosh\frac\brr\DSr&=\frac{\abs{\sin\tlr}}{\sqrt{\sin^2\tlr-\sin^2\tlt}}
  \period
\end{aligned}\notag
\end{gather}
\end{subequations}

\coorsubsection{Relation to flat cosmological family}
\begin{equation}
\begin{gathered}
  \brt
  =\frac\DSr2\log\frac{-\MAt^2+\MAr^2}{\DSr^2}
  =-\frac\DSr2\log\frac{-\MVt^2+\MVr^2}{\DSr^2}
  \commae\\
  \brr
  =\frac\DSr2\log\frac{\MAr+\MAt}{\MAr-\MAt}
  =\frac\DSr2\log\frac{\MVr-\MVt}{\MVr+\MVt}
  \commae
\end{gathered}
\end{equation}
\begin{equation}\label{apx:sttrS=MAVtr}
\begin{gathered}
  \frac\stt\DSr
  =\frac12\log\frac{-\MAt^2+\MAr^2}{\DSr^2}
  =-\frac12\log\frac{-\MVt^2+\MVr^2}{\DSr^2}
  \commae\\
  \frac\str\DSr
  =\frac\MAr\MAt
  =-\frac\MVr\MVt
  \period
\end{gathered}
\end{equation}

\coorsubsection{Relation to conformally Minkowski coordinates}
\begin{equation}
  \coth\frac\stt\DSr=\frac{2\,\DSr\MOt}{\DSr^2+\MOt^2-\MOr^2}
  \comma
  \frac\str\DSr=\frac{2\,\DSr\MOr}{\DSr^2+\MOr^2-\MOt^2}
  \commae
\end{equation}
\begin{equation}
\begin{gathered}
  \brt=\frac\DSr2\log\Bigl(-\frac{(\DSr+\MOt)^2-\MOr^2}{(\DSr-\MOt)^2-\MOr^2}\Bigr)
  \commae\\
  \brr=\frac\DSr2\log\Bigl(-\frac{(\DSr+\MOr)^2-\MOt^2}{(\DSr-\MOr)^2-\MOt^2}\Bigr)
  \commae
\end{gathered}
\end{equation}

\pagebreak
\coorsection{The hyperbolic cosmological family}\label{asc:coorHyp}
\begin{figure}[h]
\includegraphics{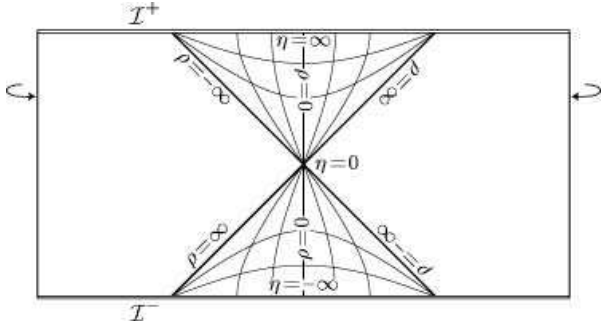}
\caption{\label{fig:coorHyp}The hyperbolic cosmological family of coordinates.}
\end{figure}

The third type of cosmological coordinates are the \defterm{hyperbolic cosmological
coordinates ${\cht,\,\chr,\,\tht,\ph}$}. The hypersurfaces ${\cht=\text{constant}}$
are homogeneous spaces with negative curvature,
coordinate lines ${\chr,\tht,\ph=\text{constant}}$ correspond to the worldlines of
cosmological observers orthogonal to these slices, and the vector
${\cvil{\cht}}$ is a timelike conformal Killing vector. The coordinates
cover spacetime only partially---they can be introduced in two disconnected
domains near the north pole, namely, in the past of the event ${\tlt=\pi/2}$,
${\tlr=0}$ (where ${\cht<0}$), and in the future
of this event (where ${\cht>0}$).

\vspace{-9pt}
\coorsubsection{The metric}
\begin{equation}
    \mtrc=-\grad\cht\formsq+{\textstyle\sinh^2\!\frac\cht\DSr}\,
    \bigl(\grad\chr\formsq+\DSr^2{\textstyle\sinh^2\!\frac\chr\DSr}\,\sphmtrc\bigr)\commae
\end{equation}
The ranges of coordinates and the signature factor $\signhyp$ are
\begin{equation}
\begin{aligned}
&\cht\in\realn^+\comma \chr\in\realn\comma\signhyp=+1&&\text{in the future patch}\commae\\
&\cht\in\realn^-\comma \chr\in\realn\comma\signhyp=-1&&\text{in the past patch}\commae
\end{aligned}
\end{equation}
with negative values of radial coordinate $\chr$ interpreted as described  in Eq.~\eqref{apx:negr}.

\vspace{-9pt}
\coorsubsection{Orthonormal tetrad}
\begin{equation}
\begin{gathered}
  \cbv{\cht}=\cv{\cht}\comma
  \cbv{\chr}=\sinh^{\!-1}\!\frac\cht\DSr\;\cv{\chr}\commae\\
  \cbv{\tht}=\sinh^{\!-1}\!\frac\cht\DSr\,\sinh^{\!-1}\!\frac\chr\DSr\;\cv{\chr}\period
\end{gathered}
\end{equation}

\vspace{-12pt}
\coorsubsection{Relation to spherical cosmological family}
\begin{equation}\label{apx:chtr=tltr}
  \tanh\frac\cht{2\DSr}=
    \signhyp\sqrt{\frac{\cos\tlr-\sin\tlt}{\cos\tlr+\sin\tlt}}\comma
  \tanh\frac\chr\DSr=
    -\frac{\sin\tlr}{\cos\tlt}\period
\end{equation}

\vspace{-12pt}
\coorsubsection{Relation to conformally Minkowski coordinates}
\begin{equation}
  \tanh\frac\cht{2\DSr}=
    \signhyp{\frac{\sqrt{\MOt^2-\MOr^2}}{\DSr}}\comma
  \tanh\frac\chr\DSr=\frac\MOr\MOt\period
\end{equation}

\pagebreak
\coorsection{The accelerated coordinate family}\label{asc:coorAcc}
\begin{figure}[h]
\includegraphics{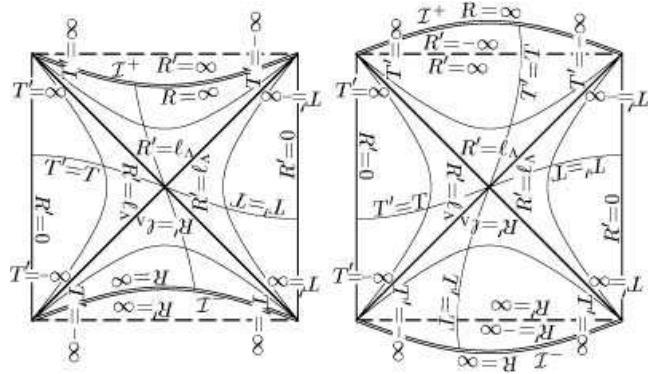}
\vspace{-6pt}
\caption{\label{fig:coorAcc}The accelerated family of coordinates.}
\vspace{-6pt}
\end{figure}

This family consists of the \defterm{accelerated coordinates ${\At,\,\Ar,\,\Ath,\,\ph}$},
and the \defterm{C-metric-like coordinates ${\Ct,\,\Cy,\,\Cx,\,\ph}$} ($\Ct$~being different
from ${\spt}$ of the standard coordinates).
Contrary to the previous cases the accelerated coordinates are
centered on uniformly accelerate origins:
${\Ar=0}$ corresponds to two worldlines with acceleration ${\abs{\accl}}$.
The transformation relations to the systems introduced above mix these three
coordinates in general.

The accelerated coordinates are closely related to the static system.
Their time coordinates coincide,~${\At\!=\!\stt}$, and coordinate lines
${\Ar,\Ath,\ph=\text{constant}}$ are the same as those with
${\str,\tht,\ph=\text{constant}}$. Both coordinate systems
are identical for ${\accl=0}$.
Sections ${\stt,\At,\ph\!=\!\text{constant}}$ with ${\str,\Ar<\DSr}$ have geometry of 2-sphere
with parallels and meridians given by the coordinate lines of the static coordinates ${\str,\,\tht}$.
The lines of coordinates ${\Ar,\,\Ath}$ are the deformed
version of static ones, their poles are shifted along meridian
${\tht=0}$ towards each other, cf.\ Fig.~\ref{fig:AccRth}.

Two conformal diagrams of sections ${\Ath,\ph=\text{constant}}$
(${\Ath<\pi/2}$ on the right, ${\Ath>\pi/2}$ on the left),
adapted to the accelerated coordinates, are depicted in Fig.~\ref{fig:coorAcc}.
The shape of the diagram varies with different values of ${\Ath}$; indeed, the position
of infinity is given by $\Ar={-\DSr^2/\Ro\,\cos^{\!-1}\Ath}$.
See also Fig.~\ref{fig:Acc} for sections ${\Ath=0,\pi}$.

The C-metric-like coordinates rescale only the values of the accelerated coordinates
and regularize the coordinate singularity ${\Ar=\pm\infty}$. de~Sitter metric in these
coordinates is a zero-mass limit of the C-metric (the metric
describing accelerated black holes; see, e.g., \cite{PodolskyOrtaggioKrtous:2003,DiasLemos:2003b}).

Finally, we use four parameters  ${\accl,\,\acp,\,\Ro,\,\Bo}$ to
parametrize the acceleration.\nopagebreak
They are related as \mbox{follows}:
\begin{alignat}{8}
  &\sinh\acp
 &\;=\;
  &\frac{\Ro}{\sqrt{\DSr^2 -\Ro^2}}
 &\;=\;
  &\frac{\Bo^2-\DSr^2}{2\,\DSr\Bo}
 &\;=\;
  &\quad\!-\accl\DSr
\commae\notag\\[-2pt]
  &\cosh\acp
 &\;=\;
  &\frac{\DSr}{\sqrt{\DSr^2 -\Ro^2}}
 &\;=\;
  &\frac{\Bo^2+\DSr^2}{2\,\DSr\Bo}
 &\;=\;
  &\sqrt{1+\accl^2\DSr^2}
\commae\notag\\[-12pt]
\\[-6pt]
  &\tanh\acp
 &\;=\;
  &\qquad\!\frac\Ro\DSr
 &\;=\;
  &\frac{\Bo^2-\DSr^2}{\Bo^2+\DSr^2}
 &\;=\;
  &-\frac{\accl\DSr}{\sqrt{1+\accl^2\DSr^2}}
\commae\notag\\[-2pt]
  &\exp\acp
 &\;=\;
  &\sqrt{\frac{\DSr+\Ro}{\DSr -\Ro}}
 &\;=\;
  &\quad\,\frac\Bo\DSr
 &\;=\;
  &\sqrt{1+\accl^2\DSr^2} - \accl\DSr
\period\notag
\end{alignat}

\coorsubsection{Metric and relation between coordinates}
\begin{gather}
    \mtrc\!=\!\Omega^2\biggl[-\Bigl(1\!-\!\frac{\Ar^2}{\DSr^2}\Bigr)\,\grad\At\formsq
          +\!\Bigl(1\!-\!\frac{\Ar^2}{\DSr^2}\Bigr)^{\!\!-1}\grad\Ar\formsq
          +\Ar^2\,\sphmtrc{}'\biggr]\!\!\commae\\
    \mtrc\!=\!\RTr^2\mspace{-1mu}\biggl[-(\Cy^2\!{-}1)\,\grad\Ct\formsq{+}\frac1{\Cy^2\!{-}1}\,\grad\Cy\formsq
          {+}\frac1{1{-}\Cx^2}\,\grad\Cx\formsq{+}(1{-}\Cx^2)\,\grad\ph\formsq\biggr]\!\!\commae\\[-12pt]
\intertext{where}
  \sphmtrc{}'=\bigl(\grad\Ath\formsq+\sin^2\Ath\,\grad\ph\formsq\bigr)\commae\\[4pt]
  \Omega = \frac{\sqrt{1-{\Ro^2}/{\DSr^2}}}{1+({\Ar\Ro}/{\DSr^2})\cos\Ath}
          =\frac\RTr\Ar=\frac{\RTr\,\Cy}\DSr\commae\\
  \RTr=\frac{\DSr}{\Cy\cosh\acp-\Cx\sinh\acp}=\Omega\,\Ar=\Omega\,\frac\DSr\Cy\period
\end{gather}
\begin{equation}\label{apx:Ctyx=Atrth}
  \Ct=\frac\At\DSr\qquad\Cy=\frac\DSr\Ar\qquad\Cx=-\cos\Ath\commae
\end{equation}

\vspace{-9pt}
\coorsubsection{Orthonormal tetrad}
\begin{gather}
  \cbv{\At}=\abs{\Omega}^{\!-1}\Bigl(1-\frac{\Ar^2}{\DSr^2}\Bigr)^{\!\!-1/2}\cv{\At}=\frac1{\RTr\sqrt{\Cy^2-1}}\,\cv{\At}\commae\notag\\
  \cbv{\Ar}=\abs{\Omega}^{\!-1}\Bigl(1-\frac{\Ar^2}{\DSr^2}\Bigr)^{1/2}\cv{\Ar}=\frac1{\RTr}\sqrt{\Cy^2-1}\,\cv{\Ar}\commae\notag\\
  \cbv{\Ath}=\frac1{\Omega\Ar}\,\cv{\Ath}=\frac1\RTr\,\cv{\Ath}\period\label{apx:AccTetrad}
\end{gather}

\vspace{-9pt}
\coorsubsection{Relation to static coordinates}
\begin{equation}
\begin{gathered}
  \stt=\At
\\[6pt]
  \str\cos\tht=\frac{\Ar\cos\Ath+\Ro}{1+({\Ar\Ro}/{\DSr^2})\cos\Ath}\commae
\\[3pt]
  \str\sin\tht=\frac{\Ar\sin\Ath\sqrt{1-\frac{\Ro^2}{\DSr^2}}}
  {1+({\Ar\Ro}/{\DSr^2})\cos\Ath}\commae
\\[3pt]
  \frac{\str^2}{\DSr^2}=
    1-\frac{\bigl(1-{\Ar^2}/{\DSr^2}\bigr)\bigl(1-{\Ro^2}/{\DSr^2}\bigr)}
    {\bigl(1+({\Ar\Ro}/{\DSr^2})\cos\Ath\bigr)^2}\commae
\\[3pt]
  \tan\tht=\frac{\Ar\sin\Ath\sqrt{1-\frac{\Ro^2}{\DSr^2}}}{\Ar\cos\Ath+\Ro}\period
\end{gathered}
\end{equation}
The inverse relations have the same form  with ${\stt,\,\str,\,\tht}$ and ${\At,\,\Ar,\,\Ath}$
interchanged only and $\acp$ replaced by $-\acp$.

\vspace{-3pt}
\begin{equation}
  \Omega
    = \frac{\sqrt{1-{\Ro^2}/{\DSr^2}}}{1+({\Ar\Ro}/{\DSr^2})\cos\Ath}
    = \frac{1-({\str\Ro}/{\DSr^2})\cos\tht}{\sqrt{1-{\Ro^2}/{\DSr^2}}}\commae
\end{equation}
\begin{gather}
  \Bigl(1+\frac{\Ar\Ro}{\DSr^2}\cos\Ath\Bigr)
  \Bigl(1-\frac{\str\Ro}{\DSr^2}\cos\tht\Bigr)
  =1-\frac{\Ro^2}{\DSr^2}
\commae\\
  \frac{1-{\Ar^2}/{\DSr^2}}{1+({\Ar\Ro}/{\DSr^2})\cos\Ath}
  =\frac{1-{\str^2}/{\DSr^2}}{1-({\str\Ro}/{\DSr^2})\cos\tht}\period
\end{gather}

\coorsubsection{Relation to Robinson-Trautman coordinates}
\begin{gather}
\At=\RTu\cosh\acp{-}
   \frac\DSr2\log\abs{\frac
   {\DSr{-}\RTr\,(\sinh\acp\cos\Ath{+}\cosh\acp)}
   {\DSr{-}\RTr\,(\sinh\acp\cos\Ath{-}\cosh\acp)}}\!\commae\notag\\[3pt]
\Ar=\frac{\RTr\,\cosh\acp}{1-(\RTr/\DSr)\sinh\acp\cos\Ath}\commae\\[6pt]
\abs{\,\tan\frac\Ath2}=\exp\Bigl(\RTp-\frac\RTu\DSr\sinh\acp\Bigr)\commae\notag
\\[12pt]
\Ct=\frac\RTu\DSr\cosh\acp{-}
   \frac12\log\abs{\frac
   {\DSr{-}\RTr\,(\sinh\acp\cos\Ath{+}\cosh\acp)}
   {\DSr{-}\RTr\,(\sinh\acp\cos\Ath{-}\cosh\acp)}}\!\commae\notag\\[3pt]
\Cy=\frac\DSr{\RTr\cosh\acp}-\tanh\acp\cos\Ath\commae\\[6pt]
\Cx=\tanh\Bigl(\RTp-\frac\RTu\DSr\sinh\acp\Bigr)\commae\notag
\end{gather}
where ${\cos\Ath=-\Cx}$ is given in terms of the Robinson-Trautman
coordinates by the last equation.

\vspace{-6pt}
\coorsubsection{Relation to flat cosmological family}

If we introduce the spherical coordinates ${\AVt,\,\AVr,\,\Ath,\,\ph}$
boosted with respect to the flat cosmological coordinates
${\MVt,\,\MVr,\,\tht,\,\ph}$ by a boost ${\acp}$
(in the sense of Minkowski space ${\MinkVspc}$),
we find that the accelerated coordinates ${\At,\,\Ar}$
are related to 
${\AAt,\,\AAr}$
in exactly the same way as
the static coordinates ${\stt,\,\str}$ are related to the coordinates
${\MVt,\,\MVr}$. The boost
${\AVt=\MVt\cosh\acp+\MVz\sinh\acp}$,
${\AVx=\MVx}$,
${\AVy=\MVy}$,
${\AVz=\MVt\sinh\acp+\MVz\cosh\acp}$,
rewritten in the spherical coordinates
${\AVr\cos\Ath=\AVz}$, ${\AVr\sin\Ath=\sqrt{\AVx^2+\AVy^2}}$,
reads
\begin{equation}
\begin{gathered}
  \AVt=\MVt\cosh\acp+\MVr\cos\tht\sinh\acp\commae\\
  \AVr\cos\Ath=\MVt\sinh\acp+\MVr\cos\tht\cosh\acp\commae\\
  \AVr\sin\Ath=\MVr\sin\tht\commae
\end{gathered}
\end{equation}
and relations analogous to Eqs.~\eqref{apx:sttrT=MAVtr} and \eqref{apx:sttrS=MAVtr} are:
\begin{equation}
  \At=-\frac\DSr2\log\Bigabs{\frac{\AVt^2-\AVr^2}{\DSr^2}}\comma
  \Ar=-\DSr\frac\AVr\AVt\period
\end{equation}

Similarly, the formulas relating the accelerated coordinates to the coordinates
${\MAt,\,\MAr,\,\tht}$ are:
\begin{equation}
\begin{gathered}
  \AAt=\MAt\cosh\acp-\MAr\cos\tht\sinh\acp\commae\\
  \AAr\cos\Ath=-\MAt\sinh\acp+\MAr\cos\tht\cosh\acp\commae\\
  \AAr\sin\Ath=\MAr\sin\tht\commae
\end{gathered}
\end{equation}
\begin{equation}
  \At=\frac\DSr2\log\Bigabs{\frac{\AAt^2-\AAr^2}{\DSr^2}}
  \comma\Ar=\DSr\frac\AAr\AAt\period
\end{equation}

The conformal factor takes the form
\begin{equation}
  \Omega=\frac\AVt\MVt=\frac\AAt\MAt
  =\cosh\acp-\frac{\str}{\DSr}\sinh\acp\cos\tht\period
\end{equation}

\pagebreak
\coorsection{The Robinson-Trautman coordinates}\label{asc:coorRT}
\begin{figure}[h]
\includegraphics{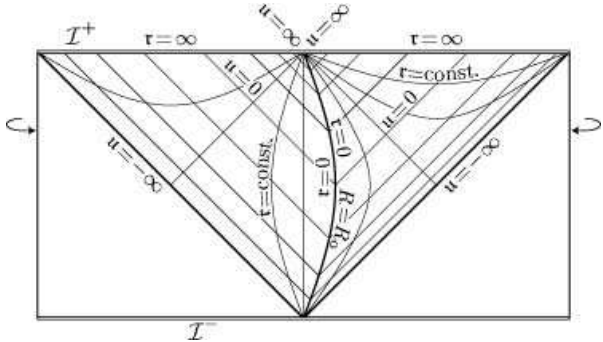}
\caption{\label{fig:coorRT}The Robinson-Trautman coordinates.}
\end{figure}

In the \defterm{Robinson-Trautman coordinates ${\RTu,\,\RTr,\,\RTp,\,\ph}$}
(or in their complex version ${\RTu,\,\RTr,\,\RTz,\,\RTb}$), de~Sitter metric
takes the standard Robinson-Trautman form \mbox{\cite{Stephanietal:book}}.
The coordinate ${\RTu}$ is null, the \vague{radial} coordinate ${\RTr}$ is an
affine parameter along coordinate lines ${\RTu,\RTp,\ph=\text{constant}}$.
These lines are null geodesics generating light cones with vertices at the origin ${\RTr=0}$.
The coordinates ${\RTp,\,\ph}$ (or 
${\RTz,\,\RTb}$) are angular coordinates, however, they are not functions
of the accelerated angular coordinates ${\Ath,\ph}$ only (cf. Eq.~\eqref{apx:Ath=RTuRTp}).
Because ${\Ath,\ph}$ have a clearer geometrical meaning, we list some
formulas also in the mixed coordinate system ${\RTu,\,\RTr,\,\Ath,\ph}$.

The origin ${\RTr=0}$ of the Robinson-Trautman coordinates is
centered on the worldline of the uniformly accelerated observer moving with
the acceleration ${\abs{\accl}=\abs{\DSr^{\!-1}\sinh\acp}}$. The coordinates are thus
closely related to the accelerated coordinates.

The coordinates ${\RTu,\,\RTr,\,\RTp,\,\ph}$ do not cover  the whole spacetime smoothly.
They can be introduced smoothly in the future of the north pole, or
in the past of the south pole. At the boundary of these two domains,
${\RTu\to\pm\infty}$.

\vspace{-6pt}
\coorsubsection{Metric and relation between coordinates}
\begin{gather}
  \mtrc=-\RTH\,\grad\RTu\formsq-\grad\RTu\stp\grad\RTr
          +\frac{\RTr^2}{\RTP^2}\,
      \bigl(\grad\RTp\formsq+\grad\ph\formsq\bigr)
      \commae\label{apx:RTppmtrc}\\
  \mtrc=-\RTH\,\grad\RTu\formsq-\grad\RTu\stp\grad\RTr
          +\frac{\RTr^2}{\RTP^2}\,\grad\RTz\stp\grad\RTb
      \commae\label{apx:RTzbmtrc}\\
\begin{split}\raisetag{42pt}\label{apx:RTthpmtrc}
  \mtrc&=-\cosh^2\!\acp\,\frac{\RTr^2}{\DSr^2}\,(\Cy^2-1)\,\grad\RTu\formsq
          -\grad\RTu\stp\grad\RTr\\&\quad
      +\cosh\acp\,\frac{\RTr^2}{\DSr}\,\sin\Ath\,\grad\RTu\stp\grad\Ath
      +\RTr^2\bigl(\grad\Ath\formsq+\sin^2\!\Ath\,\grad\ph\formsq\bigr)\commae
\end{split}
\end{gather}
\begin{align}
   \RTH&=-\frac{\RTr^2}{\DSr^2}
     +2\,\frac\RTr\DSr\,\sinh\acp\,
     \tanh\Bigl(\RTp-\frac\RTu\DSr\,\sinh\acp\Bigr)+1\notag\\
     &=-\frac{\RTr^2}{\DSr^2}
     -2\,\frac\RTr\DSr\,\sinh\acp\,\cos\Ath+1\commae\\
   \RTP&=\cosh\Bigl(\RTp-\frac\RTu\DSr\,\sinh\acp\Bigr)=\frac1{\sin\Ath}\period
\end{align}\pagebreak[2]\vspace*{-4ex}
\begin{equation}\label{apx:Ath=RTuRTp}
\begin{gathered}
   \RTp=\frac\RTu\DSr\,\sinh\acp + \log\abs{\tan\frac\Ath2}\commae\\
   \abs{\tan\frac\Ath2}=\exp\Bigl(\RTp-\frac\RTu\DSr\,\sinh\acp\Bigr)\commae
\end{gathered}
\end{equation}
\begin{equation}
\begin{aligned}
   &\RTz=\frac1{\sqrt2}(\RTp-i\ph)
   \commae&
   &\RTp=\frac1{\sqrt2}(\RTz+\RTb)
   \commae\\
   &\RTb=\frac1{\sqrt2}(\RTp+i\ph)
   \commae&
   &\ph=\frac{i}{\sqrt2}(\RTz-\RTb)
   \period
\end{aligned}
\end{equation}

\vspace{-12pt}
\coorsubsection{Null tetrad}

Since the Robinson-Trautman coordinates are closely related to the congruence of
null geodesics, it is convenient to introduce the null tetrad which is parallelly
transported along these geodesics ${\RTu,\RTp,\ph=\text{constant}}$:

\begin{equation}\label{apx:RTnulltetr}
\begin{gathered}
\kG_\cfRT=\frac1{\sqrt2}\cv{\RTr}\comma
\lG_\cfRT=-\frac1{\sqrt2}\RTH\,\cv{\RTr}+\sqrt2\,\cv{\RTu}\commae\\
\mG_\cfRT=\frac1{\sqrt2}\,\frac\RTP\RTr\,\Bigl(\cv{\RTp}-i\cv{\ph}\Bigr)\commae\\
\bG_\cfRT=\frac1{\sqrt2}\,\frac\RTP\RTr\,\Bigl(\cv{\RTp}+i\cv{\ph}\Bigr)\period
\end{gathered}
\end{equation}

\vspace{-12pt}
\coorsubsection{Relation to accelerated coordinate family}
\begin{equation}
\begin{aligned}
  \RTr&=\frac{\Ar\sqrt{1-{\Ro^2}/{\DSr^2}}}{1+({\Ar\Ro}/{\DSr^2})\cos\Ath}\commae\\
  \RTu&=\sqrt{1-\frac{\Ro^2}{\DSr^2}}\,\Biggl(\At+\frac\DSr2\log\abs{\frac{\Ar-\DSr}{\Ar+\DSr}}\Biggr)\commae\\
  \RTp&=\frac\Ro\DSr\,\Biggl(\frac\At\DSr+\frac12\log\abs{\frac{\Ar-\DSr}{\Ar+\DSr}}\Biggr)
      +\log\abs{\tan\frac\Ath2}\!\!\commae\!\!
\end{aligned}
\end{equation}
\begin{align}
  \RTr&=\frac{\DSr}{\Cy\cosh\acp-\Cx\sinh\acp}\commae\notag\\
  \RTu&=\frac\DSr{\cosh\acp}\Biggl(\Ct+\frac12\log\abs{\frac{1-\Cy}{1+\Cy}}\Biggr)\commae\\
  \RTp&=\tanh\acp\Biggl(\Ct+\frac12\log\abs{\frac{1-\Cy}{1+\Cy}}\Biggr)
      +\frac12\log\abs{\frac{1+\Cx}{1-\Cx}}\period\notag
\end{align}

\vspace{-12pt}
\coorsubsection{Relation to static family}
\begin{equation}
\RTr\!=\!\frac{\DSr}{\sqrt{1{-}\Ro^2/\DSr^2}}
     \biggl[\Bigl(1{-}\frac{\str\Ro}{\DSr^2}\cos\tht\Bigr)^2\!
     {-}\Bigl(1{-}\frac{\str^2}{\DSr^2}\Bigr)
      \Bigl(1{-}\frac{\Ro^2}{\DSr^2}\Bigr)\biggr]^{\!\frac12}
\!\!\commae
\end{equation}
\begin{equation}
\begin{aligned}
\RTr\sin\Ath&=\str\sin\tht
\comma
\RTr\cos\Ath&=\frac{\str\cos\tht-\Ro}{\sqrt{1-\Ro^2/\DSr^2}}
\period
\end{aligned}
\end{equation}
\begin{equation}
\begin{aligned}
\str\sin\tht&=\RTr\sin\Ath
\commae\\
\str\cos\tht&=\RTr\cos\Ath\sqrt{1-{\Ro^2}/{\DSr^2}}+\Ro
\period
\end{aligned}
\end{equation}

\pagebreak
\coorsection{The null family}\label{asc:coorNull}
\begin{figure}[h]
\vspace*{-3pt}
\includegraphics{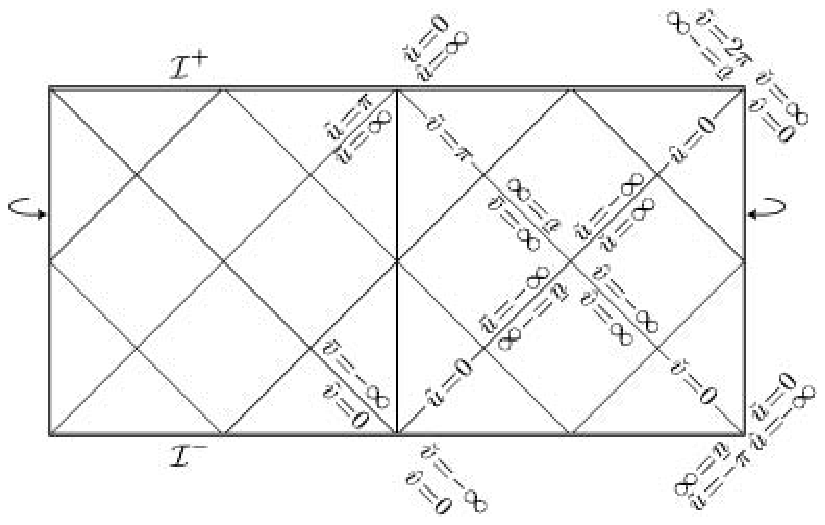}
\vspace*{-3pt}
\caption{\label{fig:coorNull}The null family of coordinates}
\end{figure}

Finally,  we return back to the coordinate systems which employ standard coordinates ${\tht}$, ${\ph}$.
Time and radial coordinates can be transformed into two null coordinates.
Such null coordinates can be associated with most coordinate families introduced above. Coordinates
${\tlu,\,\tlv}$ are related to the standard coordinates; ${\MVu,\,\MVv}$ and ${\MAu,\,\MAv}$
to the flat cosmological coordinates; ${\MOu,\,\MOv}$ to the conformally Minkowski;
and ${\bru,\,\brv}$ to the static coordinates.
Coordinate vectors ${\{\cvil{\tlu},\,\cvil{\tlv}\}}$, ${\{\cvil{\MVu},\,\cvil{\MVv}\}}$, etc.,
are the pairs of independent null vectors in the radial 2-slices ${\tht,\ph=\text{constant}}$.
We do not allow the radial coordinate to be negative in the definitions of null coordinates because this would
interchange the meaning of ${u}$ and ${v}$. The null coordinates are thus drawn in
the right half of Fig.~\ref{fig:coorNull} only.

\vspace*{-12pt}
\coorsubsection{Metric and relation to other coordinates}
\begin{gather}
    \mtrc=\frac{\DSr^2}{1-\cos(\tlu+\tlv)}
    \Bigl(-\grad\tlu\stp\grad\tlv+\bigl(1-\cos(\tlu-\tlv)\bigr)\,\sphmtrc\Bigr)\commae\\
    \mtrc=\frac{\DSr^2}{(\MAu+\MAv)^2}\,
    \Bigl(-2\,\grad\MAu\stp\grad\MAv+(\MAu-\MAv)^2\,\sphmtrc\Bigr)\comma\\
    \mtrc=\frac{\DSr^2}{(\MVu+\MVv)^2}\,
    \Bigl(-2\,\grad\MVu\stp\grad\MVv+(\MVu-\MVv)^2\,\sphmtrc\Bigr)\comma\,\\
    \mtrc=\Bigl(\frac{\DSr^2}{\DSr^2-\MOu\MOv}\Bigr)^2\,
    \Bigl(-2\,\grad\MOu\stp\grad\MOv+(\MOu-\MOv)^2\,\sphmtrc\Bigr)\commae\\
\begin{split}\raisetag{48pt}
    &\mtrc=\Bigl(\exp\frac\bru\DSr+\exp\frac\brv\DSr\Bigr)^{\!\!-2}\\
    &\,\times\!\!\Biggl(-2\exp\frac{\bru+\brv}\DSr\,\grad\bru\stp\grad\brv
    +\DSr^2\,\Bigl(\exp\frac\bru\DSr-\exp\frac\brv\DSr\Bigr)^2\,\sphmtrc\Biggr)\!\period
\end{split}
\end{gather}

The relation of time and radial coordinates ${\gent,\,\genr}$ to the corresponding null coordinates
${\genu,\,\genv}$ is given by usual formulas:
\begin{equation}
\begin{gathered}
  \gent={\textstyle\frac12}(\genv+\genu)\comma
  \genu=\gent-\genr\commae\\
  \genr={\textstyle\frac12}(\genv-\genu)\comma
  \genv=\gent+\genr\period
\end{gathered}
\end{equation}
Here ${\{\gent,\,\genr\}}$ stands for
${\{\tlt,\,\tlr\}}$, ${\{\MVt,\,\MVr\}}$, ${\{\MAt,\,\MAr\}}$, ${\{\MOt,\,\MOr\}}$, and ${\{\brt,\,\brr\}}$ respectively;
similarly with ${\{\genu,\,\genv\}}$.

\vspace*{-15pt}
\coorsubsection{Relation between null coordinates}

The coordinates  ${\MAu,\,\MAv}$,\; ${\MOu,\,\MOv}$, and ${\MVu,\,\MVv}$
can be viewed~as null coordinates in the conformally related Minkowski
spaces ${\MinkAspc}$, ${\MinkOspc}$, and ${\MinkVspc}$; 
these are shifted with respect to each other by ${\frac\pi2}$ in the direction
of the conformally Einstein time coordinate ${\tlt}$,
or associated null coordinates:\\[-3ex]
\begin{equation}\label{apx:Muv=tluv}
\begin{aligned}
\frac\MAu\DSr&=\tan\frac\tlu2\commae&
\frac\MAv\DSr&=\tan\frac\tlv2\commae\\
\frac\MOu\DSr&=\tan\Bigl(\frac\tlu2-\frac\pi4\Bigr)\commae&
\frac\MOv\DSr&=\tan\Bigl(\frac\tlv2-\frac\pi4\Bigr)\commae\\
\frac\MVu\DSr&=\tan\Bigl(\frac\tlu2-\frac\pi2\Bigr)\commae&
\frac\MVv\DSr&=\tan\Bigl(\frac\tlv2-\frac\pi2\Bigr)\period
\end{aligned}
\end{equation}
The remaining coordinates ${\bru,\,\brv}$ are related to
the conformally Einstein null coordinates ${\tlu,\,\tlv}$ by the \vague{compactification transformation}:
\begin{equation}\label{apx:tluv=bruv}
  \tan\frac\tlu2=\signu\exp\frac\bru\DSr\comma
  \tan\frac\tlv2=\signv\exp\frac\brv\DSr\period
\end{equation}
Here the sign factors ${\signu}$ and ${\signv}$ are given by
\begin{equation}\label{apx:signuv}
  \signu=\sign\tan\frac\tlu2\comma \signv=\sign\tan\frac\tlv2\period
\end{equation}
Relations \eqref{apx:Muv=tluv}, \eqref{apx:tluv=bruv} between null coordinates
can also be rewritten as follows:
\begin{gather}
  \tan\frac\tlu2
    =\signu\exp\frac\bru\DSr
    =\frac\MAu\DSr
    =-\frac\DSr\MVu
    =\frac{\DSr+\MOu}{\DSr-\MOu}
    \commae\\
  \tan\tlu
    =-\signu\sinh^{\!-1}\!\frac\bru\DSr
    =\frac{2\MAu\DSr}{\DSr^2-\MAu^2}
    =\frac{2\MVu\DSr}{\DSr^2-\MVu^2}
    =\frac{\MOu^2-\DSr^2}{2\MOu\DSr}
    \!\commae\notag\\
  \sin\tlu
    =\signu\cosh^{\!-1}\!\frac\bru\DSr
    =\frac{2\MAu\DSr}{\DSr^2+\MAu^2}
    =\frac{-2\MVu\DSr}{\DSr^2+\MVu^2}
    =\frac{\DSr^2-\MOu^2}{\DSr^2+\MOu^2}
    \commae\notag\\
  \cos\tlu
    =-\tanh\frac\bru\DSr
    =\frac{\DSr^2-\MAu^2}{\DSr^2+\MAu^2}
    =\frac{\MAu^2-\DSr^2}{\MAu^2+\DSr^2}
    =\frac{-2\MOu\DSr}{\DSr^2+\MOu^2}
    \commae\notag
\end{gather}
\begin{equation}
\frac\MAu\DSr=\tan\frac\tlu2=\signu\exp\frac\bru\DSr=-\frac\DSr\MVu=\frac{\DSr+\MOu}{\DSr-\MOu}\commae
\end{equation}
\begin{equation}
-\frac\MVu\DSr=\cot\frac\tlu2=\signu\exp\Bigl(-\frac\bru\DSr\Bigr)=\frac\DSr\MAu=\frac{\DSr-\MOu}{\DSr+\MOu}\commae
\end{equation}
\begin{equation}
\begin{aligned}
\frac\MOu\DSr&
  =-\frac{1-\sin\tlu}{\cos\tlu}
  =-\frac{\cos\tlu}{1+\sin\tlu}\\&
  =\Bigl(\tanh\frac\bru{2\DSr}\Bigr)^\signu
  =\frac{\MAu-\DSr}{\MAu+\DSr}
  =\frac{\DSr+\MVu}{\DSr-\MVu}
\commae
\end{aligned}
\end{equation}
\begin{equation}
\begin{aligned}
  \frac\bru\DSr&
    =\log\abs{\tan\frac\tlu2}
    =\log\abs{\frac\MAu\DSr}
    =\log\abs{\frac\DSr\MVu}\\&
    =\log\abs{\frac{\DSr+\MOu}{\DSr-\MOu}}
    =2\arctanh\Bigl(\frac\MOu\DSr\Bigr)^\signu
  \period
\end{aligned}
\end{equation}\\[-4ex]
\begin{equation}
\MAu\MVu=-\DSr^2\comma
\frac\MAu\DSr+\frac\DSr\MVu=0\period
\end{equation}
The same relations hold for coordinates ${v}$.

\pagebreak

%
%
%
%
%
%
%



\end{document}